\documentclass[%
reprint,
superscriptaddress,
longbibliography,
bibnotes,
 amsmath,amssymb,
 aps,
prb,
]{revtex4-2}
\usepackage[utf8]{inputenc}
\usepackage[english]{babel}
\usepackage[plainpages = false, pdfpagelabels, 
                 bookmarks,
                 bookmarksopen = true,
                 bookmarksnumbered = true,
                 breaklinks = true,
                 linktocpage,
                 colorlinks = true,
                 linkcolor = blue,
                 urlcolor  = blue,
                 citecolor = blue,
                 anchorcolor = green,
                 hyperindex = true,
                 hyperfigures
                 ]{hyperref} 
\usepackage{amsmath}
\usepackage{enumerate}
\usepackage{makecell,tabularx}
    
\usepackage{multirow}
\usepackage{tikz}
    \usetikzlibrary{patterns}
\usepackage{slashed}
\usepackage{physics}
\usepackage{bbold}
\usepackage{bm}
\usepackage{array,bigdelim,booktabs}
\usepackage[caption=false,singlelinecheck=off]{subfig}
\usepackage{blindtext}
\usepackage{bbm}
\usepackage{graphicx}
\usepackage{dcolumn}
\usepackage{bm}
\usepackage{romannum}
\AtBeginDocument{\pagenumbering{arabic}}
\usepackage{float}
\usepackage{mathrsfs}
\usepackage{mathtools}
\setcellgapes{3pt}
\definecolor{ballblue}{rgb}{0.13, 0.67, 0.8}
\definecolor{cborange}{HTML}{E69F00}
\definecolor{cbcyan}{HTML}{56B4E9}
\definecolor{cbgreen}{HTML}{009E73}
\definecolor{cbyellow}{HTML}{F0E442}
\definecolor{cbblue}{HTML}{0072B2}
\definecolor{cbred}{HTML}{D55E00}
\definecolor{cbmagenta}{HTML}{CC79A7}

\newcolumntype{Y}{>{\centering\arraybackslash}X}

\begin{document}
\title{\texorpdfstring{Thermal conductance and noise of Majorana modes along interfaced $\nu=5/2$ fractional quantum Hall states}{}}
\author{Michael Hein}
\affiliation{Department of Physics, University of Konstanz, D-78457 Konstanz, Germany}
\affiliation{Department of Microtechnology and Nanoscience (MC2),Chalmers University of Technology, S-412 96 G\"oteborg, Sweden}
\author{Christian Sp\r{a}nsl\"{a}tt}
\affiliation{Department of Microtechnology and Nanoscience (MC2),Chalmers University of Technology, S-412 96 G\"oteborg, Sweden}
  
\begin{abstract}
Identifying the topological order of the fractional quantum Hall state at filling $\nu=5/2$ is an important step towards realizing non-Abelian Majorana modes in condensed matter physics. However, to unambiguously distinguish between various proposals for this order is a formidable challenge. Here, we present a detailed study of transport along interfaced edge segments of fractional quantum Hall states hosting non-Abelian Majorana modes. With an incoherent model approach, we compute, for edge segments based on Pfaffian, anti-Pfaffian, and particle-hole-Pfaffian topological orders,  thermal conductances, voltage biased charge current noise, and delta-$T$ noise. We determine how the thermal equilibration of edge modes impacts these observables and identify the temperature scalings of transitions between regimes of differently quantized thermal conductances. In combination with recent experimental data, we use our results to estimate thermal and charge equilibration lengths in real devices.  We also propose an experimental setup which permits measuring several transport observables for interfaced fractional quantum Hall edges in a single device. It can, e.g., be used to rule out edge reconstruction effects. In this context, we further point out some subtleties in two-terminal thermal conductance measurements and how to remedy them. Our findings are consistent with recent experimental results pointing towards a particle-hole-Pfaffian topological order at filling $\nu=5/2$ in GaAs/AlGaAs, and provide further means to pin-point the edge structure at this filling and possibly also other exotic fractional quantum Hall states.
\end{abstract}
\date{\today}
\maketitle

\section{Introduction}
\label{sec:Intro}
Non-Abelian anyons are exotic excitations in two-dimensional condensed matter systems with no counterpart in particle physics~\cite{Stern2010Mar}. A system with non-Abelian anyons has a ground state degeneracy, and interchanging -- or ``braiding -- the anyons can shift the system between different ground states. This shift depends only on the order of exchanges and information about the braiding process is stored globally in the ground state wavefunction. Besides signifying a strongly correlated phase of matter, non-Abelian braiding forms the basis for the appealing, but so-far speculative, idea of topologically protected quantum computations~\cite{Kitaev2003Jan,Nayak2008}.

A most promising candidate system to host non-Abelian anyons is the fractional quantum Hall (FQH) state~\cite{Stormer1982,Laughlin1983} at filling $\nu=5/2$~\cite{Willet1987,Moore1991,Wen1991Feb,ReadGreen2000}. The quantum state that is realized at this filling is however still not fully understood and poses a long standing question in condensed matter physics (see Ref.~\cite{Ma2022Aug} for a recent overview). 

The purpose of this paper is to model FQH edge transport involving various candidate states at this filling, in light of recent experiments~\cite{dutta2022sep,Dutta2022} that have brought progress towards identifying the $5/2$ state. To date, the three most prominent candidate states are the Pfaffian (Pf)~\cite{Moore1991}, anti-Pfaffian (aPf)~\cite{Levin2007,Lee2007},
and particle-hole-Pfaffian (phPf)~\cite{Fidkowski2013,Son2015,Zucker2016,Antonic2018}. Whereas numerical calculations have seemed to favor the aPf state~\cite{Morf1998,Storni2010,Wojs2010Aug,Rezayi2011Mar,Zaletel2015Jan,Rezayi2017,Simon2020Jan}, tunneling experiments are more consistent with the aPf, or the Abelian SU(2)$_2$, 331, or 113 states~\cite{Radu2008,Lin2012,Lin2014}. 

\begin{figure}[t!]
    \centering
    \includegraphics[width=0.9\columnwidth]{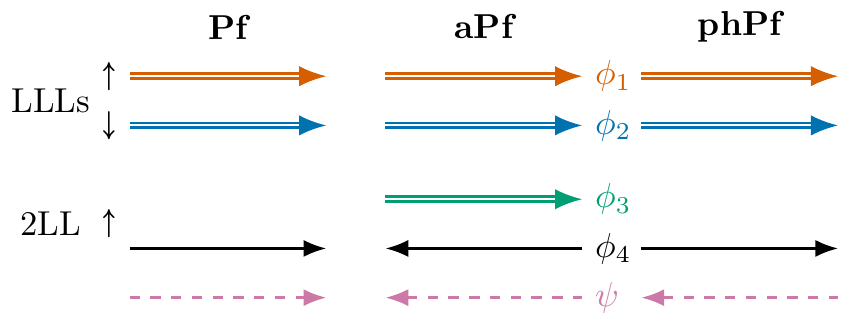}
    \caption{Edge structures for the Pfaffian (Pf), anti-Pfaffian (aPf), and particle-hole-Pfaffian (phPf) candidate states at filling $\nu=5/2$. Solid, double arrows (in red, blue, and green) describe integer modes labelled $\phi_1$, $\phi_2$, and $\phi_3$. Black, solid lines describe fractional $1/2$ bosonic modes $\phi_4$, and purple, dashed arrows are Majorana modes $\psi$. The edge modes belong to the two lowest Landau levels (LLLs) or the second (2LL) Landau level. Vertical arrows depict the expected Landau level spin polarizations.} 
    \label{fig:edge_structures}
\end{figure}

Although all three non-Abelian candidates are (by construction) compatible with the observed Hall conductance $G_H/(e^2/h)=\nu=5/2$, they have distinct bulk topological orders~\cite{Wen1990a}. Via the bulk-boundary correspondence, this order is in turn manifested by different edge structures, i.e., the number, chirality, and type of channels (or ``modes'') propagating around the FQH edge. As depicted in Fig.~\ref{fig:edge_structures},  the proposed edge structures of the Pf, aPf, and phPf states differ only in the second Landau level (2LL). This distinction can be quantified by the topological quantum number
\begin{align}
    \label{eq:nuQ}
    \nu_Q \equiv c-\bar{c},
\end{align}
where $c$ and $\bar{c}$ are the total central charges for the chiral and anti-chiral sectors, respectively, in the underlying conformal field theory~\cite{francesco2012conformal}. For Abelian FQH edges, $c$ ($\bar{c}$) equals the number of ``downstream'' (``upstream'') channels (where the downstream direction is defined as that of the equilibrated charge flow). For more exotic edge structures, $c$ and $\bar{c}$ may take rational values, e.g., $c=1/2$ for a single non-Abelian Majorana mode (MM). As can be seen in Fig.~\ref{fig:edge_structures}, $\nu_Q^{\rm Pf}=7/2$, $\nu_Q^{\rm aPf}=3/2$, and $\nu_Q^{\rm phPf}=5/2$. 

Quite remarkably, the abstract quantity~\eqref{eq:nuQ} can be related to the experimentally accessible edge thermal conductance $G^Q$ according to the relation~\cite{Kane1997,Capelli2002} 
\begin{align}
    \label{eq:GQ}
    G^Q = \nu_Q \kappa_0 \bar{T},
\end{align}
in which $\kappa_0 \bar{T} \equiv  \pi^2 k_{\rm B}^2/(3h) \bar{T}$ is the quantum of heat (with $k_{\rm B}$ and $h$ the Boltzmann and Planck constants, respectively, and $\bar{T}$ is the temperature).
\begin{table*}[t!]
    \centering
    \caption{Summary of results for the two terminal thermal conductance $G^Q_{\rm 2T}$, the thermal Hall conductance $G^Q_H$ as well as charge current noise and delta-$T$ noises for FQH interfaces phPf-$n$, aPf-$n$, and Pf-$n$, where $n \in \{2,3\}$. Expressions are given in the limits of vanishing, $\alpha, \beta \to 0$, and fully developed, $\alpha, \beta \to \infty$, thermal equilibration. We always assume full charge equilibration: $\delta \gg 1$ and no edge reconstruction. The units of noise are given in terms of voltage bias $\Delta V$ or large thermal bias $\Delta T\gg \bar{T}$. The two-terminal charge conductance $G_{\rm 2T}/(e^2/h)=1/2$ for all interfaces.}
    \label{tab:summary}
    \makegapedcells
    \setlength\tabcolsep{6pt}
    \begin{tabularx}{\linewidth}{*{7}{Y}}
        \Xhline{1pt}
        \multicolumn{7}{c}{\textbf{Interfaces with three modes:} equilibration parameters $\alpha$ and $\beta$} \\
        Interface & \multicolumn{4}{c}{$G^Q_{\rm 2T}$ $[\kappa_0 \Bar{T}]$} & \multicolumn{2}{c}{$G^Q_{\rm H}$ $[\kappa_0 \Bar{T}]$} \\
        \Xhline{0.2pt}
        & \makecell{$\alpha, \beta \to 0$} & \makecell{$\alpha \to \infty$ \\ $\beta \to 0$} & \makecell{$\alpha \to 0$ \\ $\beta \to \infty$} & \makecell{$\alpha, \beta \to \infty$} & \multicolumn{2}{c}{Universal value}\\
        \Xhline{0.8pt}
        phPf-$3$ & 5/2 & 1/2 & 3/2 & 1/2 & \multicolumn{2}{c}{1/2} \\
        aPf-$2$ and Pf-$3$ & 5/2 & 1/2 & 3/2 & 1/2 & \multicolumn{2}{c}{1/2} \\
        \Xhline{0.8pt}
        & \multicolumn{2}{c}{\makecell{Voltage biased charge current noise \\ $[\Delta V e^3/h]$}} & \multicolumn{2}{c}{\makecell{Downstream excess delta-$T$ noise \\ $[2 G_{\rm 2T} k_{\rm B} \Delta T]$}} & \multicolumn{2}{c}{\makecell{Upstream excess delta-$T$ noise \\ $[2 G_{\rm 2T} k_{\rm B} \Delta T]$}} \\
        \Xhline{0.2pt}
        & \makecell{$\alpha,\beta \to 0$} & \makecell{$\alpha,\beta \to \infty$} & \makecell{$\alpha, \beta \to 0$} & \makecell{$\alpha,\beta \to \infty$} & \makecell{$\alpha,\beta \to 0$} & \makecell{$\alpha,\beta \to \infty$}\\
        \Xhline{0.8pt}
        phPf-$3$ & $\simeq 0.113$ & $ \simeq \exp(-\delta)$ & $\simeq 0.865 $ & $\simeq 1$ & $\simeq 0.365$ & $\simeq \exp(-\delta)$\\
        aPf-$2$ and Pf-$3$ & $\simeq 0.152$ & $ \simeq 0.195$ & $\simeq 1.361$ & $ \simeq 1$ & $\simeq 0.438 $ & $\simeq 0.5$\\
        \Xhline{1pt}
        \multicolumn{7}{c}{\textbf{Interfaces with two modes:} equilibration parameter $\alpha$} \\
        & \multicolumn{4}{c}{$G^Q_{\rm 2T}$ $[\kappa_0 \Bar{T}]$} & \multicolumn{2}{c}{$G^Q_{\rm H}$ $[\kappa_0 \Bar{T}]$} \\
        \Xhline{0.2pt}
         & \multicolumn{2}{c}{$\alpha \to 0$} & \multicolumn{2}{c}{$\alpha \to \infty$} & \multicolumn{2}{c}{Universal value}\\
        \Xhline{0.8pt}
        phPf-$2$ & \multicolumn{2}{c}{3/2} & \multicolumn{2}{c}{1/2} & \multicolumn{2}{c}{1/2} \\
        aPf-$3$ and Pf-$2$ & \multicolumn{4}{c}{Universal: 3/2} & \multicolumn{2}{c}{3/2} \\
        \Xhline{0.8pt}
        & \multicolumn{2}{c}{\makecell{Voltage biased charge current noise \\ $[\Delta V e^3/h]$}} & \multicolumn{2}{c}{\makecell{Downstream excess delta-$T$ noise \\ $[2 G_{\rm 2T} k_{\rm B} \Delta T]$}} & \multicolumn{2}{c}{\makecell{Upstream excess delta-$T$ noise  \\ $[2 G_{\rm 2T} k_{\rm B} \Delta T]$}} \\
        \Xhline{0.2pt}
         & $\alpha \to 0$ & $\alpha \to \infty$ & $\alpha \to 0$ & $\alpha \to \infty$ & $\alpha \to 0$ & $\alpha \to \infty$ \\
        \Xhline{0.8pt}
        phPf-$2$ & $\simeq 0.086$ & $\simeq \exp(-\delta_{23})$   & $\simeq 0.899$ & $\simeq 1$ & $\simeq 0.338$ & $\simeq \exp(-\delta_{23})$ \\
        aPf-$3$ and Pf-$2$ & 0 & 0 & $1$ & $1$ & 0 & 0 \\
        \Xhline{1pt}
    \end{tabularx}
\end{table*}
Two-terminal conductance measurements for QH edges, first demonstrated in Ref.~\cite{Jezouin2013}, have subsequently been performed for many FQH states in both GaAs/AlGaAs~\cite{Banerjee2017,Banerjee2018,dutta2022sep,Melcer2022Jan,Melcer2023Jan} and graphene~\cite{Srivastav2019Jul,Srivastav2021May,LeBreton2022Sep,Srivastav2022Sep}. In particular, Ref.~\cite{Banerjee2018} reported $G^Q/(\kappa_0 \bar{T}) \approx 5/2$ at filling $\nu=5/2$, which fits well with Eq.~\eqref{eq:nuQ} for the phPf state (see Fig.~\ref{fig:edge_structures}). This value further rules out Abelian candidate states, since those are incompatible with half-integer $G^Q$. However, Eq.~\eqref{eq:GQ} is valid for the two-terminal conductance only when the heat transport is fully equilibrated, i.e., when edge channels exchange energy efficiently~\cite{Kane1995,Protopopov2017,Nosiglia2018,Aharon2019,Ma2020Jul,Srivastav2022Sep}. This equilibration can be quantified by a characteristic thermal equilibration length $\ell^Q_{\rm eq}$, so that the condition for Eq.~\eqref{eq:GQ} to hold reads $\ell^Q_{\rm eq} \ll L$, where $L$ is the edge length. Importantly, $\ell^Q_{\rm eq}$ is non-universal and depends on microscopic details such as inter-channel interactions, the edge disorder strength, and the temperature. It is worth pointing out that an analogous condition, $\ell^C_{\rm eq}\ll L$, with $\ell^C_{\rm eq}$ a characteristic charge equilibration length, is needed also for robust charge conductance quantization for FQH edges with counter-propagating modes ($c,\bar{c}\neq 0$). However, almost all FQH experiments to date indicate that this condition is normally well fulfilled (see Refs.~\cite{Lafont2019,Cohen2019} for exceptions).

The interpretation of the experiment in Ref.~\cite{Banerjee2018} as revealing a thermally equilibrated phPf edge was therefore questioned, since $G^Q/(\kappa_0 \bar{T}) \approx 5/2$ can be obtained also for a partially equilibrated aPf edge~\cite{Simon2018,Feldman2018,Asasi2020,Simon2020,Park2020Oct}. The same value can, under certain conditions, be obtained from models with random puddles of alternating non-Abelian orders~\cite{Mross2018,Wang2018,Biao2018,Zhu2020,Hsin2020,Fulga2020}, or from reentrant states due to Landau level mixing~\cite{Das2022Jun}. 

\begin{figure}[t!]
  \centering
    \includegraphics[scale=1]{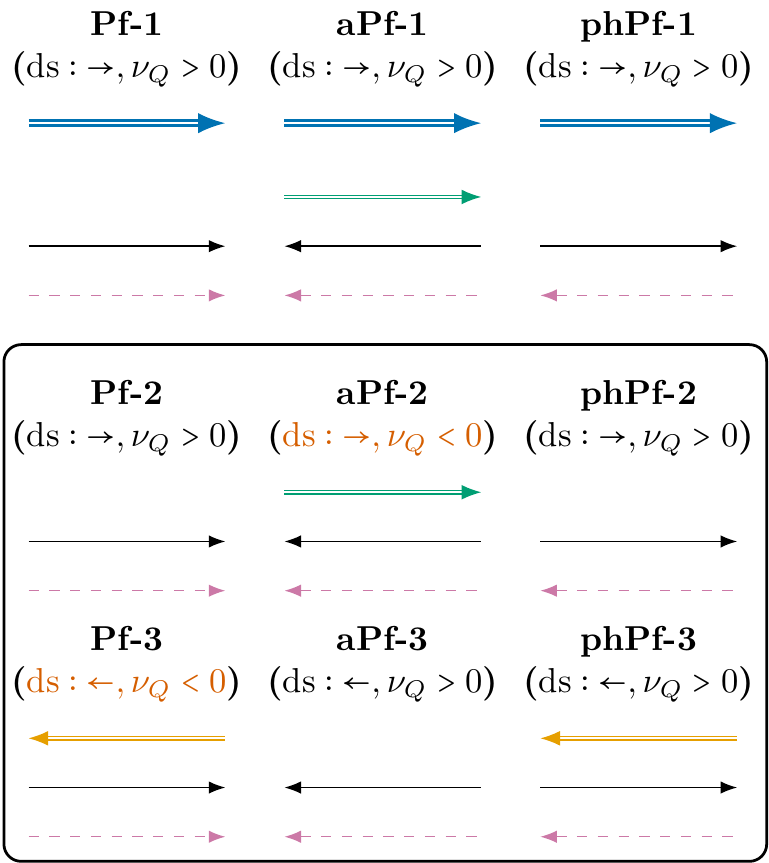}
    \caption{Effective structures of interfaces between non-Abelian $\nu=5/2$ candidate states and integer states $n=1,2,3$. In this work, we focus on $n=2,3$ (indicated by the frame), which are the integers that expose the second Landau level. The interfaces with orange text below their name are those for which the directions of equilibrated charge and heat transport are opposite.}
\label{fig:interfaced_edge_structures}
\end{figure}

Further progress has been made recently in GaAs/AlGaAs devices where the $\nu=5/2$ state is interfaced with integer QH states~\cite{dutta2022sep,Dutta2022}. This results in an effective $\nu=5/2-n$ edge, where $n=1,2,3$ (see Fig.~\ref{fig:interfaced_edge_structures}). The basic idea is that since all candidate states share two Abelian, integer edge channels (coming from two filled LLs), successive elimination of them exposes the remaining non-Abelian ``$\nu=1/2$ structure'' for which the states differ. This elimination occurs either due to Anderson localization~\cite{Giamarchi1988Jan,Gornyi2007Feb,Spanslatt2023Feb} or efficient equilibration of the edge states. For $n=3$, a possible particle-hole symmetry of the 2LL can be tested. In Ref.~\cite{dutta2022sep}, the thermal conductance for the $\nu=5/2-2$ and $\nu=5/2-3$ edges were both measured to $G^Q/(\kappa_0 \bar{T}) \approx 1/2$. It was argued that this is only possible if the underlying $\nu=5/2$ state is the phPf, thereby further strengthening the case for this state. It has also been proposed that charge conductance measurements of $5/2-n$ interfaces could distinguish between non-Abelian candidate states~\cite{Yutushui2022Jan}.

For the same type of interfaced structures but in another device, the authors in Ref.~\cite{Dutta2022} measured the excess noise for a current biased edge segment. Previously, it was proposed~\cite{Park2019,Spanslatt2019,Spanslatt2020,Park2020Oct} that such noise discloses important properties of the edge. More specifically, for full thermal and charge equilibration of any edge structure, the noise $S$ scales with $L$ in one of three possible classes: $S\simeq \exp(-L/\ell^C_{\rm eq})$ for $\nu_Q>0$, $S\simeq \sqrt{\ell^C_{\rm eq}/L}$ for $\nu_Q=0$, or $S\simeq const.$ for $\nu_Q<0$~\cite{Spanslatt2019}. This classification holds under conditions where heat leakage into the QH bulk is negligible. If the heat leakage is efficient, the noise is strongly suppressed in $L/\ell^C_{\rm eq}$ also for $\nu_Q=0$ and $\nu_Q<0$. By contrast, under conditions where the thermal equilibration between downstream and upstream modes (if present) is negligible, the noise scales as $S\simeq const.$, and absence of upstream modes implies identically $S=0$. This noise classification is the result of the chiral nature of the edge: For equilibrated charge transport, the voltage drop across a biased edge segment occurs only close to one of the contacts (in the so-called hot spot) whereas partitioning of charges, i.e., excess noise, is dominantly produced close to the other contact (in the so-called noise spot), see Fig.~\ref{fig:incoherent}. This partitioning is enhanced with respect to the equilibrium noise only in the presence of upstream modes that transport heat from the hot spot to the noise spot. In turn, this heat transport depends strongly on $\nu_Q$ which leads to the correspondence between $S$ and $\nu_Q$. 

For an interfaced Pf edge, the classification above implies that only the $5/2-3$ interface can generate excess noise, since only that interfacing results in upstream modes. For the aPf, both $5/2-1$ and $5/2-2$ can result in noise whereas $5/2-3$ cannot. Finally, for the phPf, all three interfaces host counter-propagating modes and can therefore produce finite noise. However, since $\nu_Q>0$ for the phPf interfaces~\cite{Comment}, the noise is exponentially suppressed in $L/\ell^C_{\rm eq}$ and finite noise should only emerge for poor thermal equilibration, i.e., either for small $L$ and/or large $\ell^C_{\rm eq}$. 

Indeed, Ref.~\cite{Dutta2022} reported finite excess noise for both $5/2-2$ and $5/2-3$ interfaces for short $L$, but the noise weakened significantly for larger $L$ (see Fig.~\ref{fig:lengths} below). At the same time, the two-terminal charge conductance of the interface, $G_{\rm 2T}$, was always measured to $G_{\rm 2T}/(e^2/h)=1/2$, indicating a well established charge equilibration between downstream and upstream (if present) modes.

Taken together, the two experiments~\cite{dutta2022sep,Dutta2022} suggest that the $\nu=5/2$ state in GaAs/AlGaAs is of phPf type. However, a detailed model of how the upstream mode mediated noise is generated for interfaced $\nu=5/2$ edges remains lacking. Moreover, the above interpretation of the noise measurements hinges on the absence of edge reconstruction~\cite{Wan2006,Wan2008,Overbosch2008,Zhang2014}. This effect introduces non-topological pairs of counter-propagating modes which complement the edge structure from the bulk boundary correspondence. The addition of such modes and conditions with poor thermal equilibration can result in noise generation for any FQH edge, which complicates experimental interpretations. 

In this work, we incorporate the qualitative noise and conductance analysis above in a comprehensive theoretical model which further permits a quantitative comparison with experimental data. To this end, we study transport along interfaces between non-Abelian and integer $n$ edges with the incoherent edge approach, recently developed in Refs.~\cite{Nosiglia2018,Park2019,Spanslatt2019,Spanslatt2020,Park2020Oct,Manna2022Dec}. We review the basics of this model in Sec.~\ref{sec:Model_Setup}. In Secs.~\ref{sec:Conductance_results} and Sec.~\ref{sec:Noise_Results}, we use the model to compute the thermal conductance and the noise, respectively, for interfaced edge structures.  We focus on $n=2,3$, since it is those integers that expose the 2LL structure, see Fig.~\ref{fig:interfaced_edge_structures}. A summary of the calculations in these sections is given in Tab.~\ref{tab:summary}. In Sec.~\ref{sec:thermal_equilibration_lengths}, we analyze the temperature scalings of the thermal equilibration lengths. This scaling permits us to analyze the temperature dependence of the noise and thermal conductance. In Sec.~\ref{sec:Discussion}, we compare our results to the experiments in Refs.~\cite{dutta2022sep,Dutta2022} and provide estimates on thermal and charge equilibration lengths. We then propose in Sec.~\ref{sec:Device_proposal} a unified experimental setup allowing several independent experiments for probing FQH edge structures in a single device. We argue that this device is beneficial for ruling out edge reconstruction effects as well as possible sample-to-sample differences between separate devices probing noise and the thermal conductance. We summarize and conclude our work in Sec.~\ref{sec:Summary_Outlook}. A number of technical calculations are delegated to Appendices~\ref{sec:MajoranaAppendix}-\ref{sec:Conversion}.

\section{Model of edge transport}
\label{sec:Model_Setup}
\subsection{Charge and energy transport}
\begin{figure}[t!]        
    \centering
    \includegraphics{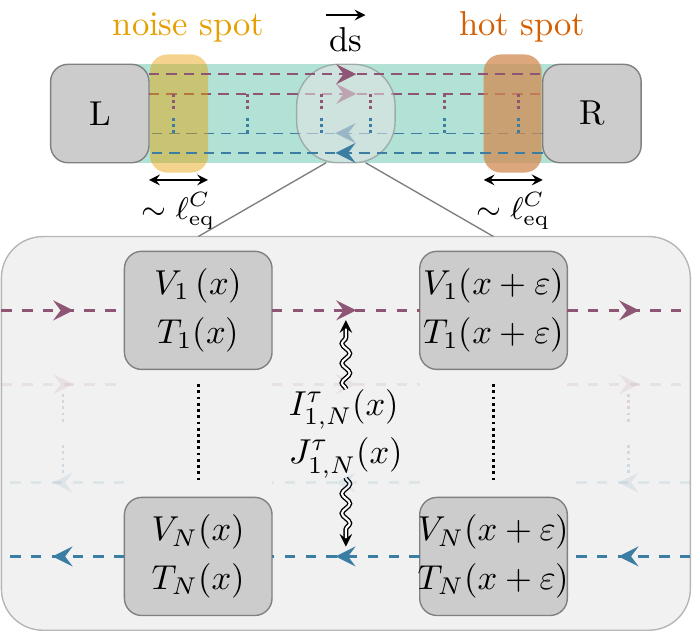}
    \caption{Schematics of the incoherent edge transport model for a single edge segment with two attached contacts, L and R. A set of $N$ edge channels are characterized by reservoirs with local voltages, $V_n(x)$, and temperatures, $T_n(x)$, for $n=1,\dots, N$. Differences in these quantities drive inter-channel charge and energy currents ($I_{n,m}^\tau$ and $J_{n,m}^\tau$ respectively) between the channels, which equilibrate the edge. For equilibrated, chiral charge transport (defining the downstream, ds, direction), Joule heating occurs only in the hot spot (red area) while noise is dominantly produced in the noise spot (yellow area). Equations~\eqref{eq:charge_transport_equation}-\eqref{eq:energy_transport_equation} are obtained in the continuum limit $N\rightarrow \infty$, $\varepsilon \rightarrow 0$, with $N/\epsilon = const.$}
    \label{fig:incoherent}
\end{figure}

Our starting point is the generic edge segment with two attached contacts, depicted in Fig.~\ref{fig:incoherent}. Charge transport along this segment is described by~\cite{Nosiglia2018,Spanslatt2020}
\begin{align}
    \label{eq:charge_transport_equation}
    \partial_x \vec{V}(x) = \mathcal{M}_V \vec{V}(x).
\end{align}
Here, $\vec{V}(x) \equiv \left[V_1(x),\hdots,V_N(x)\right]^T$ (with $[...]^T$ denoting vector transpose) describes the local voltages of $N$ edge channels and the matrix
\begin{align}
    \label{eq:MV}
    \mathcal{M}_V =  \begin{pmatrix} -\frac{\sum\limits_{n\neq1} (l^C_{1,n})^{-1}}{\chi_1 \nu_1} & \frac{(l^C_{1,2})^{-1}}{\chi_1 \nu_1} & \dots & \frac{(l^C_{1,N})^{-1}}{\chi_1\nu_1} \\[3pt] \frac{(l^C_{2,1})^{-1}}{\chi_2 \nu_2}  & -\frac{\sum\limits_{n\neq2} (l^C_{2,n})^{-1}}{\chi_2 \nu_2}  & \dots & \frac{(l^C_{2,N})^{-1}}{\chi_2 \nu_2} \\[3pt] \vdots & \vdots & \ddots & \vdots \\ \frac{(l^C_{N,1})^{-1}}{\chi_N\nu_N}  & \frac{(l^C_{N,2})^{-1}}{\chi_N \nu_N} & \dots & -\frac{\sum\limits_{n\neq N} (l^C_{N,n})^{-1}}{\chi_N \nu_N} \end{pmatrix}
\end{align}
describes couplings between edge channels in terms of the channel chiralities $\chi_i=\pm 1$ (with $+1$ and $-1$ corresponding to downstream and upstream directions, respectively),  filling factor discontinuities $\nu_i$, and charge equilibration lengths $l^C_{i,j}=l^C_{j,i}$ between modes $i$ and $j$. The microscopic content affecting these lengths can be obtained within a chiral Luttinger liquid approach, see e.g., Refs.~\cite{Ma2020Jul,Asasi2020,Srivastav2021May}. As follows, we always label the modes of all edge structures in Fig.~\ref{fig:interfaced_edge_structures} with $i=1$ starting from the top. 
The local electric currents $\vec{I}(x)\equiv\left[I_1(x),\hdots,I_N(x)\right]^T$, corresponding to $\vec{V}(x)$, obey a similar equation
\begin{equation}
    \label{eq:CurrentEquation}
    \partial_{x}\vec{I}(x) = \mathcal{M}_I \vec{I}(x), \qquad \mathcal{M}_I =   \mathcal{D}^{}_I \mathcal{M}_V  \mathcal{D}^{-1}_I,
\end{equation}
 with the diagonal matrix $\mathcal{D}_I = \text{diag}(\chi_1 \nu_1,\hdots,\chi_n \nu_n )$.
This description of edge currents was presented also in Ref.~\cite{Asasi2020} and can further be mapped onto the capacitive circuit model in Ref.~\cite{Fujisawa2022Apr}. Coherence effects between successive inter-channel scattering in all these descriptions are neglected (e.g., due to thermal dephasing), which is a prerequisite for robust conductance quantization~\cite{Kane1995,Protopopov2017,Nosiglia2018,Spanslatt2020}. The present model therefore describes incoherent transport along FQH edges, hence the name.

Similarly to the charge transport, edge energy transport is described by
\begin{equation}
    \label{eq:energy_transport_equation}
    \partial_x\vec{T^2}(x) = \mathcal{M}_T \vec{T^2}(x) + \delta\vec{V}(x),
\end{equation}
where $\vec{T^2}(x)=\left[T_1^2(x),\hdots,T_N^2(x)\right]^T$ are the local temperatures (squared), and the matrix
\begin{align}
    \label{eq:MT}
    \mathcal{M}_T =  \begin{pmatrix} -\frac{\sum\limits_{n\neq1} (l^Q_{1,n})^{-1}}{\chi_1 n_1} & \frac{(l^Q_{1,2})^{-1}}{\chi_1 n_1} & \dots & \frac{(l^Q_{1,N})^{-1}}{\chi_1n_1} \\[3pt] \frac{(l^Q_{2,1})^{-1}}{\chi_2 n_2}  & -\frac{\sum\limits_{n\neq2} (l^Q_{2,n})^{-1}}{\chi_2 n_2}  & \dots & \frac{(l^Q_{2,N})^{-1}}{\chi_2 n_2} \\[3pt] \vdots & \vdots & \ddots & \vdots \\ \frac{(l^Q_{N,1})^{-1}}{\chi_Nn_N}  & \frac{(l^Q_{N,2})^{-1}}{\chi_N n_N} & \dots & -\frac{\sum\limits_{n\neq N} (l^Q_{N,n})^{-1}}{\chi_N n_N} \end{pmatrix}
\end{align}
in which $l^Q_{i,j}=l^Q_{j,i}$ are the thermal equilibration lengths (see Appendix~\ref{sec:MajoranaAppendix} for an example). The final term in Eq.~\eqref{eq:energy_transport_equation} is the Joule heating contribution
\begin{align}
    \label{eq:Vdiff2}
    & \delta \vec{V}(x) = \\\notag &\frac{e^2}{h \kappa_0} \sum_{n=1}^N \left( \frac{[V_1(x)-V_n(x)]^2}{l_{1,n}^C \chi_1}, \dots, \frac{[V_N(x)-V_n(x)]^2}{l_{N,n}^C \chi_N} \right)^T,
\end{align}
 and originates from the voltage drops between the edge channels. In the matrix $\mathcal{M}_T$, the thermal conductance of the edge modes are described by the numbers $n_i$, which equal $1$ for Abelian edge channels and $1/2$ for Majorana edge channels. These numbers are related to Eq.~\eqref{eq:nuQ} as
\begin{align}
    & c = \sum_{i:\chi_i=+1} n_i, \\
    & \bar{c} = \sum_{i:\chi_i=-1} n_i,\\
    & \nu_Q = \sum_i \chi_i n_i.
\end{align}
The local temperatures are further related to local heat currents as
\begin{align}
    \label{eq:heat_currents}
    J_i(x)=\frac{n_i \kappa_0}{2}T^2_i(x).
\end{align}

The transport equations~\eqref{eq:charge_transport_equation} and \eqref{eq:energy_transport_equation} must be supplemented by boundary conditions that depend on the setup. Boundary conditions are discussed in more detail in Sec.~\ref{sec:Conductance_results}.


\subsection{\texorpdfstring{Charge current noise}{}}
We now focus on the regime of efficient charge equilibration, which we denote as $\delta \equiv L/\ell^C_{\rm eq}\gg 1$, where $L$ is the edge length and
\begin{align}
    \ell^C_{\rm eq}\equiv \text{max}(\ell^C_{i,j}),\quad \chi_i \times \chi_j = -1,
\end{align}
is defined as the largest equilibration length in the set of pairs of counter-propagating channels. As mentioned in Sec.~\ref{sec:Intro}, this condition is normally fulfilled on all FQH edges. Then, the excess charge current noise $S$, due to a voltage bias, in any of the two contacts (equal due to charge conservation), see Fig.~\ref{fig:incoherent}, is to good approximation given by~\cite{Park2019,Spanslatt2019,Kumar2022Jan}
\begin{align}
    \label{eq:noise_generation}
    S \simeq \frac{2e^2}{h \ell^C_{\rm eq}}\frac{ \nu_-}{ \nu_+} (\nu_+-\nu_-) \int_0^L dx\;e^{-\frac{2x}{\ell^C_{\rm eq}}} \Lambda(x).
\end{align}
A detailed derivation this formula can be found in Ch. 3 of Ref.~\cite{Hein2022}. In Eq.~\eqref{eq:noise_generation}, ``$\simeq$'' should here, and below, be understood as ``equal in the limit of very large $\delta$''.Furthermore,
\begin{align}
    & \nu_+ = \sum_{i:\chi_i=+1} \nu_i, \\
    & \nu_- = \sum_{i:\chi_i=-1} \nu_i,
\end{align}
are the total filling factor discontinuities of the downstream $(+)$ and upstream $(-)$ edge modes respectively. They satisfy the relations $\nu_+ > \nu_-$ and $\nu=\nu_+-\nu_-$, where $\nu$ is the effective filling factor of the edge structure. For a $5/2-n$ interface, we simply have $\nu=5/2-n$ for $\delta\gg 1$. The exponential suppression in Eq.~\eqref{eq:noise_generation} follows from the chiral nature of the edge~\cite{Park2019}. It implies that the noise generation is predominantly influenced by the region of size $\sim \ell^C_{\rm eq}$ close to the most upstream contact. We call this the noise spot.   

The key quantity in Eq.~\eqref{eq:noise_generation} is the local noise kernel
\begin{align}
    \label{eq:Lambda}
    \Lambda(x) \equiv \frac{S_{\rm loc}[\delta V(x),T_+(x),T_-(x)]}{2g_{\rm loc}[\delta V(x),T_+(x),T_-(x)]},
\end{align}
where $S_{\rm loc}$ and $g_{\rm loc}$ is the local dc noise and the (dimensionless) tunneling conductance, respectively. It is assumed that all downstream and upstream edge channels charge-equilibrate separately very efficiently, e.g., by emanating from the same (ideal) contact~\cite{Spanslatt2021Sep}. Importantly, both $S_{\rm loc}$ and $g_{\rm loc}$ depend on microscopic details of the edge such as inter-channel interactions, the edge disorder strength, the local voltage difference between the modes $\delta V(x)$, and the thermal equilibration-induced effective temperatures $T_\pm$ of downstream and upstream edge modes. Appendix~\ref{sec:NoisekernelAppendix} outlines how noise kernels are obtained from a bosonization approach. 

Our procedure to find the charge and heat flows along an edge segment is as follows. We first solve Eqs.~\eqref{eq:charge_transport_equation} and~\eqref{eq:energy_transport_equation} with suitable boundary conditions. These will depend on the type of setup. We then use these solutions  to compute charge and thermal conductances, or insert them
first into Eqs.~\eqref{eq:Lambda} and then \eqref{eq:noise_generation} to obtain the noise.  

\section{Thermal conductance}
\label{sec:Conductance_results}
\subsection{Two-terminal thermal conductance}
\label{sec:TTGQ}
\begin{figure}[t!]
  \centering
    \includegraphics[width=0.8\columnwidth]{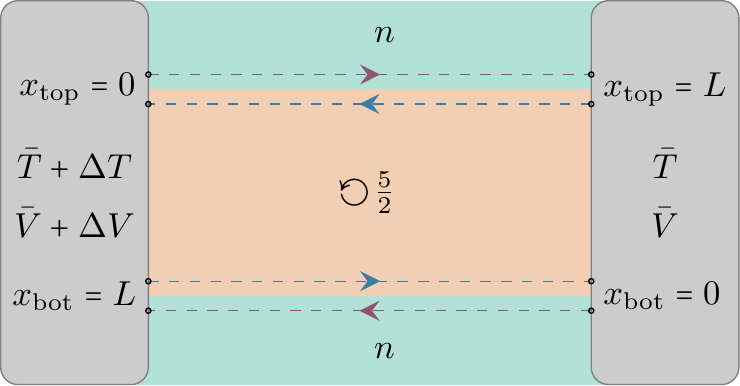}
    \caption{Schematic two-terminal setup for an interface between regions at filling $5/2$ and integers $n$. For $n<5/2$, the round arrow depicts the downstream direction (in which the charge current flows) while for $n>5/2$, it depicts the upstream direction. Swapping the magnetic field direction in the latter case restores the arrow direction to indicate the downstream direction.}
    \label{fig:conductance_setup}
\end{figure}

\begin{figure*}[t!]
    \centering
    \subfloat[]{\label{fig:Qconductance_phPfm2}\includegraphics[scale=1]{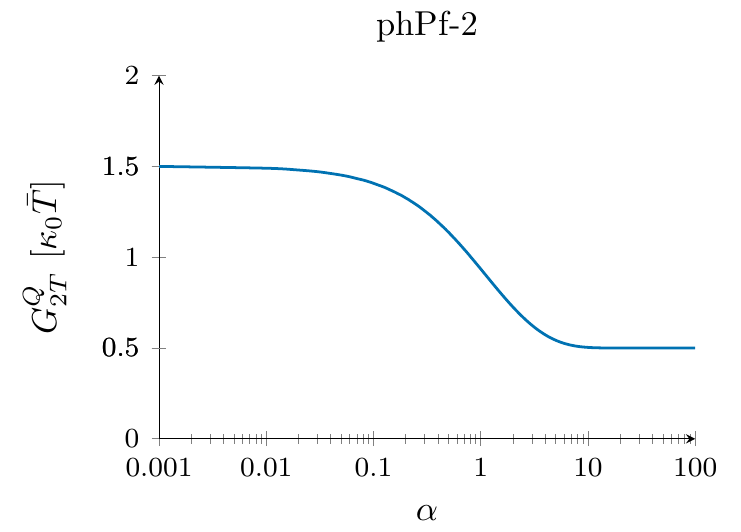}}
    \hspace{1.5cm}
    \subfloat[]{\label{fig:Qconductance_map_three}\includegraphics[scale=1]{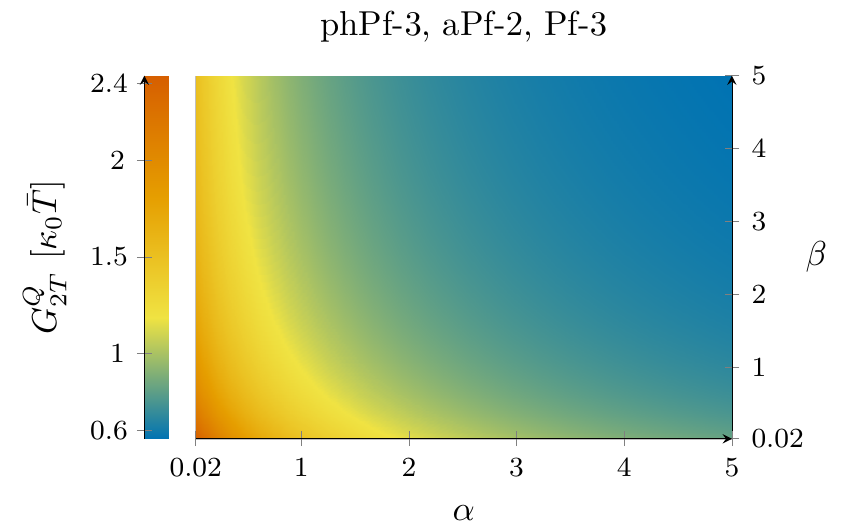}} 
    \caption{(a) Two-terminal thermal conductance $G^Q_{\rm 2T}$ (in units of the thermal conductance quantum $\kappa_0\bar{T}$) for the phPf-$2$ interface as a function of the degree of pairwise thermal equilibration $\alpha$ [see Eq.~\eqref{eq:alpha_beta_1}].  (b) $G^Q_{\rm 2T}$ as a function of the two equilibration parameters $\alpha$ and $\beta$ [see Eqs.~\eqref{eq:alpha_beta_1}-\eqref{eq:alpha_beta_2}] for interfaces phPf-$3$, aPf-$2$ and Pf-$3$.}
    \label{fig:GQPlots}
\end{figure*}

In this section, we compute the two-terminal thermal conductance for various $5/2-n$ (with $n=2,3$) interfaces by applying Eq.~\eqref{eq:energy_transport_equation} to the setup in Fig.~\ref{fig:conductance_setup}. To this end, we set the voltages in both contacts equal to zero: $\bar{V}=\Delta V=0$. The solutions to Eq.~\eqref{eq:charge_transport_equation} are then trivial: $\vec{V}(x)=0$. For the thermal transport, the boundary conditions for the  top edge segment read
\begin{subequations}
    \label{eq:DSBCS}
    \begin{align}
        &T_i(0)=\bar{T}+\Delta T, \ &\text{for} \quad &\chi_i=+1, \label{eq:BC1}\\
        &T_i(L)=\bar{T}, \ &\text{for} \quad &\chi_i=-1, \label{eq:BC2}
    \end{align}
\end{subequations}
and for the bottom segment we have
\begin{subequations}
    \label{eq:USBCS}
    \begin{align}
        &T_i(0)=\bar{T}, \ &\text{for} \quad &\chi_i=+1, \label{eq:BC3} \\
        &T_i(L)=\bar{T}+\Delta T, \ &\text{for} \quad &\chi_i=-1. \label{eq:BC4}
    \end{align}
\end{subequations}
We obtain the heat currents on the top and bottom edge segments by solving Eq.~\eqref{eq:energy_transport_equation} for
\begin{align}
    & J^{\rm top}_Q\equiv \sum_{i:\chi_i=+1} J_i(L) - \sum_{i:\chi_i=-1} J_i(L), \label{eq:JQ_top}\\
    & J^{\rm bot}_Q\equiv \sum_{i:\chi_i=+1} J_i(0) - \sum_{i:\chi_i=-1} J_i(0)\label{eq:JQ_bot}, 
\end{align}
with $J_i(x)$ given in Eq.~\eqref{eq:heat_currents} [see also Appendix A.9 in Ref.~\cite{Hein2022} for details in solving Eq.~\eqref{eq:energy_transport_equation}]. Next, by differentiating the total edge current with respect to the temperature difference $\Delta T$,  we obtain the two-terminal thermal conductance
\begin{align}
    \label{eq:GQ_2T}
    G^Q_{\rm 2T} \equiv \lim_{\Delta T \to 0} \left(\frac{d (J^{\rm top}_Q +J^{\rm bot}_Q ) }{d \Delta T} \right).
\end{align}
Let us start with the edge structures Pf-$2$, and aPf-$3$, which have only co-propagating modes. For both these interfaces, we readily find $G^Q_{\rm 2T}/(\kappa_0 \bar{T})=3/2$, independently of $L$ and $\ell^Q_{i,j}$, which follows from the fact that the heat exchanged between co-propagating channels never backscatters, and such processes can therefore not affect $G^Q_{\rm 2T}$.

For the other four interfaces aPf-$2$, Pf-$3$, phPf-$2$, and phPf-$3$, we consider only pairwise thermal equilibration between the counter-propagating channels, see Fig. \ref{fig:interfaced_edge_structures}. Thermal equilibration between co-propagating channels can be ignored, as discussed in the previous paragraph. For example, for the Pf-$3$ interface, we consider only the thermal equilibration between channel pairs $1$-$2$ and $1$-$3$. As we will show, the thermal conductance then depends on the degrees of thermal equilibration $L/\ell^Q_{i,j}$ between these pairs. To parameterize the equilibration,  we introduce two dimensionless equilibration parameters $\alpha$ and $\beta$ as
\begin{align}
    \label{eq:alpha_beta_1}
    \alpha = \frac{L}{\ell^Q_{1,2}} \quad \text{and} \quad \beta = \frac{L}{\ell^Q_{1,3}}
\end{align}
for the phPf-$2$, Pf-$3$, and aPf-$2$ interfaces and
\begin{align}
    \label{eq:alpha_beta_2}
    \alpha = \frac{L}{\ell^Q_{1,2}} \quad \text{and} \quad \beta = \frac{L}{\ell^Q_{2,3}}
\end{align}
for the phPf-$3$ structure (note the slight difference in the definition of $\beta$, due to their different structures). We plot the thermal conductances as functions of $\alpha$ and $\beta$ in Fig.~\ref{fig:GQPlots}. For the phPf-$2$ interface, Fig.~\ref{fig:Qconductance_phPfm2} shows that $G^Q_{\rm 2T}$ has a step-like behavior and transitions from $G^Q_{\rm 2T}/(\kappa_0 \bar{T})=1/2\to 3/2$ with decreasing $\alpha$ (there is no $\beta$-parameter for this interface). For large $\alpha$ the thermal transport is effectively mediated by a single, ``collective mode'' with thermal quantum number $n_{\rm tot}=1-1/2=1/2$. For small $\alpha$, the two edge channels are essentially decoupled and heat transport occurs in both directions along the edge. The thermal quantum numbers of the channels then add up, $n_{\rm tot}=1+1/2=3/2$, in their contribution to $G^Q_{\rm 2T}$.

The thermal conductance of aPf-$2$, Pf-$3$ and phPf-$3$, is depicted in Fig.~\ref{fig:Qconductance_map_three}. For these interfaces, we see that, depending on which pair of modes that equilibrates most efficiently, the thermal conductance approaches different limits. More specifically we have
\begin{align}
    \label{eq:GQintermediate}
    \begin{aligned}
        \lim \limits_{\substack{\alpha \to \infty \\ \beta \to \infty}} &G^Q_{\rm 2T} = \frac{1}{2} \kappa_0 \bar{T}, \quad \lim \limits_{\substack{\alpha \to \infty \\ \beta \to 0}} &G^Q_{\rm 2T} = \frac{1}{2} \kappa_0 \bar{T}, \\
        \lim \limits_{\substack{\alpha \to 0 \\ \beta \to \infty}} &G^Q_{\rm 2T} = \frac{3}{2} \kappa_0 \bar{T}, \quad \lim \limits_{\substack{\alpha \to 0 \\ \beta \to 0}} &G^Q_{\rm 2T} = \frac{5}{2} \kappa_0 \bar{T} .
    \end{aligned}
\end{align} 
Similar to the phPf-$2$ interface, the maximum value of the conductance is obtained when there is no thermal equilibration and the contributions of all edge channels along the interface add up: $n_{\rm tot}=1+1+1/2=5/2$, as expected. The value $G^Q_{\rm 2T}/(\kappa_0 \bar{T}) = 1/2$ is generated when all edge channels being fully equilibrated ($\alpha,\beta\rightarrow \infty$) and $n_{\rm tot}=1-1+1/2=1/2$, in accordance with Eq.~\eqref{eq:GQ}, up to corrections exponentially small in $L$. Alternatively, $G^Q_{\rm 2T}/(\kappa_0 \bar{T}) = 1/2$ is produced for $\alpha\rightarrow \infty, \beta\rightarrow 0$, where the edge structure becomes a  decoupled MM and a pair of two strongly equilibrated bosons: $n_{\rm tot}=1/2+0$. The zero here corresponds to a diffusive correction which vanishes as $ \sim \alpha^{-1}$ (see, e.g., the Supplemental Material of Ref.~\cite{Spanslatt2019} for a detailed discussion). In contrast, $G^Q_{\rm 2T}/(\kappa_0 \bar{T}) = 3/2$ is produced by two decoupled collective modes generating $n_{\rm tot} = 1/2+1=3/2$. The limits in Eq.~\eqref{eq:GQintermediate} are clearly idealized, and real devices have finite values of  $\alpha,\beta$. However to estimate the relative magnitude of $\alpha$ and $\beta$ requires detailed microscopic information about inter-channel energy exchange mechanisms, which generically depend on e.g., spin, orbital, or valley degrees of freedom of the LLs. Incorporating these effects is a challenging problem which, however, lies beyond the scope of the present work. 

We end this subsection by pointing out that the experimental value of $G^Q_{\rm 2T}/(\kappa_0 \bar{T}) = 1/2$ for both $5/2-2$ and $5/2-3$ can, at least in principle, be generated from a state other than the phPf. Consider an edge structure similar to the aPf edge but with the MM direction reversed. Then, for full thermal equilibration $G^Q_{\rm 2T}/(\kappa_0 \bar{T}) = 1/2$ is indeed produced for both $n=2$ and $n=3$ ($n_{\rm tot}=1-1+1/2=1/2$ and $n_{\rm tot}=1-1/2=1/2$, respectively). The non-equilibrated limits for those interfaces are however $G^Q_{\rm 2T}/(\kappa_0 \bar{T}) = 5/2$ and $G^Q_{\rm 2T}/(\kappa_0 \bar{T}) = 3/2$, respectively, which stand in contrast to the phPf which has $G^Q_{\rm 2T}/(\kappa_0 \bar{T}) = 3/2$ and $G^Q_{\rm 2T}/(\kappa_0 \bar{T}) = 5/2$ in this limit. 

\subsection{Thermal Hall conductance}
Equation~\eqref{eq:GQ_2T} describes the two-terminal thermal conductance which we have shown to significantly depend on the degree of thermal equilibration. Only for efficient thermal equilibration does $G^Q_{\rm 2T}$ take the universal value as specified in Eq.~\eqref{eq:GQ}. Note that in Eq.~\eqref{eq:GQ_2T}, the two edge current contributions are added (cf., Fig.~\ref{fig:conductance_setup}). By instead subtracting the two edge currents, one can define a thermal Hall conductance as
\begin{align}
    \label{eq:GQ_H}
    G^Q_{\rm H} \equiv \lim_{\Delta T \to 0} \left(\frac{d (J^{\rm top}_Q -J^{\rm bot}_Q ) }{d \Delta T} \right).
\end{align}
In the most general case, the top and bottom edge heat currents depend on all mutual equilibration lengths between all pairs of counter-propagating modes. However, within the model in Sec.~\ref{sec:Model_Setup}, we prove (see Appendix~\ref{sec:thermal_Hall_appendix} for the proof) that for any edge structure, universality emerges
\begin{align}
    \label{eq:GQH}
    G^Q_{\rm H}/(\kappa_0 \bar{T}) = c-\bar{c}=\nu_Q, 
\end{align}
provided the degrees of \textit{equilibration on the top and bottom edge segments are equal}, even if they are poor. Such a situation can, e.g., be achieved by designing devices with equal top and bottom edge segment lengths. Indeed, such a setup was presented in Ref.~\cite{Melcer2023Jan}, in which the thermal Hall conductance was measured at $\nu=2/3$ and was found to be in good agreement with the expected value $G^Q_{\rm H}/(\kappa_0 \bar{T})=0$. At the same time, it was found that, $G^Q_{\rm 2T}/(\kappa_0 \bar{T})\approx 0.82$, indicating an incomplete thermal equlibration. In the present context, Figs.~\ref{fig:edge_structures}-\ref{fig:interfaced_edge_structures} suggests that a Hall type of thermal conductance measurement for interfaces $5/2-n$ would give $G^Q_{\rm H}/(\kappa_0 \bar{T})=7/2-n$, $3/2-n$, and $5/2-n$ for the Pf, aPf, and phPf edges, respectively. A thermal Hall measurement therefore unambiguously distinguishes between these three edge structures. This conclusion holds even in the absence of interfacing, since already $n=0$ results in different $G^Q_{\rm H}$.

Let us end our treatment of the thermal conductance by emphasizing that for edge states with co-propagating channels, there is no significant advantage gained from a thermal Hall measurement. By contrast, for edges with counter-propagating channels, $G^Q_{\rm H}$ provides information directly related to the state's topological order, i.e., $\nu_Q$, independent of the thermal equilibration (which might be hard to accurately control experimentally). Finally, we point out that it is simple to check, that within our model, corresponding conclusions hold for a similar definition of the charge Hall conductance as well, i.e.,
\begin{align}
    \label{eq:G_H_def}
    G_{\rm H} \equiv \frac{I_{\rm top}-I_{\rm bot}}{\Delta V} = \nu\frac{e^2}{h},
\end{align}
when top and bottom charge equilibrations are identical. Here, $I_{\rm top/bottom}$ are defined in perfect analogy with Eqs.~\eqref{eq:JQ_top}-\eqref{eq:JQ_bot}.

\label{sec:Noise_generation}
\section{Noise generation on interfaced edges}
\label{sec:Noise_Results}
\begin{figure}[t!]
    \centering
    \begin{tabular}{c}
        \subfloat{\label{fig:noise_setup_1} \includegraphics[scale=1]{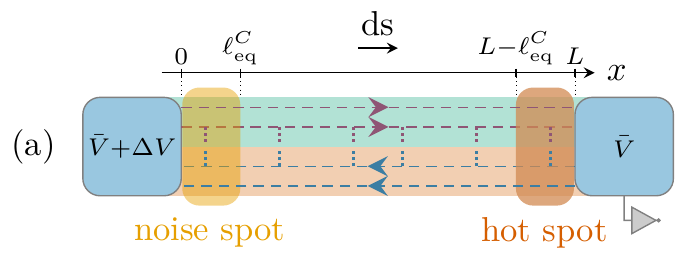}} \\
        \subfloat{\label{fig:noise_setup_2} \includegraphics[scale=1]{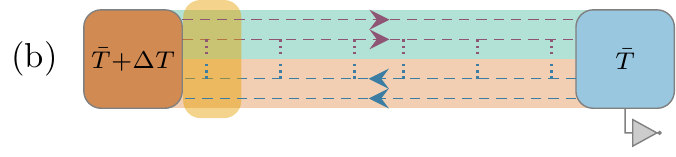}} \\
        \subfloat{\label{fig:noise_setup_3} \includegraphics[scale=1]{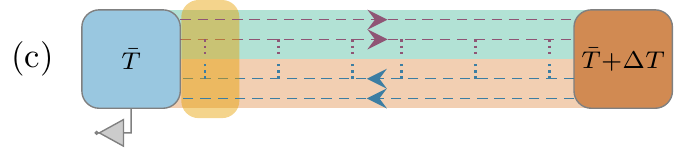}}
    \end{tabular}
    \caption{Schematic setups for measuring charge current noise (in the amplifiers depicted in gray) of a single edge segment in response to (a) a voltage bias, (b) a temperature bias applied in the ``downstream'' (ds) or (c)  ``upstream'' (us) direction.}
    \label{fig:noise_setups}
\end{figure}

In this section, we compute the noise generated on a single interfaced edge segment with two attached contacts. Such a setup has been realized in experiments, see, e.g., Refs.~\cite{Kumar2022Jan,Melcer2022Jan,Dutta2022}. We are here interested in three cases: First, in Sec.~\ref{sec:deltaV_noise}, we take the two contacts to have equal temperatures and impose a voltage bias. This is illustrated in Fig~\ref{fig:noise_setup_1}. We then move on to the case with no applied voltage bias, but let the two contacts have different temperatures. Here, we first take the most upstream contact (the left one in the figure) to be the hot one, see Fig~\ref{fig:noise_setup_2}, and compute in Sec.~\ref{sec:deltaT_noise_DS} the excess noise in the colder, downstream contact. We call this downstream delta-$T$ noise. Finally, in Sec.~\ref{sec:deltaT_noise_US} we consider the situation in which the hot contact lies most downstream and the noise is measured in the upstream, colder contact, see Fig~\ref{fig:noise_setup_3}. We call this upstream delta-$T$ noise.
\subsection{Voltage biased edge segment}
\label{sec:deltaV_noise}
For an applied voltage bias $\Delta V$, the injected downstream charge current dissipates heat only close to one of the contacts, when the charge equilibration is efficient (the hot spot location is independent of the voltage bias direction~\cite{Park2019}). We assume this in the following, which further implies that in the noise spot region, all charged edge modes equilibrate to the same electrochemical potential~\cite{Park2019,Spanslatt2019}. We can then set $\delta \vec{V}(x\lesssim \ell^C_{\rm eq})\approx \vec{0}$ [see Eq.~\eqref{eq:Vdiff2}], with negligible corrections $\sim \exp[-L/\ell^C_{\rm eq}]\ll 1$~\cite{Park2019}. We further assume that $e\Delta V\gg k_{\rm B}\bar{T}$, so that in this subsection, we may set the base temperature $\bar{T}$ to zero. Our computed noise then amounts to the excess noise. 
\begin{figure*}[ht!]
    \centering
    \begin{tabular}{ccc}
        \hspace{1cm}\textbf{Efficient Thermal Equilibration} & & \hspace{1cm}\textbf{No Thermal Equilibration} \\
        \subfloat[]{\label{fig:Noise_Vcomp_full}\includegraphics{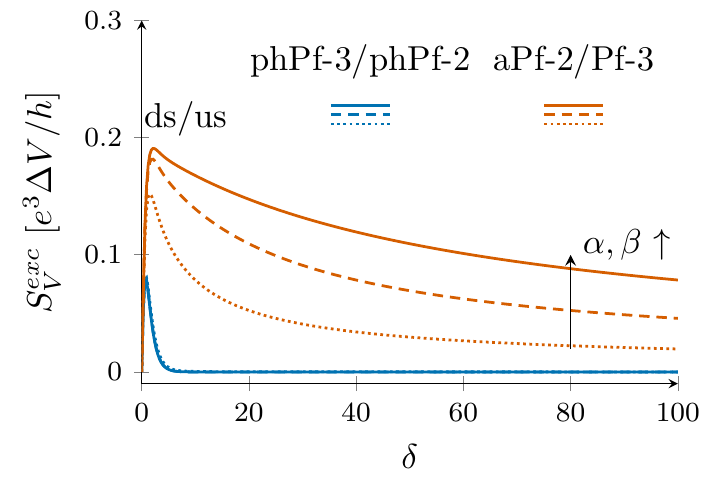}} & \hspace{1cm} & \subfloat[]{\label{fig:Noise_Vcomp_abs}\includegraphics{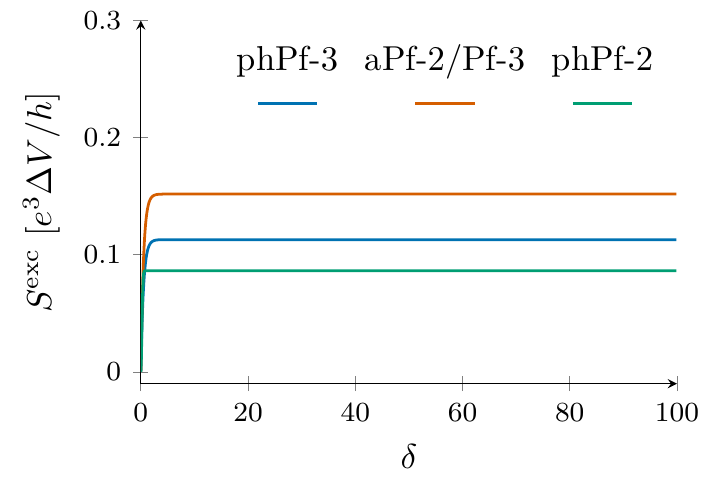}} \\
        \subfloat[]{\label{fig:Noise_Tcomp_full}\includegraphics[scale=1]{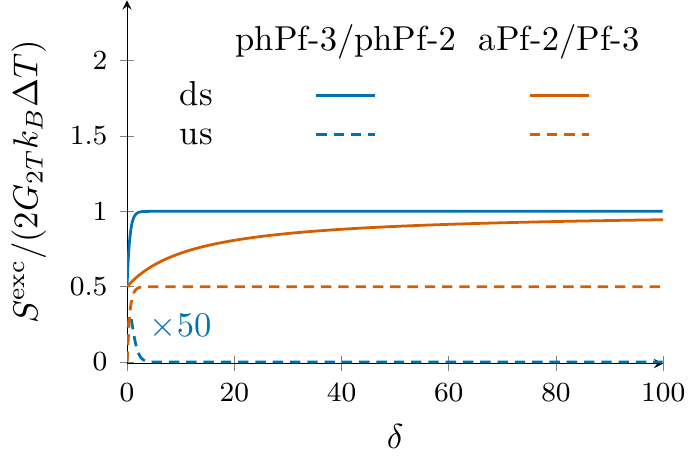}} & \hspace{1cm} & \subfloat[]{\label{fig:Noise_Tcomp_abs}\includegraphics[scale=1]{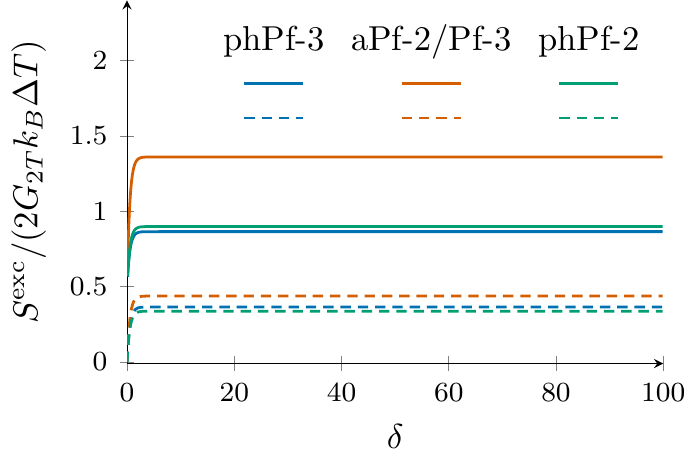}}
    \end{tabular}
    \caption{ (a) Excess (basis temperature $\bar{T}=0$) noise $S^{\rm exc}$ in units of $e^3\Delta V/h$, where $\Delta V$ is the bias voltage. The noise is computed vs the charge equilibration parameter $\delta = L/\ell^C_{\rm eq}$ for parameters causing efficient efficient thermal equilibration: $(\alpha, \beta) = \{(250, 250), (100, 100), (25, 25) \}$ [see Eqs.~\eqref{eq:alpha_beta_1}-\eqref{eq:alpha_beta_2}], for solid, dashed, and dotted lines, respectively. (b) same as (a) but in the limit of absent thermal equilibration $\alpha,\beta \rightarrow 0$. (c) Excess noise in the presence of a temperature gradient $\Delta T$ in downstream (ds) or upstream (us) direction for very efficient thermal equilibration with $(\alpha,\beta) = (100,100)$. (d) same as (c) but for absent thermal equilibration. For the phPf-$2$ interface, we have used $\delta=\delta_{23}$, see Fig.~\ref{fig:noise_phPfm2}}
    \label{fig:noisecomp}
\end{figure*}
Such excess noise is generated only if heat from the hot spot can propagate upstream (see Fig.~\ref{fig:noise_setup_1}). This possibility depends in turn on the edge structure and on how well the edge channels thermally equilibrate. This feature is captured within our model in Eq.~\eqref{eq:noise_generation}. As follows, we consider the two limits of either very efficient or very poor thermal equilibration. 

\subsubsection{Efficient thermal equilibration}
\label{sec:DeltaVThermalEq}
For efficient thermal equilibration, we start by solving Eq.~\eqref{eq:charge_transport_equation} for the boundary conditions $V_i(0)=\Delta V$ (we set $\bar{V}=0$) for $\chi_i=+1$ and $V_i(L)=0$ for $\chi_i=-1$. This is done for all the charged edge channels. From the solution, we extract the voltage drops~\eqref{eq:Vdiff2}, which we then insert into Eq.~\eqref{eq:energy_transport_equation}, which is finally solved with the boundary conditions $T_i(0)=T_i(L)=0$. We give examples of voltage and temperature profiles in Appendix~\ref{sec:Profile_Appendix}. The resulting temperature profiles are used in Eq.~\eqref{eq:noise_generation}. This integral is dominated by the temperatures at the noise spot, i.e., $T_i(x\lesssim\ell^C_{\rm eq})$. Moreover, due to the efficient thermal equilibration, the channel temperatures are similar in this region. Performing the integration in Eq.~\eqref{eq:noise_generation}, these two features result in noise on Nyquist-Johnson (NJ) form~\cite{Nyquist1928Jul,Johnson1928Jul}
\begin{align}
   S \simeq \frac{2e^2}{h}\frac{ \nu_-}{ \nu_+} (\nu_+-\nu_-)k_{\rm B} T_{\rm ns},
\end{align}
in terms of an effective noise spot temperature $T_{\rm ns}$. This temperature depends on the voltage bias $\Delta V$ and the way $T_{\rm ns}$ depends on $L$ stands is in one-to-one correspondence to the three possible cases of a thermally equilbirated edge:  $\nu_Q>0$, $\nu_Q<0$, and $\nu_Q=0$~\cite{Spanslatt2019}. This classification of the noise was described in detail in Sec.~\ref{sec:Intro}.

We first analyze the noise generated on the Pf-$2$, and aPf-$3$ interfaces. For these, we have that the noise vanishes identically, $S=0$, for any value of $\delta$ and $\Delta V$. This happens simply because there are no upstream modes present that can transport heat to the noise spot, and thus both $\nu_-=0$ and $\Lambda(x)=0$ (the latter equality holds because we set $\bar{T}=0$).

The other four interfaces have $\bar{c}\neq 0$, but the aPf-$2$ and Pf-$3$ have $\nu_Q<0$ whereas phPf-$2$ and phPf-$3$ have $\nu_Q>0$. For these four interfaces, we plot the noise $S$ vs $\delta$ in Fig.~\ref{fig:Noise_Vcomp_full}. We see that the noise approaches a constant value $S\simeq const.$ for aPf-$2$ and Pf-$3$ but decays exponentially $S\simeq \exp(-\delta)$ for phPf-$2$ and phPf-$3$. We also see that with increasing thermal equilibration (increasing $\alpha,\beta$) the noise is overall suppressed for $\nu_Q>0$, because the heat that reaches the noise spot is suppressed even further. In contrast, when $\nu_Q<0$, increasing thermal equilibration leads to enhanced noise since, in this case, there is more heat propagating upstream to the noise spot. 

\subsubsection{Absent thermal equilibration}\label{eq:absent_thermal}
For absent thermal equilibration, downstream and upstream edge modes are generally at very different temperatures at the noise spot. To estimate these temperatures, we follow the approach in Ref.~\cite{Melcer2022Jan} and model the hot spot as a point-like heat source with power~\cite{Spanslatt2020}
\begin{align}
    \label{eq:Joule_Power}
    P_0 = \frac{e^2 \Delta V^2}{2 h} \frac{\nu_- (\nu_+ - \nu_-)}{\nu_+}.
\end{align}
We next assume that $P_0$ is equally divided among all edge channels. The temperature of the modes propagating upstream from the hot spot is thus approximated as
\begin{align}
    \label{eq:TmApprox}
    T_- \approx \sqrt{\frac{\bar{c}}{c+\bar{c}}} \sqrt{\frac{2 P_0}{\kappa_0}}.
\end{align}
Since we have set the temperature in the contacts to $\bar{T}=0$, the downstream modes will have zero temperature at the noise spot
\begin{align}
    \label{eq:TpApprox}
    T_+\approx 0.
\end{align}
Within the approximations~\eqref{eq:TmApprox}-\eqref{eq:TpApprox}, we compute noise kernels $\Lambda(x)$ in terms of finite and zero temperature bosonized Green's functions. Details of these calculations are given in Appendix~\ref{sec:NoisekernelAppendix}. 

We present the thermally non-equilibrated noise in Fig.~\ref{fig:Noise_Vcomp_abs}. Similarly to the case of efficient thermal equilibration, the Pf-$2$, and aPf-$3$ do not generate any excess noise. For the other interfaces, the noises are essentially constant as functions of $\delta$, since the heat reaching the noise spot is now independent of this parameter. The relative magnitude between the interfaces follow from different noise spot temperatures due to differing pre-factors in Eqs.~\eqref{eq:noise_generation} and~\eqref{eq:TmApprox}. 
\begin{figure}[b]
    \centering
    \includegraphics{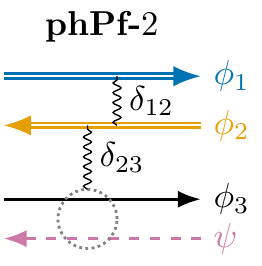}
    \caption{Edge structure of phPf-$2$ with an additional pair of counter-propagating integer modes. These modes are responsible for Joule-heating in the hot spot and charge partitioning in noise spot.}
    \label{fig:noise_phPfm2}
\end{figure}

The case of the phPf-$2$ interface warrants here an extra discussion. The noise of this interface was a crucial ingredient in the interpretation of the experiment in Ref.~\cite{Dutta2022}. Figure.~\ref{fig:edge_structures} shows that this structure only has a single downstream charge mode. Hence, no partitioning would be possible and no excess noise could be generated. However, by recalling that on sufficiently small length-scales, $x \ll \ell^C_{\rm eq}$, there are in fact counter-propagating integer modes close to both source and drain contacts. These modes are responsible both for Joule-heating and charge partitioning. To correctly describe the noise, we include these modes, see Fig.~\ref{fig:noise_phPfm2}.  This procedure introduces two charge equilibration lengths between each pair of counter-propagating modes. We parametrize them as $\delta_{12} \equiv L/\ell^C_{12}$ and $\delta_{23} \equiv L/\ell^C_{23}$ in the channel basis $\{\phi_1$,$\phi_2,\phi_3,\psi\}$ with $\chi=\{+1,-1,+1,-1\}$ and $\nu=\{1,1,1/2,0\}$.

As our next step, we assume that the charge equilibration of the integer channels is much faster than their equilibration with the bosonic 1/2 mode. In terms of our equilibration parameters, this amounts to taking $\delta_{12} \gg \delta_{23}$. The main contribution to the Joule-heating then comes from the equilibration of integer channels. This limit is consistent with a quantized expected charge conductance, since we have that
\begin{align}
    \lim\limits_{\begin{subarray}{l} \delta_{12} \to \infty \\ \delta_{23} \to 0 \end{subarray}} G_{\rm 2T} = \frac{1}{2} \frac{e^2}{h},
\end{align} 
which was indeed observed experimentally for the $5/2-2$ interface~\cite{Dutta2022}. With this implementation, our model describes how the phPf-$2$ interface generates noise in the thermally non-equilibrated limit, see Fig.~\ref{fig:Noise_Tcomp_abs}. We note that this noise is similar in magnitude as the phPf-$3$, in accordance with experimental observations~\cite{Dutta2022}, see also the discussion in Sec.~\ref{sec:Discussion} below. We further note that the outlined procedure bares a similarity with implementing edge-reconstruction~\cite{Wan2006,Wan2008,Overbosch2008,Zhang2014}. Effects of edge-reconstruction are discussed in more detail in Sec.~\ref{sec:Device_proposal}.

\subsection{Downstream delta-T noise}
\label{sec:deltaT_noise_DS}
Here, we consider the situation where the two contacts are at the same potential, but at different temperatures. Excess charge current noise generated only by a temperature gradient and in the absence of an average charge current, is known as ``thermally activated shot noise'' or ``delta-$T$ noise''~\cite{Lumbroso2018}. This noise has attracted considerable attention lately (for recent work on delta-$T$ noise for weak tunneling between FQH edges, see Refs.~\cite{RechMartinPRL20,Schiller2022,Zhang2022,Rebora2022Dec}). Moreover, delta-$T$ noise is, in fact, at play in two-terminal thermal conductance measurements when the central contact is at zero potential (see  Sec.~\ref{sec:Device_proposal} below).

We first take the hot contact to lie upstream from the cold contact, see Fig.~\ref{fig:noise_setup_2}. The boundary conditions are then given in Eq.~\eqref{eq:DSBCS}. We are interested in the noise in the right, cold contact, which we call downstream delta-$T$ noise (because the noise is measured downstream from the heat source). Note however that the noise spot lies close to the hot contact (given that the equilibrated charge flow is from left to right). The noise is given by a modification of Eq.~\eqref{eq:noise_generation}, namely
\begin{align}
    \label{eq:noise_generation_wBC}
    S&\simeq \frac{2e^2}{h \ell^C_{\rm eq}}\frac{ \nu_-}{ \nu_+} (\nu_+-\nu_-) \int_0^L dx\;e^{-\frac{2x}{\ell^C_{\rm eq}}} \Lambda(x) \notag \\
    &+ \frac{2e^2}{h}\frac{(\nu_+-\nu_-)^2}{\nu_+}k_{\rm B} (\bar{T}+\Delta T).
\end{align}
Here, the second term takes into account thermal fluctuations coming from the hot, left contact~\cite{Melcer2022Jan}. The thermal noise coming from the cold, right contact is suppressed by a factor that is exponential in $\delta=L/\ell^C_{\rm eq}\gg 1$~\cite{Park2019,Hein2022}. This term can thus be safely neglected due to the efficient charge equilibration. In contrast to the voltage biased edge segment (where we assumed $e\Delta V\gg k_{\rm B} \bar{T}\approx 0$), we focus here on excess noise obtained by subtracting from Eq.~\eqref{eq:noise_generation_wBC} the equilibrium noise in the cold contact, i.e., 
\begin{align}
    \label{eq:S_excess}
    S^{\rm excess} \equiv S-2G_{\rm 2T}k_{\rm B}\bar{T},
\end{align}
where $G_{\rm 2T}=e^2(\nu_+-\nu_-)/h$ is the two-terminal charge conductance. 

\subsubsection{Efficient thermal equilibration}
For efficient thermal equilibration, we find that the first term in Eq.~\eqref{eq:noise_generation_wBC} for any edge structure, reduces to $2e^2\nu_-(\nu_+-\nu_-)k_{\rm B}(T+\Delta T)/(h\nu_+)$. Adding this to the second term, we obtain the downstream delta-$T$ noise as
\begin{align}
    \label{eq:DeltaTDSEq}
    S_{\rm ds}&\simeq 2\frac{e^2}{h}(\nu_+-\nu_-)k_{\rm B} (\bar{T}+\Delta T)\notag \\ &= 2G_{\rm 2T}k_{\rm B} (\bar{T}+\Delta T).
\end{align}
The corresponding downstream excess noise~\eqref{eq:S_excess} reduces to
\begin{align}
    \label{eq:S_ds_full}
    S^{\rm excess}_{\rm ds} \simeq 2G_{\rm 2T}k_{\rm B} \Delta T.
\end{align}
Hence, the excess downstream delta-$T$ noise for strong thermal equilibration equals the excess NJ-like noise emanating from the hot contact. This result is independent of $\nu_Q$. We also note that for $\Delta T=0$, the entire edge segment and the contacts are in thermal equilibrium and $S_{\rm ds}=2G_{\rm 2T}k_{\rm B}\bar{T} \Rightarrow S^{\rm excess}_{\rm ds}=0$ as expected.
In Fig.~\ref{fig:Noise_Tcomp_full}, we plot, for phPf-$3$, phPf-$2$, Pf-$3,$ and aPf-$2$, $S^{\rm excess}_{\rm ds}/(2G_{\rm 2T}k_{\rm B})$ vs $\delta$. We see that with increasing $\delta$ this ratio approaches unity as expected. For aPf-$3$ and Pf-$2$, where $\bar{c}=0$, we have trivially $S^{\rm excess}_{\rm ds} = 2G_{\rm 2T}k_{\rm B} \Delta T$ for all $\delta$.  

\subsubsection{Absent thermal equilibration}
In case of absent thermal equilibration, edge modes can have very different temperatures at the noise spot. When computing the noise in Eq.~\eqref{eq:noise_generation_wBC} for this case, we find that the excess noise still takes a form similar to the NJ noise~\eqref{eq:S_ds_full}, namely
\begin{align}
    \label{eq:S_ds_abs}
    S^{\rm excess}_{\rm ds} = 2\lambda_{\rm ds} G_{\rm 2T}k_{\rm B} \Delta T.
\end{align}
Here, $\lambda_{\rm ds} \sim \mathcal{O}(1)$ is a correction factor~\cite{Melcer2022Jan}, which reflects the poor thermal equilibration. Its origin is the first term in Eq.~\eqref{eq:noise_generation_wBC}: For poor thermal equilibration, this term does take the simple form $\sim k_{\rm B}(T+\Delta T) $ that produced Eq.~\eqref{eq:DeltaTDSEq}. Instead, we compute (see Appendix~\ref{sec:NoisekernelAppendix}) the noise kernels with distinct temperatures at the noise spot and insert these kernels in Eqs.~\eqref{eq:noise_generation_wBC}-\eqref{eq:S_excess}. We then obtain Eq.~\eqref{eq:S_ds_abs} with the correction factors $\lambda_{\rm ds}$.

In Tab.~\ref{tab:lambda_ds}, we give the values of $\lambda_{\rm ds}$ for weak temperature bias $\Delta T\ll\bar{T}$ and strong bias $\Delta T \gg \bar{T}$ (see Appendix~\ref{subsec:Noisekernel_ds_abs} for the derivation). Note that the values are obtained in the extreme limit of no thermal equilibration, while full equilibration amounts to $\lambda_{\rm ds}=1$ according to Eq.~\eqref{eq:S_ds_full}. Partial equilibration then corresponds to values larger than those in Tab.~\ref{tab:lambda_ds}, but still below $1$. We use the strong bias values for the plots in Fig.~\ref{fig:Noise_Tcomp_abs}.

\begin{table}[t!]
    \centering
    \caption{Correction factors for thermally non-equilibrated downstream (ds) excess delta-$T$ noise [see Eq.~\eqref{eq:S_ds_abs}] for small and large applied bias $\Delta T$.}
    \label{tab:lambda_ds}
    \makegapedcells
    \setlength\tabcolsep{6pt}
    \begin{tabular}{c|ccc}
        $\lambda_{\rm ds}$ & phPf-$3$ & aPf-$2$ & phPf-$2$  \\[2pt]
        \hline
        $\Delta T \ll \bar{T}$  & $\displaystyle \frac{3}{4}$ & $\displaystyle \frac{5}{8}$ & $\displaystyle \frac{4}{5}$\\ 
        $\Delta T\gg \bar{T}$ & $0.865$ & $0.774$ & $0.899$ \\ 
    \end{tabular}
\end{table} 

\subsection{Upstream delta-T noise}
\label{sec:deltaT_noise_US}
We now analyze the situation where the hot contact lies downstream to the cold contact, as depicted in Fig.~\ref{fig:noise_setup_3}. The noise is measured in the cold, upstream contact and we call this noise upstream noise (since noise is measured upstream of the heat source). Note that also in this case, the noise spot is close to the left contact. The boundary conditions are now given in Eq.~\eqref{eq:USBCS} and the noise reads
\begin{align}
    \label{eq:noise_generation_wBC_2}
    S&\simeq \frac{2e^2}{h \ell^C_{\rm eq}}\frac{ \nu_-}{ \nu_+} (\nu_+-\nu_-) \int_0^L dx\;e^{-\frac{2x}{\ell^C_{\rm eq}}} \Lambda(x) \notag \\
    &+ \frac{2e^2}{h}\frac{(\nu_+-\nu_-)^2}{\nu_+}k_{\rm B} \bar{T}.
\end{align}
The difference between Eq.~\eqref{eq:noise_generation_wBC} and Eq.~\eqref{eq:noise_generation_wBC_2} lies only in the temperature entering the last term: now the thermal charge fluctuations from the \textit{hot, right contact} are exponentially suppressed in $\delta=L/\ell^C_{\rm eq}$. The definition of excess noise~\eqref{eq:S_excess} holds also for the upstream delta-$T$ noise.

\subsubsection{Efficient thermal equilibration}
Whether the noise spot acquires a temperature larger than $\bar{T}$ depends, for efficient thermal equilibration, strongly on the thermal quantum number $\nu_Q$ in Eq.~\eqref{eq:nuQ}. For $\nu_Q>0$, the upstream heat flow is exponentially small in $\delta$ and the resulting noise is to excellent approximation given as the NJ noise corresponding to the cold contact
\begin{align}
    \label{eq:DeltaUSSEq}
    S_{\rm us}\simeq 2G_{\rm 2T}k_{\rm B} \bar{T}.
\end{align}
This implies $S^{\rm excess}_{\rm us}=0$ for all edges with $\nu_Q>0$. For $\nu_Q\leq 0$, the situation is similar to the voltage biased and thermally equilibrated edge in Sec.~\ref{sec:DeltaVThermalEq}. The only difference is that instead of heat coming from dissipation in the hot spot, it comes from a heated contact. Hence, the classification presented in Ref.~\cite{Spanslatt2019} (this classification was outlined in Sec.~\ref{sec:Intro}) carries over to thermally equilibrated, upstream delta-$T$ noise.  

We plot the upstream, equilibrated delta-$T$ noise in Fig.~\ref{fig:Noise_Tcomp_full}. The interfaces Pf-$2$, aPf-$3$, phPf-$2$, and phPf-$3$ all have $\nu_Q>0$ (relative to their charge flows) and their upstream delta-$T$ noises therefore either vanish identically $S=0$ (aPf-$2$ and Pf-$3$) or decay exponentially in $\delta$ (aPf-$3$ and Pf-$2$). Due to the single Majorana mode on all interfaces, none of them have $\nu_Q=0$. For aPf-$2$ and Pf-$3$, $\nu_Q<0$ and $S^{\rm excess}_{\rm us}$ reaches a constant value with increasing $\delta$. 

\subsubsection{Absent thermal equilibration}
For very poor thermal equilibration, upstream excess delta-$T$ noise becomes possible in the presence of upstream modes, $\bar{c} \neq 0$. For $\bar{c}=0$, we have that $S_{\rm us}\simeq 2G_{\rm 2T}\bar{T} \Rightarrow S^{\rm excess}_{\rm us}\simeq 0$. This is the case for Pf-$2$ and aPf-$3$ interfaces. 

All other interfaces have $\bar{c}\neq 0$. We compute the noise kernels for those in Appendix~\ref{sec:NoisekernelAppendix}. The resulting noise
can be written as
\begin{align}
    \label{eq:S_us_abs}
    S^{\rm excess}_{\rm us} = 2 \lambda_{\rm us} G_{\rm 2T}k_{\rm B} \Delta T,
\end{align}
with $ \lambda_{\rm us}$ given in Tab.~\ref{tab:lambda_us}, for both weak, $\Delta T \ll \bar{T}$, and strong temperature bias $\Delta T \gg \bar{T}$. Since the thermally equilibrated $S^{\rm excess}_{\rm us}$ depends on $\nu_Q$, we do not expect a continuous limit for $\lambda_{\rm us}$, as was the case for $S^{\rm excess}_{\rm ds}$. The strong bias values for $\lambda_{\rm us}$ are used for the plots in Fig.~\ref{fig:Noise_Tcomp_abs}: The $\bar{c}\neq 0$ interfaces display constant noise vs $\delta$.

\begin{table}[htb]
    \centering
    \caption{Correction factors for thermally non-equilibrated upstream (us) excess delta-$T$ noise [see Eq.~\eqref{eq:S_us_abs}] for small and large applied bias $\Delta T$.}
    \label{tab:lambda_us}
    \makegapedcells
    \setlength\tabcolsep{6pt}
    \begin{tabular}{c|ccc}
        $\lambda_{\rm us}$ & phPf-$3$ & aPf-$2$ & phPf-$2$ \\[2pt]
        \hline
        $\Delta T \ll \bar{T} $ & $\displaystyle \frac{3}{4}$ & $\displaystyle \frac{7}{8}$ & $\displaystyle \frac{7}{10}$\\ 
        $\Delta T \gg \bar{T}$ & $0.365$ & $0.438$ & $0.338$ \\ 
    \end{tabular}
\end{table} 

The results for the various types of noise computed in Sec.~\ref{sec:Noise_Results} are summarized in Tab.~\ref{tab:summary}.

\subsection{\texorpdfstring{Downstream-upstream symmetry of phPf-$3$ for absent thermal equilibration}{}}
\label{eq:dsus_symmetry}

\begin{figure}[b!]
    \centering
    \includegraphics[scale=1]{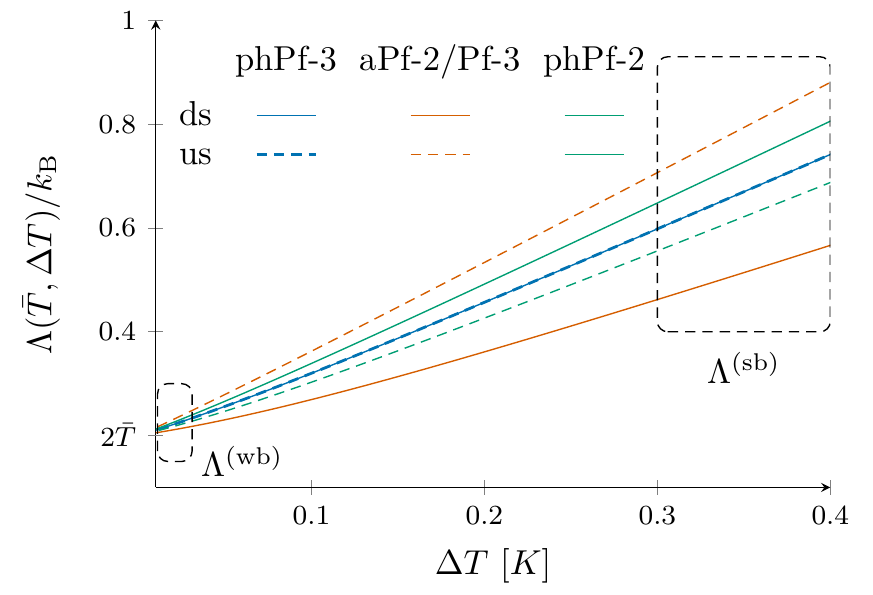}
    \caption{Noise kernel $\Lambda$ dependencies on the base temperature $\bar{T}$ (here, we set $\bar{T} = 0.1$ for concreteness) and applied bias $\Delta T$, for poor thermal equilibration. The boxed areas correspond to regions in which the approximation for weak or strong biases (wb and sb, respectively) are valid, see Appendix~\ref{sec:fulTdepLambda}. The downstream and upstream noise kernels are identical for all $\Delta T,\bar{T}$ for the phPf-$3$ interface. }
    \label{fig:Lambda_dT}
\end{figure}

Examination of $\lambda_{\rm ds}$ and $\lambda_{\rm us}$ exposes a unique relation for the phPf-$3$ interface. For strong applied biases, we have for this interface that $\lambda_{\rm ds} \equiv 0.5 + \lambda_{\rm us}$. The constant offset follows from Eqs.~\eqref{eq:noise_generation_wBC} and~\eqref{eq:noise_generation_wBC_2} together with the fact that for the phPf-$3$ the noise kernels for excess delta-$T$ noise in the down- and upstream bias configurations are identical, see Appendix~\ref{sec:fulTdepLambda}. By computing the noise kernels for general $\Delta T / \bar{T}$, we find that the above downstream-upstream symmetry holds for all $\Delta T$ (as long as the thermal equilibration is poor). In Fig.~\ref{fig:Lambda_dT}, this result is visualized by overlapping solid and dashed lines for all $\Delta T$ at constant $\bar{T}$ only for the phPf-$3$ (blue). We present an experimental method to test this symmetry in Sec.~\ref{sec:Device_proposal}.

\section{Temperature dependence of thermal equilibration lengths}
\label{sec:thermal_equilibration_lengths}
Here, we discuss the impact of the base temperature $\bar{T}$ on the thermal equilibration. The degree of this equilibration is determined by parameters on the form $L/\ell^Q_{\rm eq}$. Thus, to investigate phenomena related to thermal equilibration, one may either tune $L$, which requires advanced devices (see, e.g.,  Ref.~\cite{Melcer2022Jan}), or tune $\ell^Q_{\rm eq}$ through its temperature dependence. Hence, this dependence is an important edge characteristic, as it determines
\begin{enumerate}[i)]
    \item The value of the two-terminal thermal conductance $G^Q_{\rm 2T}$ for a given temperature $\bar{T}$ and edge length $L$~\cite{Srivastav2022Sep}.
    \item The sharpness of transitions between plateaus of quantized two-terminal thermal conductances (see Sec.~\ref{sec:TTGQ}).
    \item The magnitude of the noise generated on a single edge segment (see Sec.~\ref{sec:Noise_generation}).
\end{enumerate}
In the remainder of this section, we therefore discuss the temperature dependence of various $\ell^Q_{i,j}$ for our considered interfaces. 

In Ref.~\cite{Ma2020Jul}, the authors argued that most thermal equilibration lengths scale as
\begin{align}
    \label{eq:ellQT2}
    \ell^Q_{\rm eq} \sim \bar{T}^{-2},
\end{align}
due to tunneling of particles between the edge channels. The  exponent in~\eqref{eq:ellQT2} comes from the fact that in most cases, the most relevant (in the renormalization group sense) tunneling operators $\mathcal{O}$ take the form 
\begin{align}
    \mathcal{O} \sim e^{i \sum_j m_j\phi_j} \text{ or } \sim \psi e^{i \sum_j m_j\phi_j}.
\end{align}
Here, $\psi$ is a MM, $\phi_j$ are bosons, and $m_j$ are real-valued numbers indicating the number and charges of particles that tunnel. When the edge is  sufficiently close to a strong disorder fixed point~\cite{Kane1994,Kane1995b, Moore1998,Protopopov2017}, the $m_j$ take on integer values which results in an integer scaling dimension of $\mathcal{O}$, denoted $\Delta$~\cite{francesco2012conformal}, and the exponent $\alpha=2(\Delta-1)$ in~\eqref{eq:ellQT2} is thus also an integer. Away from such a point, however, the exponents are expected to take on smaller, non-universal values which depend on the inter-edge interaction strength~\cite{Protopopov2017}. 

An important exception from Eq.~\eqref{eq:ellQT2} is the equilibration length between a counter-propagating boson and Majorana mode. The most relevant operator coupling these modes is instead on the form~\cite{Ma2020Jul}
\begin{align}
    \label{eq:densdens}
    \mathcal{O} \sim \partial_x \phi \psi i \partial_x \psi.
\end{align}
Essentially, \eqref{eq:densdens} follows from the fact that it is impossible to construct tunneling operators between a counter-propagating boson $\phi$ and a MM $\psi$ (the point is that only their combination can create electrons), and the simplest operator exchanging energy between the modes is instead of density-density type. This feature leads to the temperature scaling
\begin{align}
    \label{eq:ellQT4}
    \ell^Q_{\rm eq} \sim \bar{T}^{-4}.
\end{align}
We present a detailed derivation of this result in Appendix~\ref{sec:MajoranaAppendix}. Equation~\eqref{eq:ellQT4} suggests that, for edges with a counter-propagating boson and a MM, which is the case for phPf-$2$ and phPf-$3$ interfaces (see Fig.~\ref{fig:interfaced_edge_structures}), there can be unusually sharp transitions between thermal conductance plateaus. They are important signatures in thermal conductance measurements. For example, the phPf-$3$ interface produces a thermal conductance transition from $G^Q_{\rm 2T}/(\kappa_0 \bar{T})=1/2$ to $G^Q_{\rm 2T}/(\kappa_0 \bar{T})=3/2$ with decreasing temperature $\bar{T}$. The scaling of this transition is of the sharper form~\eqref{eq:ellQT4}, which might enhance the prospects of its experimental observation at not too low $\bar{T}$.

Temperature scalings of the thermal equilibration lengths enter the incoherent model (summarized in Sec.~\ref{sec:Model_Setup}), as~\cite{Srivastav2021May}
\begin{subequations}
    \begin{align}
        \label{eq:eqthermlength_general}
        &\ell^Q_{i,j} = \frac{a}{g_{i,j}\gamma_{i,j} (n_i - n_j)} = \frac{a \kappa_0 \bar{T}}{G_{i,j}^Q \gamma_{i,j} (n_i - n_j)},\quad  \ n_i \neq n_j\\
        &\ell^Q_{i,j} = \frac{a}{g_{i,j}\gamma_{i,j}} = \frac{a \kappa_0 \bar{T}}{G_{i,j}^Q \gamma_{i,j}}, \quad n_i = n_j
    \end{align}
\end{subequations}
Here, $n_i$ and $n_j$ are the central charges of the two modes, $\gamma_{i,j}$ is a parameter characterizing deviations from the Wiedemann-Franz law~\cite{Nosiglia2018,Asasi2020,Srivastav2021May}, $a$ is the typical lengthscale for inter-channel heat exchange, and $G^Q_{i,j}/(\kappa_0 \bar{T})\equiv g_{i,j}$ are the dimensionful and dimensionless thermal conductances of this exchange. The latter quantities can be computed within the chiral Luttinger liquid model (see e.g., Refs.~\cite{Ma2020Jul,Asasi2020,Srivastav2021May}).

For the special case of the counter-propagating boson and MM, denoted $\phi_2$ and $\psi$, respectively, we have
\begin{align}
    \label{eq:eqlength}
    & \ell^{Q}_{\phi,\psi} = \frac{a}{g (n_\phi - n_\psi) \gamma_{\phi,\psi}} = \frac{2 a \kappa_0 \bar{T}}{G_{\rm int}^Q}
\end{align}
with $\gamma_{\phi,\psi}=1$ and
\begin{align}
    G_{\rm int}^Q = \frac{8 b^2 k_{\rm B}^4 \pi^5 \Gamma_0^2 }{35 \hbar^6 v_{\phi}^2 v_{\psi}^4}  \bar{T}^4 \kappa_0 \bar{T},
\end{align}
as computed in Appendix~\ref{sec:MajoranaAppendix}. Here, $\Gamma_0$ is the bare coupling strength between the channels, which we assume to be weak, and $v_{\phi}$, $v_{\psi}$ are the propagation velocities of $\phi$ and $\psi$, respectively. The parameter $b$ is an ultra-violet cutoff, with dimension of length.

\begin{figure}[t!]
    \centering
    \includegraphics{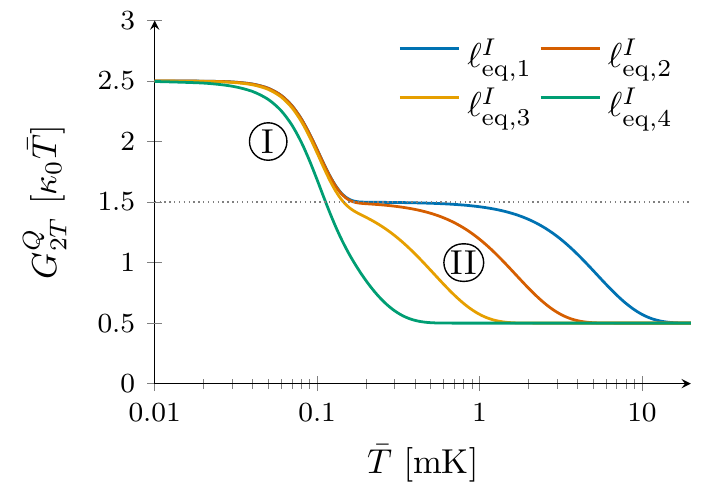}
    \caption{Two-terminal thermal conductance $G^Q_{\rm 2T}$ (in units of the thermal conductance quantum $\kappa_0\bar{T}$) for $5/2-3$ vs $\bar{T}$ for FQH interfaces with three modes and two distinct thermal equilibration lengths. For concreteness, we have chosen numerical values $\ell^{\Romannum{1}}_{\rm eq,i} \sim 5 \cdot \{10^{-5}, 10^{-6}, 10^{-7}, 10^{-8} \} \cdot \bar{T}^{-2}$ and $\ell^{\Romannum{2}}_{\rm eq} \sim \{ 10^{-16}\} \cdot \bar{T}^{-4}$, such that the transitions take place at temperatures below $10$ mK.}
    \label{fig:Qconductance_three}
\end{figure}

Knowledge of the temperature scalings of $\ell^Q_{i,j}$, permits us to analyze $G^Q_{\rm 2T}$ vs  $\bar{T}$, rather than $\alpha,\beta$ (as we did in Sec.~\ref{sec:Conductance_results}). As an instructive example, we show in Fig.~\ref{fig:Qconductance_three} such a plot for the phPf-$3$ interface. The characteristic feature is the presence of plateaus with different quantized values of  $G^Q_{\rm 2T}$. These plateaus are associated to different regimes of equilibrated and non-equilibrated edge modes. In this particular example, there are two inter-plateau transitions, which we label $\Romannum{1}$ and $\Romannum{2}$. These transitions are related to thermal equilibration by electron tunneling between $\phi_1$ and $\psi \times \phi_2$, transition $\Romannum{1}$, and by a density-density interaction between $\phi_2$ and the MM $\psi$: transition $\Romannum{2}$. From the above analysis, we have different temperature scalings for these transitions, namely $\ell^{\Romannum{1}}_{\rm eq} \sim \bar{T}^{-2}$ and $\ell^{\Romannum{2}}_{\rm eq} \sim \bar{T}^{-4}$. 
Figure~\ref{fig:Qconductance_three} shows that for full thermal equilibration (high $\bar{T}$), $G^Q_{\rm 2T}/(k_{\rm B}\bar{T})=1-1+1/2=1/2$ which transitions at lower $\bar{T}$ to the value for absent thermal equilibration: $G^Q_{\rm 2T}/(k_{\rm B}\bar{T})=1+1+1/2=5/2$. However, an intermediate plateau emerges in a temperature regime when there is a separation of the two equilibration length scales $\ell^{\Romannum{1}}_{\rm eq} \ll \ell^{\Romannum{2}}_{\rm eq}$. This plateau vanishes for $\ell^{\Romannum{1}}_{\rm eq} \approx \ell^{\Romannum{2}}_{\rm eq}$ within the range of the transition temperature. A similar analysis can be performed for any edge structure, with its own specific set of plateaus and transitions. The set of these two features can be viewed as a ``pin-code'' of the FQH edge structure~\cite{Srivastav2022Sep}. 

In the above analysis, we have have neglected interference effects from scattering of bosonic plasmon waves on contact-edge interfaces~\cite{Safi1995,Krive1998,Protopopov2017}. This amounts to assuming $L\gg L_T\sim\bar{T}^{-1}$, where $L_T$ is a characteristic thermal length scale. Including the interference effects can potentially generate additional plateaus~\cite{Protopopov2017,Melcer2022Jan}.

\section{Comparison to experiments}
\label{sec:Discussion}
In this section, we compare the results from our model to the recent experimental findings in Refs.~\cite{dutta2022sep,Dutta2022}. We start with the thermal conductance. In Ref.~\cite{dutta2022sep}, the two-terminal thermal conductance at the interface between the $5/2$ state and integer states $n$ was observed to obey $G^Q_{\rm 2T}/(\kappa_0 \bar{T}) \approx \left| 5/2 - n \right|$. We see that this result is incompatible with both the Pf and aPf edge structures,  since for Pf-$2$ and aPf-$3$, we have $G^Q_{\rm 2T}/(\kappa_0 \bar{T}) = 3/2$, independently of the thermal equilibration (see Sec.~\ref{sec:TTGQ}). The measured conductances can further be compared with Fig.~\ref{fig:GQPlots}, where we note that the measured values are consistent with a phPf edge structure, if we assume an efficient thermal equilibration when interfaced with integer modes. Indeed, among the three non-Abelian candidates in Fig.~\ref{fig:edge_structures}, the phPf is the only one compatible with the expected values of $G^Q_{\rm 2T}$ for all interfaced structures. 

We further note that with decreasing temperature, and thus decreasing degree of equilibration (see Sec.~\ref{sec:thermal_equilibration_lengths}), the thermal conductance should increase and saturate at $G^Q_{\rm 2T}/(\kappa_0 \bar{T}) = 3/2$ and $5/2$ for the phPf-$2$ and phPf-$3$ edges, respectively. As mentioned in the end of Sec.~\ref{sec:TTGQ} the non-equilibrated limit is an important check for uniquely pin-pointing the edge structure. This could be tested in future experiments similar to that in Ref.~\cite{Srivastav2022Sep}. 

We next move on to the noise. Ref.~\cite{Dutta2022} reported measurements of excess noise at interfaces of $5/2-n$ for an applied voltage bias. From our analysis in Sec.~\ref{sec:Noise_generation}, we note that for both Pf-$2$, and aPf-$3$, which consist of only co-propagating modes over long length scales, no excess noise is expected for any degree of thermal equilibration. As finite, but with $L$ and $\bar{T}$ decreasing, noise was found for both $5/2-2$ and $5/2-3$, the most compatible edge structure is, just as in Ref.~\cite{Dutta2022}, the phPf, this time assuming the thermal equilibration to be neither full nor absent but rather in an intermediate regime.

Our model allows a comparison on a quantitative level: We compare the measured noise data with our Figs.~\ref{fig:Noise_Vcomp_full}-\ref{fig:Noise_Vcomp_abs}. Due to the the well-established charge conductance quantization in Ref.~\cite{Dutta2022}, we can safely assume efficient charge equilibration $\delta \gg 1$. From our calculations,  we find that the slope of the noise vs bias voltage curve for the phPf-$3$ interface approximately evaluates to
\begin{align}
    0\leq \partial_{\Delta V} S^{\rm exc}_{\rm phPf-3}\leq 0.113 \frac{e^3}{h} \approx 0.70 \times 10^{-30}\frac{A^2}{\mu V \ \rm Hz}, 
\end{align}
between the two limits of absent and full thermal equilibration (see Appendix~\ref{sec:Conversion} for conversion between units). For this comparison we have chosen $\delta > 10$. The only constraint on $\delta$ we have is to reproduce a correctly quantized charge conductance of $G_{\rm 2T}/(e^2/h) \approx 1/2$ to good accuracy. Since this is still the case $5 < \delta < 10$, our upper bound for the noise is slightly enhanced for such choices. Access to more accurate estimates of $\delta$ from experimental data would improve the accuracy of our noise magnitudes. For the phPf-$2$ interface, our noise approximately evaluates to
\begin{align}
    0\leq \partial_{\Delta V} S^{\rm exc}_{\rm phPf-2}\leq 0.086 \frac{e^3}{h} \approx 0.54 \times 10^{-30}\frac{A^2}{\mu V \ \rm Hz}, 
\end{align}
i.e., it is of similar magnitude to the one for phPf-$3$. In comparison, the experimentally observed noise characteristics for both $5/2-2$ and $5/2-3$ were~\cite{Dutta2022}
\begin{align}
    \partial_{\Delta V} S^{\rm exc}_{\rm measured}\approx 0.1 \times 10^{-30}\frac{A^2}{\mu V \ \rm Hz}, 
\end{align}
which lies within our estimate and is moreover compatible with interfaced phPf edges that are not fully thermal equilibrated. 
\begin{figure}[t!]
    \centering
    \includegraphics{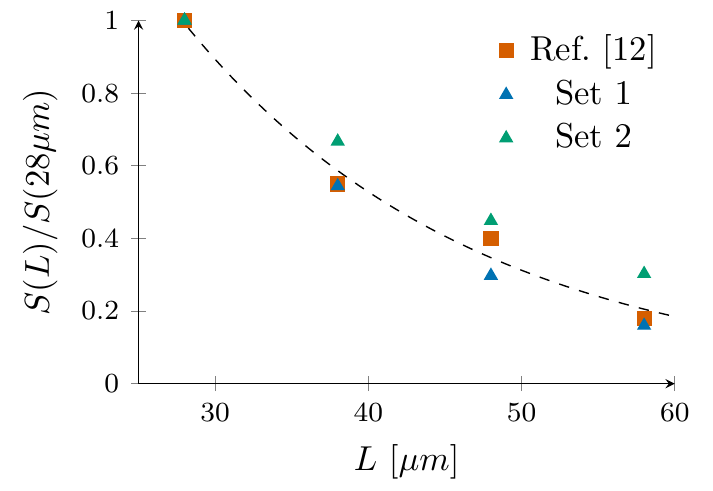}
    \caption{Length dependent noise $S(L)$ at the $5/2-3$ interface. The red squares and dashed, black line are measured noise data and a fit, respectively, from  Ref.~\cite{Dutta2022}. The data is obtained for interface lengths $L= \{28,38,48,58\} \mu m$ and was normalized to $S(L=28\mu m)$. Blue and green triangles is our computed noise for a phPf-$3$ interface. For the sets $1$ and $2$, we used parameters $(\alpha, \beta, \delta) = (6,10,100)$ and $(\alpha, \beta, \delta) = (4,8,25)$ at $L= 28 \mu m$ respectively, to obtain the normalizing constants $S(L=28\mu m) = \{ 0.023, 0.026\} \Delta V \cdot 10^{-30} A^2/(\mu V Hz)$ and thus $S(L)/S(L=28 \mu m)$.}
    \label{fig:lengths}
\end{figure}

We finally estimate the thermal equilibration lengths from noise data measured at lengths $L=\{28,38,48,58\} \mu m$ in Ref.~\cite{Dutta2022}, see Fig.~\ref{fig:lengths}. For concreteness, we use the phPf-$3$ edge structure and compute, for a given set of $\alpha, \beta, \delta$, the temperature profiles for a given bias $\Delta V\gg \bar{T}\approx 0$. These profiles are then used with Eq.~\eqref{eq:noise_generation} to obtain the noise $S(\alpha,\beta,\delta)$. In Fig.~\ref{fig:lengths}, we present $S(L)$-profiles for two sets of model parameters that reasonably reproduce the observed length dependent noise. For set $1$ (blue triangles), we have taken $(\alpha, \beta, \delta) = (6,10,100)$ and for set $2$ (green triangles), we took $(\alpha, \beta, \delta) = (4,8,25)$. For the shortest length $L=28\mu m$, the values $\delta=100$ and $\delta = 25$ correspond to $ \ell^C_{\rm eq} \approx 0.28 \mu m$ and $ \ell^C_{\rm eq} \approx 1.12 \mu m$, respectively. Both these charge equilibration lengths accurately produce $G_{\rm 2T}/(e^2/h)=1/2$, as observed experimentally. 

For both parameter sets, we find that thermal equilibration lengths
\begin{align}
    & 3 \mu m \leq \{\ell^Q_{12}, \ell^Q_{23}\} \leq 7 \mu m
\end{align}
fits the data well.

\section{Combined conductance and noise measurements in a single device}
\label{sec:Device_proposal}
\begin{figure}[t!]
  \centering
    \includegraphics[width=0.9\columnwidth]{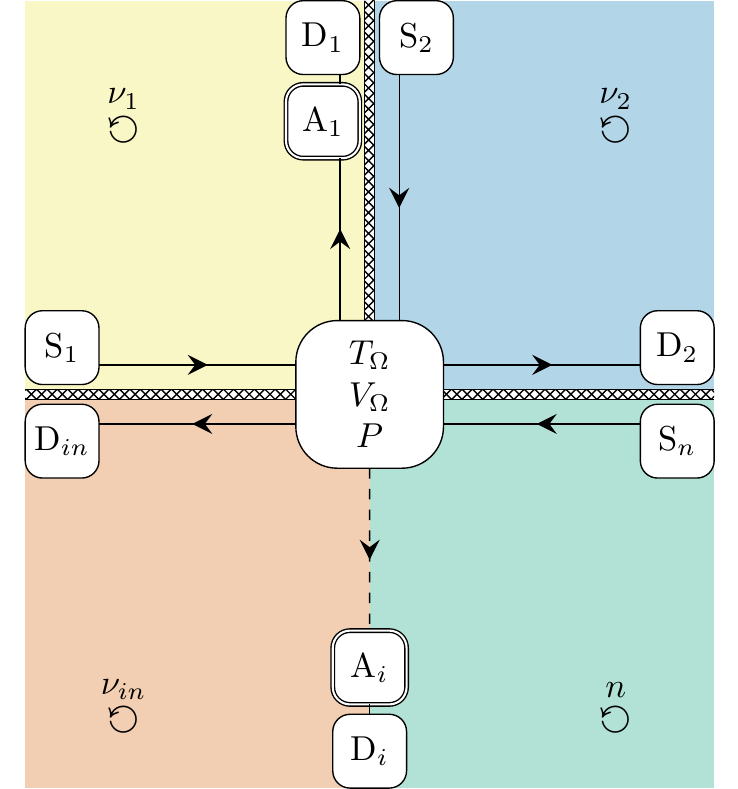}
    \caption{Schematic setup to measure both the two-terminal thermal conductance and noise for FQH interfaces. The source contacts $S_1$, $S_2$ and $S_n$ feed currents into the central contact $\Omega$, dissipating power $P$ and establishing the electrical potential $V_\Omega$, see Eqs.~\eqref{eq:V_Omega}-\eqref{eq:P_diss}. The dissipated power results in an increased temperature $T_\Omega$ which can be extracted from excess noise in the downstream contact $A_{1}$. Heat is carried away from $\Omega$ by edge modes along interfaces between gated regions (with fillings $\nu_1$, $\nu_2$, $n$, and $\nu_{in}$) or with the vacuum (thick, black lines). Upstream noise for the $\nu_{in}-n$ interface can be measured for either $P=0$, $V_\Omega\neq 0$ or $P\neq0$, $V_\Omega= 0$, see the text in Sec.~\ref{sec:Device_proposal}.}
    \label{fig:proposed_setup}
\end{figure}

In this section, we propose a device designed for measurements of both $G^Q_{\rm 2T}$ and various types of noise in a single device, see Fig.~\ref{fig:proposed_setup}. This device is beneficial for two main reasons: First, it can be used to rule out possible sample-to-sample differences between separate devices targeted towards noise and conductance measurements. Second, it can be used to exclude effects of edge reconstruction~\cite{Wan2006,Wan2008,Overbosch2008,Zhang2014}. To do so is particularly important, since in its presence, and under conditions of poor thermal equilibration, \textit{any} edge structure can produce excess upstream noise. The benefit with our device is that the two-terminal thermal conductance at poor equilibration gives access to the total number of edge channels (with MM's counted as a ``half-channel''). This number can  can be used to ascertain that in the noise measurements there is no upstream heat transport in non-topological, spurious upstream modes from edge reconstruction. 
In describing our device below, we will also point out some subtleties in $ G^Q_{\rm 2T}$ measurement schemes for states with counter-propagating edge channels, and how to remedy them. 

In Fig.~\ref{fig:proposed_setup}, four QH regions are connected to a central, floating contact, $\Omega$, which can act as a hot reservoir when electrical currents impinge on it. We take the three regions with fillings $\nu_1$, $\nu_2$, and $n$ as the source regions: the source contacts $S_1$, $S_2$ and $S_n$ inject currents $I_1 = \nu_1 V_1 e^2/h$, $I_2 = \nu_2 V_2 e^2/h$, and $I_n=n V_n e^2/h$, respectively, and $V_1$, $V_2$, and $V_n$ are the corresponding three bias voltages with respect to ground. The drain contacts $D_{1}$, $D_{2}$, $D_{i}$, and $D_{in}$ are grounded, and the amplifiers $A_{1}$ and $A_{i}$ are floating. Note that the currents that emanate from $\Omega$ and enter the amplifiers are emitted to grounded contacts. Thus, these currents do not propagate back into source contacts. The filling in the region $\nu_{\rm in}$ is here tuned to $5/2$ (or to some other state of interest), and this region is interfaced with a region with integer filling $n$.

The contact potential $V_\Omega$ and the dissipated electrical power $P$ can be tuned independently. They are given as
 \begin{align}
    & V_\Omega = \frac{h}{e^2}\frac{I_1+I_2+I_3}{\nu_1+\nu_2+\text{max}(\nu_{in},n)},\label{eq:V_Omega}\\
    & P = \frac{h}{2e^2}\Big(\frac{I_1^2}{\nu_1} + \frac{I_2^2}{\nu_2} + \frac{I_n^2}{n}  - \frac{ \left(I_1 + I_2+I_n\right)^2}{\nu_1+\nu_2+\text{max}(\nu_{in},n)}  \Big)\label{eq:P_diss}.
\end{align}
Strictly speaking, Eq.~\eqref{eq:P_diss} holds when the sourced edges have $\nu_Q>0$. If this is not the case, it is not evident that all dissipated power heats the central contact. Instead, some Joule heat can leak out back on the source edge via upstream modes, without heating the contact. This produces a correction to Eq.~\eqref{eq:P_diss} which will influence how the thermal conductance is extracted from the experimental data. We refer to Ref.~\cite{Melcer2022Jan} for a detailed discussion on this issue. However, this possible complication can be avoided by tuning the fillings $\nu_1$ and $\nu_2$ to, e.g., integer values. Then, it is certain that Eq.~\eqref{eq:P_diss} is the dissipated power in the contact. In the following, we assume that this is the case. 

Two operational modes with the device are of particular interest for this paper:

i) $P=0$ and $V_\Omega\neq 0$ and the noise in the amplifier $A_i$ is measured vs $V_\Omega$. This configuration realizes the setup in Fig.~\ref{fig:noise_setup_1}, upon the identification $V_\Omega=\Delta V$ and assuming $n<5/2$. For $n>5/2$, the charge flow direction along the interface is from $A_i$ to $\Omega$ but this flow can be reversed by swapping the magnetic field direction~\cite{Dutta2022}. Since $A_i$ is floating, no net current is injected into $\Omega$ for $n>5/2$.
Note that the charge fluctuations emanating from the hot contact in Fig.~\ref{fig:proposed_setup} is split between four edges. This requires a minor modification of the downstream noise as compared to Secs.~\ref{sec:deltaV_noise}-\ref{sec:deltaT_noise_DS}. Specifically, the device geometry is taken into account by substituting in Eqs.~\eqref{eq:DeltaTDSEq},\eqref{eq:S_ds_full},~\eqref{eq:DeltaUSSEq}, and \eqref{eq:S_us_abs} 
\begin{align}
\label{eq:G_substitution}
& G_{\rm 2T}\rightarrow G^{*}_m,\notag\\
    &\frac{1}{G^*_m} = \frac{1}{G_m}+ \frac{1}{\sum_{k\neq m}G_k}.
\end{align}
Here, $G_m$ is the equilibrated charge conductance of edge $m$, where $m=1,(\text{in}-n)$ labels the two outgoing edges with amplifiers and $k=1,2,\text{in},(\text{in}-n)$ labels all outgoing edges.

ii) $P\neq 0$ and $V_\Omega =0$. This produces the situations considered in Figs.~\ref{fig:noise_setup_2}-\ref{fig:noise_setup_3}. A measurement of the excess noise downstream  of the central contact (see Fig.~\ref{fig:noise_setup_2}) allows to determine its excess temperature $T_\Omega=\bar{T}+\Delta T$. In turn, access to this noise allows extraction of the two-terminal thermal conductance~\cite{Jezouin2013}. Let us mention that $V_\Omega=0$ is not a necessary condition when measuring the thermal conductance. However, finite $V_\Omega$ leads to additional hot spots in the device (close to all drain contacts downstream of $\Omega$). In the presence of upstream heat flow, the generated heat at these points could potentially effect the heating of the central contact. Such complications are absent for $V_\Omega=0$.  Ideally, the downstream noise for this purpose is probed on an edge with $\bar{c}=0$ (e.g., at contact $A_1$ with $\nu_1$ tuned to an integer). This choice avoids possible corrections to the downstream thermal noise, compare Eq.~\eqref{eq:DeltaTDSEq} and Eq.~\eqref{eq:S_ds_abs}. In fact, it is very useful to compare the downstream noises in $A_1$ and $A_i$ to investigate the validity of the NJ relation~\cite{Nyquist1928Jul,Johnson1928Jul} for QH states with counter-propagating channels and poor thermal equilibration. To illustrate this, we consider $\nu_1$ having only downstream modes. The excess noise in $A_1$ is then obtained from Eq.~\eqref{eq:DeltaTDSEq} as
\begin{align}
\label{eq:noiseA1}
    \frac{S^{\rm excess,1}_{\rm ds}}{2G^*_{1}} = k_{\rm B} \Delta T.
\end{align}
With the assumption of no upstream modes above, this expression holds independently of any equilibration. In contrast, the excess noise in $A_i$ reads
\begin{align}
\label{eq:noiseAi}
    \frac{S^{\rm excess,i}_{\rm ds}}{2G^*_{\rm in-n}} = \lambda_{\rm ds}k_{\rm B} \Delta T,
\end{align}
where $\lambda_{\rm ds}$ does depends on the thermal equilibration of the $\nu_{\rm in}-n$ interface. Only for full thermal equilibration is $\lambda_{\rm ds}=1$.

The ratio between Eq.~\eqref{eq:noiseA1} and Eq.~\eqref{eq:noiseAi} allows extraction of $\lambda_{\rm ds}$:
\begin{align}
    \lambda_{\rm ds}=\Bigg(\frac{S^{\rm excess,i}_{\rm ds}}{2G^*_{\rm in-n}}\Bigg)\times\Bigg(\frac{S^{\rm excess,1}_{\rm ds}}{2G^*_1}\Bigg)^{-1}.
\end{align}

We next move on to upstream noise measurements. By swapping $\nu_{in}$ and $n$, the noise in $A_i$ corresponds to upstream delta-$T$ noise as presented in Fig.~\ref{fig:noise_setup_3} (again, the substitution~\eqref{eq:G_substitution} is needed). Similarly to the downstream delta-$T$ noise, the upstream delta-$T$ can be measured vs $\Delta T$ where the latter is obtained from Eq.~\eqref{eq:noiseA1}.

Finally, we discuss measurements of the phPf-$3$ downstream-upstream symmetry of the delta-$T$ noise at poor thermal equilibration, as presented in Sec.~\ref{eq:dsus_symmetry}. To detect this symmetry, we propose to first measure the downstream delta-$T$ noise~\eqref{eq:noise_generation_wBC}, but in contrast to the excess noise~\eqref{eq:S_excess}, one subtracts in this case the \textit{hot contact thermal noise}, i.e.,
\begin{align}
    \label{eq:S_excess_ds}
    S^{\rm ds}_{\Delta T} \equiv S-2G^*_{\rm in-n}k_{\rm B}(\bar{T}+\Delta T) = \frac{G^*_{\rm in-n}}{2}\Lambda^{\rm ds}.
\end{align}
This quantity can then be compared with the upstream delta-$T$ noise~\eqref{eq:noise_generation_wBC_2} for which the \textit{cold contact thermal noise} is subtracted
\begin{align}
    \label{eq:S_excess_us}
    S^{\rm us}_{\Delta T} \equiv S-2G^*_{\rm in-n}k_{\rm B}\bar{T} = \frac{G^*_{\rm in-n}}{2}\Lambda^{\rm us}.
\end{align}
In going from Eqs.~\eqref{eq:noise_generation_wBC} and~\eqref{eq:noise_generation_wBC_2} to Eqs.~\eqref{eq:S_excess_ds} and~\eqref{eq:S_excess_us}, we have again used the conductance substitution~\eqref{eq:G_substitution}. The downstream-upstream symmetry is now exposed in the ratio
\begin{align}
    \label{eq:dTnoiseratio}
    \frac{S^{\rm ds}_{\Delta T}}{S^{\rm us}_{\Delta T}} = \frac{\Lambda^{\rm ds}}{\Lambda^{\rm us}}
\end{align}
which uniquely equals unity for the phPf-$3$ interface. Comparison of delta-$T$ noises for thermally non-equilibrated interfaces thus provides another experimental signature to distinguish between non-Abelian candidate states.

\section{Summary and conclusions}
\label{sec:Summary_Outlook}
We have presented a comprehensive theoretical description of quantum transport along interfaces formed between integer ($n=2$ and $n=3$) QH states and non-Abelian Pf, aPf, and phPf candidates for the $\nu=5/2$ state. Such interfaces isolate the ``non-Abelian part'' of the edge and was used in recent experiments~\cite{dutta2022sep,Dutta2022}, to distinguish between different candidate theories.

For such interfaces and experiments, we have here in detail determined the impact of thermal equilibration on the edge thermal conductance as well as on excess noise on voltage or temperature biased edge segments (so-called delta-$T$ noise). In contrast to Abelian edges, non-Abelian $5/2-n$ interfaces feature Majorana modes, which although charge neutral, may influence the charge current noise generated on the edge. A major finding is that non-Abelian interfaces with counter-propagating modes are highly sensitive to thermal equilibration. 
This feature produces a significant length and temperature dependence for the two-terminal thermal conductance. For the noise, the degree of thermal equilibration influences the noise magnitude, the length dependence of the noise, as well as the Nyquist-Johnson relation for charge current noise emanating from a heated contact. Our more quantitative results are summarized in Tab.~\ref{tab:summary}. 

In contrast to the two-terminal thermal conductance and the noise, we have proved that for the thermal Hall conductance, as it was defined in Eq.~\eqref{eq:GQH}, the thermal equilibration dependency drops out for a generic edge structure, \textit{provided the two involved edge segments have the same degree of thermal equilibration}. Such experimental conditions were put forward in Ref.~\cite{Melcer2023Jan}. Our results suggest that a similar experiment performed at $\nu=5/2$ provides a clear distinction between all candidate states, without any interfacing. 

The authors of Refs.~\cite{dutta2022sep,Dutta2022}, interpreted their noise and thermal conductance measurements  as pointing towards the phPf as the realized state in GaAs/AlGaAs. On the qualitative level, our findings favor such an interpretation as well. We further computed the voltage biased noise for phPf-$2$ and phPf-$3$ interfaces, and our results are quantitatively consistent with the experimentally obtained values. Importantly, our calculations show that the measured noise magnitude is indeed consistent with a FQH edge that is not fully thermally equilibrated, which was a crucial feature for the interpretation in Ref.~\cite{Dutta2022}. 

We would like to further emphasize that the interpretation in Ref.~\cite{Dutta2022} favoring the phPf state is based on the absence of edge reconstruction. If this effect would be present, any edge can generate noise for poor thermal equilibration. This severely complicates the interpretation of the experimental data. Ideally, one would therefore like to confidently rule out such effects. To address this problem, we proposed a device to do precisely that, by allowing thermal conductance and noise measurements in the same sample. Combining such measurements permits an unambiguous determination of the edge structure. We also pointed out some potential issues (highlighted previously in Ref.~\cite{Melcer2022Jan}) with the standard two-terminal thermal conductance setup~\cite{Jezouin2013} for edges with counter-propagating modes. Here, we proposed concrete improvements of the setup in order to mitigate these issues.

While our work targeted the famous $\nu=5/2$ state, we envision that it can be generalized for the purpose of pin-pointing edge structures of other exotic FQH states, such as the even-denominator states in graphene~\cite{Ki2014Apr,Kim2015Nov,Zibrov2017Sep,Zibrov2018Sep,Li2017,Kim2019Feb, Huang2022Jul} or the state at $\nu=12/5$~\cite{Read1999,Xia2004,Kumar2010}.  

\begin{acknowledgments}
We thank Janine Splettstoesser for numerous enlightening discussions during the course of this work, Therese Karmstrand and Oliver Hahn for their input on Appendix~\ref{sec:thermal_Hall_appendix}, and  Matteo Acciai for his insights on the calculations in Appendix~\ref{sec:MajoranaAppendix}. We are also grateful to Jinhong Park for critical comments on the manuscript.

C.S. acknowledges support from the Excellence Initiative Nano at Chalmers University of Technology. This project has received funding from the European Union's Horizon 2020 research and innovation programme under grant agreement No. 101031655 (TEAPOT). \\

\textit{Note:} Some of the results in this paper were reported in the Master thesis Ref.~\cite{Hein2022}. 
 
\end{acknowledgments}

\appendix

\section{Heat exchange between counter-propagating boson and Majorana modes}
\label{sec:MajoranaAppendix}
The phPf-$2$ edge structure, depicted in Fig.~\ref{fig:interfaced_edge_structures}, consists of two modes: One bosonic mode $\phi$ and one Majorana mode $\psi$. This structure is valid for lengthscales $L\gg \ell^C_{\rm eq}$, where the contribution of integer edge modes can be neglected. A low energy effective description of the two counter-propagating modes is then given by the free Lagrangian density
\begin{align}
    \mathcal{L}_0 = \frac{2}{4\pi}\partial_x \phi (\partial_t - v_\phi \partial_x)\partial_x\phi + i\psi(-\partial_t -v_\psi\partial_x)\psi,
\end{align}
where $v_\phi$ and $v_\psi$ are the mode velocities. Our goal is to compute the heat exchange between the two modes due to a coupling between them. The edge electron operator reads $\psi e^{i2\phi}$, so we cannot introduce electron or quasiparticle tunneling operators to couple these modes. Instead, the simplest, most relevant (in the renormalization group sense) operator coupling the channels is given by the density-density operator~\cite{Ma2020Jul}
\begin{align}
    \label{eq:heatinteraction}
    \mathcal{O} = \partial_x \phi \psi i \partial_x \psi.
\end{align}
We therefore add to $\mathcal{L}_0$ the point-like perturbation
\begin{align}\label{eq:LTPert}
    \mathcal{L}_T = \Gamma_0 \delta(x) \partial_x \phi \psi i \partial_x \psi,
\end{align}
where $\Gamma_0$ is the coupling constant, assumed to be weak. We thus seek the heat current and inter-mode thermal conductance induced by $\mathcal{L}_T$, treated as a weak perturbation. 

To do so, we consider the unperturbed bosonic energy current, which is given in terms of the stress energy tensor of the upstream bosonic field $\mathscr{T}_\phi \equiv \left( \partial_x \phi \right)^2$, as~\cite{Capelli2002}
\begin{align}
    J_{Q,\phi}^{(0)}(d,t) = \hbar \frac{v_\phi^2}{2 \pi} \mathscr{T}_\phi(\Tilde{t}).
\end{align}
Here, $\Tilde{t} = t - \frac{d}{v_\phi}$ is a shifted time referring to the transport of energy a small distance $d$ away from the point $x=0$. In the interaction picture, the average heat current in the presence of $\mathcal{L}_T$, can be written as
\begin{align}
    \langle J_{Q,\phi}(t) \rangle = \langle \mathcal{T} e^{\frac{i}{\hbar} \int_t H^\prime(t)} J_{Q,\phi}^{(0)}(\Tilde{t}) e^{-\frac{i}{\hbar} \int_t H^\prime(t)} \rangle,
\end{align}
where $H^\prime$ is the Hamiltonian corresponding to $\mathcal{L}_T$, and $\mathcal{T}$ denotes time ordering. We next expand the time-evolution operators up to $\mathcal{O}(\Gamma_0^2)$. Collecting terms corresponding to the same order in $\Gamma_0$, the first and second order correction to the heat current become
\begin{align}
    & J_{Q,\phi}^{(1)}(t) = \frac{i}{\hbar} \int_{-\infty}^{t} \,dt^\prime [H^{\prime}(t^\prime), J_{Q,\phi}^{(0)}(\Tilde{t})] 
\end{align}
and
\begin{align}
    & J_{Q,\phi}^{(2)}(t) = \frac{i^2}{\hbar^2} \int_{-\infty}^{t} \,dt^\prime \int_{-\infty}^{t^\prime} d t^{\prime \prime} [H^{\prime}(t^{\prime \prime}),[H^{\prime}(t^\prime),J_{Q,\phi}^{(0)}(\Tilde{t})]],
\end{align}
respectively. The commutators in these expressions can be computed with operator product expansions with the stress energy tensor~\cite{francesco2012conformal}. After some algebra, we find the corrections to the average energy current in terms of the modes Green's functions as
\begin{align}
    & \langle J_{Q,\phi}^{(1)}(t) \rangle = 0 \label{eq:firstordercurrent} 
    \intertext{and}
    & \langle J_{Q,\phi}^{(2)}(t) \rangle = \frac{\Gamma_0^2}{\hbar} \int_{-\infty}^{\infty} \big(\partial_{\tau} G_\phi(\tau)\big) \mathcal{G}_\psi(\tau) \,d\tau \label{eq:secondordercurrent}.
\end{align}
In Eq.~\eqref{eq:secondordercurrent}, the integrand is given in terms of the finite temperatures Green's functions, $G_\phi(\tau)$ and $\mathcal{G}_\psi(\tau)$~\cite{francesco2012conformal}, which read
\begin{align}
    G_\phi(\tau) &\equiv \langle \partial_x \phi(\tau,0) \partial_x \phi(0,0) \rangle \notag \\
    &= \left(\frac{\pi b k_{\rm B} T_\phi}{\hbar v_\phi} \csc\left( \frac{\pi k_{\rm B} T_\phi}{\hbar v_\phi} (b - i v_\phi\tau) \right) \right)^2 \label{eq:GF_bosheat}.
\end{align}
with $b$ the UV cutoff, and
\begin{align}
    \mathcal{G}_\psi(\tau) &\equiv \langle \mathscr{T}_\psi(\tau,0) \mathscr{T}_\psi(0,0) \rangle \notag \\
    &= \left(\frac{\pi b k_{\rm B} T_\psi}{\hbar v_\psi} \csc\left( \frac{\pi k_{\rm B} T_\psi}{\hbar v_\psi} (b - i v_\psi \tau) \right) \right)^4 \notag \\
    &+ \frac{c^2}{36} \left( \frac{\pi k_{\rm B} T_\psi}{\hbar  v_\psi} \right)^4 \label{eq:GF_MMheat},
\end{align}
where $c=1/2$ and $\mathscr{T}_\psi = \psi i \partial_x \psi$ for the MM $\psi$. 

For completeness, we give also the correlation function for a tunneling operator of the form $\mathcal{O}\sim \psi e^{i 2 \phi_1 + \phi_2}$. From the statistical independence of the involved fields we obtain
\begin{align}
    \langle \mathcal{O}(\tau)\mathcal{O}^{\dagger} (0)\rangle \propto G_{\phi_1}(\tau)G_{\phi_2}(\tau) G_{\psi}(\tau),
\end{align}
with
\begin{align}
    G_\phi(\tau) &\equiv \langle e^{i \nu_\phi \phi(\tau,0)} e^{i \nu_\phi \phi(0,0)} \rangle \notag \\
    &= \left(\frac{\pi b k_{\rm B} T_\phi}{\hbar v_\phi} \csc\left( \frac{\pi k_{\rm B} T_\phi}{\hbar v_\phi} (b - i v_\phi\tau) \right) \right)^{1/\nu_\phi} \label{eq:GF_bos}
    \intertext{in which $\nu_\phi \in \{1/2,1\}$, and}
    G_\psi(\tau) &\equiv \langle \psi(\tau,0) \psi(0,0) \rangle \notag \\
    &= \left(\frac{\pi b k_{\rm B} T_\psi}{\hbar v_\psi} \csc\left( \frac{\pi k_{\rm B} T_\psi}{\hbar v_\psi} (b - i v_\psi \tau) \right) \right) \label{eq:GF_MM}.
\end{align}
Next, we insert the expressions for $G_\phi(\tau)$ from Eq.~\eqref{eq:GF_bosheat} and $\mathcal{G}_\psi(\tau)$ from Eq. \eqref{eq:GF_MMheat} into Eq.~\eqref{eq:secondordercurrent} and shift variables $\{ \tau \to \tau + i \frac{b}{v_\phi} - i \frac{\hbar \beta_\phi}{2} \}$. Following Ref.~\cite{Martin2005}, the integral boundaries are switched back for a properly introduced cut-off $b$ that satisfies $\hbar v_\phi \beta_\phi > b$. Setting the temperatures $T_\psi = \bar{T} + \frac{\Delta T}{2}$ and $T_\phi = \bar{T} - \frac{\Delta T}{2}$, gives to leading order in $\Delta T$ the interaction induced heat current
\begin{align}
    \langle J_{Q,\phi}^{(2)}(t) \rangle = \frac{8 b^2 k_{\rm B}^6 \pi^7 \Gamma_0^2}{105 \hbar^7 v_\phi^2 v_\psi^4} \bar{T}^5 \Delta T \ .
\end{align}
The corresponding interaction thermal conductance thus reads
\begin{align}
    \label{eq:heatinteractioncond}
    G_{\rm int}^Q = \lim_{\Delta T \to 0} \left(\frac{d}{d \Delta T} \langle J_{Q,\phi}^{(2)} \rangle\right) = \frac{8 b^2 k_{\rm B}^4 \pi^5 \Gamma_0^2 }{35 \hbar^6 v_\phi^2 v_\psi^4}  \bar{T}^4 \kappa_0 \bar{T}.
\end{align}
As our final step, we consider an array of points, distanced with the length $a$, with couplings on the form~\eqref{eq:LTPert} and take the continuum limit (see Refs.~\cite{Park2019,Spanslatt2020,Asasi2020,Srivastav2021May}). This procedure relates the conductance~\eqref{eq:heatinteractioncond} to the thermal equilibration length as
\begin{align}
    \ell_{\rm eq}^Q = \frac{a}{g\gamma_{\phi,\psi} (n_\phi - n_\psi)} = \frac{2 a}{g \gamma_{\phi,\psi}} = \frac{2 a \kappa_0 \bar{T}}{G_{\rm int}^Q \gamma} \sim \bar{T}^{-4},
\end{align}
which is Eq.~\eqref{eq:eqlength}. For the interaction~\eqref{eq:LTPert}, we have $\gamma_{\phi,\psi}=1$.

\section{Computation of noise kernels}
\label{sec:NoisekernelAppendix}
Our approach to compute noise kernels $\Lambda(x)=S^{\rm loc}(x)/[2g^{\rm loc}(x)]$ follows that in Ref.~\cite{Kumar2022Jan}. The local noise $S^{\rm loc}$ and local tunneling conductance $g^{\rm loc}$ generically read
\begin{align}
    & S^{\rm loc}(x) \approx 4 \int_{- \infty}^{\infty} \langle \mathcal{O}(\tau, 0) \mathcal{O}^\dagger(0,0) \rangle \,d\tau
    \intertext{and}
    & g^{\rm loc}(x) \approx 2 i \int_{-\infty}^{\infty} \tau \langle \mathcal{O}(\tau, 0) \mathcal{O}^\dagger(0,0) \rangle \,d\tau \ .
\end{align}
Here, we have assumed zero voltage difference between the edge channels, since at the noise spot, edge channels equilibrate to the same electrochemical potential. Furthermore, $\mathcal{O}$ denotes the most relevant tunneling operator coupling charged edge channels, and the corresponding correlation function can be expressed as a product of Green's functions [see Eqs.~\eqref{eq:GF_bos}-\eqref{eq:GF_MM}] as
\begin{align}
   \langle \mathcal{O}(\tau, 0) \mathcal{O}^\dagger(0,0) \rangle = \frac{\Gamma_0^2}{(2 \pi b)^2} \prod_k G_{k}(\tau,0). \label{eq:correlator}
\end{align}
Here, $k \in \{\psi, \phi_{i \in \{1,2\}} \}$, $b$ is a short distance cut-off, and $\Gamma_0$ is the bare coupling amplitude. As follows, we compute $\Lambda(x)$ for various interfaces in the two limiting cases of efficient and absent thermal equilibration. We emphasize that all $x$-dependence in $\Lambda(x)$ enters in the mode temperature profiles $T_{k}(x)$. 

\subsection{\texorpdfstring{Voltage biased charge current noise for absent thermal equilibration}{}}
\label{sec:ChargeNoiseAbsEqdV}
\subsubsection{\texorpdfstring{phPf-$3$}{}}
For this interface, equilibrated charge transport is from right to left (see Fig.~\ref{fig:interfaced_edge_structures}). The hot spot and noise spot are therefore inter-changed in comparison to  Fig.~\ref{fig:noise_setup_1}. Heat from the hot spot is transported upstream by the bosonic 1/2 mode. Absence of thermal equilibration leads, with the procedure outlined in Sec.~\ref{eq:absent_thermal}, to the temperatures $T_{\phi_1} = T_{\psi} = T_+= 0$ and $T_{\phi_2} = T_- = \sqrt{\frac{4 P_0}{5 \kappa_0}}$. Following the same approach as in the previous section, using Eqs.~\eqref{eq:GF_bos} and \eqref{eq:GF_MM}, we arrive at the noise kernel
\begin{align}
    & \Lambda^{\Delta V}_{{\rm phPf-}3} = \frac{12 \zeta(3)}{\pi^2} k_{\rm B} T_-,
\end{align}
where $\zeta(z)$ is the Riemann zeta-function. 

\subsubsection{\texorpdfstring{aPf-$2$ and Pf-$3$}{}}
In treating the interfaces aPf-$2$ and Pf-$3$, we notice that they are constructed by the same set of modes but in opposite directions. They are therefore expected to generate the same charge current noise and thus to have the same form of $\Lambda(x)$, also in absence of thermal equilibration. With use of Eqs.~\eqref{eq:TmApprox}-\eqref{eq:TpApprox}, the modes' temperatures are given by $T_{\phi_1} = 0$, $T_{\phi_2} \approx T_{\psi} = T_- = \sqrt{\frac{6 P_0}{5 \kappa_0}}$,  we arrive at
\begin{align}
    & \Lambda^{\Delta V}_{{\rm aPf-}2} = \Lambda^{\Delta V}_{{\rm Pf-}3} \approx 1.604 k_{\rm B} T_- .
\end{align}

\subsubsection{\texorpdfstring{phPf-$2$}{}}
For the phPf-$2$ interface, we take into account an additional pair of integer modes which are not thermally equilibrated (see the discussion in Sec.~\ref{eq:absent_thermal} and Fig.~\ref{fig:noise_phPfm2}). The charge transport equation~\eqref{eq:charge_transport_equation} then takes the form
\begin{align}
    & \partial_x \Vec{V}(x) = \frac{\delta_{12}}{L} \begin{pmatrix} -\chi_1/\nu_1 & \chi_1/\nu_1 & 0 \\ \chi_2/\nu_2 & - \chi_2/\nu_2 & 0 \\ 0 & 0 & 0 \end{pmatrix} \begin{pmatrix} V_1(x) \\ V_2(x) \\ V_3(x) \end{pmatrix} \notag \\
    & \phantom{\partial_x V(x)} + \frac{\delta_{23}}{L} \begin{pmatrix} 0 & 0 & 0 \\ 0 & -\chi_2/\nu_2 & \chi_2/\nu_2 \\ 0 & \chi_3/\nu_3 & - \chi_3/\nu_3 \end{pmatrix} \begin{pmatrix} V_1(x) \\ V_2(x) \\ V_3(x) \end{pmatrix} 
\end{align}
with $\nu_1=\nu_2=1$, $\nu_3=1/2$ and $\chi_1=-\chi_2=\chi_3=1$. The boundary conditions are
\begin{align}
    & V_1(0) = \Delta V ,\quad V_2(1) = 0 \ \ \text{, and} \ \ V_3(0) = \Delta V .
\end{align}
To ensure a quantized charge conductance, we choose $\delta_{12} \gg \delta_{23}$. In general it is not possible to give accurate estimates on $\delta_{12}$ and $\delta_{23}$ without further experimental data. For our purpose it is sufficient to adjust them such that $G_{\rm 2T}/(e^2/h) = 1/2$ to good accuracy. Within this description, charge partitioning occurs mainly within the region $0 \lesssim x \lesssim \delta_{12}$. Furthermore, the Joule heating contribution is non-zero and leads to a total dissipated power of
\begin{align}
    & P_0 = \frac{3}{\pi^2} \frac{e^2 V_0^2}{h \kappa_0} \int_{0}^{\delta_{12}} \big[ \delta_{12} (V_1 - V_2)^2 + \delta_{23} (V_2 - V_3)^2 \big] \,dx,
\end{align}
where we omitted the $x$-dependence of $V_i$ for notational ease. The process of interest for the noise characteristics on the phPf-$2$ edge is the charge tunneling between the bosonic channels of filling $\nu_1 = 1$ and the electronic mode comprised of $\nu_2 = 1/2$ and the MM $\psi$. Then, the noise kernel for absent thermal equilibration is the same as in the phPf-$3$ case with the temperature $T_-=\sqrt{\frac{2 P_0}{3 \kappa_0}}$. Following the same steps as before, the noise is computed vs the charge equilibration length $\delta_{23}$, giving a noise of similar magnitude as for absent thermal equilibration on the phPf-$3$ edge for a ratio $3 \lesssim \delta_{12}/\delta_{23} \lesssim 5$. The deviation from the expected $G_{2 T}$, is less than $0.5 \%$ if $\delta_{23} \gtrsim 8$ which is what we assume here.

\subsection{\texorpdfstring{Downstream delta-$T$ noise for absent thermal equilibration}{}}
\label{subsec:Noisekernel_ds_abs}   
For this type of noise, the appropriate boundary conditions are given in Eq.~\eqref{eq:DSBCS}. We consider the limiting cases of a weak bias (wb), $\Delta T \ll\bar{T}$ and strong bias (sb), $\Delta T \gg \bar{T}$ by expanding the expressions to first order in $\Delta T/\bar{T}$  and setting $\bar{T}=0$, respectively. We refer to the corresponding noise kernels as $\Lambda^{\rm (wb)}(x)$ and $\Lambda^{\rm (sb)}(x)$, respectively. Following the same steps as in Appendix~\ref{sec:MajoranaAppendix}, using a shift of variables such that the mode with the largest temperature, say $T_m$ and thus smallest $\beta_m \equiv [k_{\rm B} (\bar{T} + \Delta T)]^{-1}$ fulfills $\hbar v_m \beta_m > b$ and performing the integrals, we arrive at our desired expressions. In case of weak applied bias $\bar{T} \gg \Delta T$, we find to first order in $\Delta T$ 
\begin{subequations}
    \label{eq:lambda_ds_wb}
    \begin{align}
        & \Lambda^{\rm (wb)}_{{\rm phPf-}3}(x) = 2 k_{\rm B} \bar{T} + k_{\rm B} \Delta T,\label{eq:Lphpf3wbds}\\
        & \Lambda^{\rm (wb)}_{{\rm aPf-}2}(x) = 2 k_{\rm B} \bar{T} + \frac{k_{\rm B}}{2} \Delta T,\\
        & \Lambda^{\rm (wb)}_{{\rm phPf-}2}(x) = 2 k_{\rm B} \bar{T} + \frac{6 k_{\rm B}}{5} \Delta T,
    \end{align}
\end{subequations}
which reduce to the equilibrium NJ form for vanishing applied bias $\Delta T$ as expected. For large bias $\Delta T \gg \bar{T}$ we obtain instead
\begin{subequations}
    \label{eq:lambda_ds_sb}
    \begin{align}
        & \Lambda^{\rm (sb)}_{{\rm phPf-}3}(x) = \frac{12  \zeta(3)}{\pi^2} k_{\rm B}\Delta T,\label{eq:Lphpf3sbds}\\
        & \Lambda^{\rm (sb)}_{{\rm aPf-}2}(x) = \frac{9 \zeta(3)}{\pi^2} k_{\rm B}\Delta T, \\
        & \Lambda^{\rm (sb)}_{{\rm phPf-}2}(x) = \frac{17 \pi^4}{60 (\pi^2 \log(4) + 3 \zeta(3))} k_{\rm B}\Delta T.
    \end{align}
\end{subequations}

\subsection{\texorpdfstring{Upstream delta-$T$ noise for absent thermal equilibration}{}}
For the upstream delta-$T$ noise, we use the boundary conditions from Eq.~\eqref{eq:USBCS}. The noise kernels are computed in a similar manner as for the downstream delta-$T$ noise. To first order in $\Delta T$ we find for weak applied biases $\Delta T\ll\bar{T}$
\begin{subequations}
    \label{eq:lambda_us_wb}
    \begin{align}
        & \Lambda^{\rm (wb)}_{{\rm phPf-}3}(x) = 2 k_{\rm B} \bar{T} + k_{\rm B}  \Delta T,\label{eq:Lphpf3wbus}\\
        & \Lambda^{\rm (wb)}_{{\rm aPf-}2}(x) = 2 k_{\rm B} \bar{T} + \frac{3 k_{\rm B}}{2} \Delta T,\\
        & \Lambda^{\rm (wb)}_{{\rm phPf-}2}(x) = 2 k_{\rm B} \bar{T} + \frac{4 k_{\rm B}}{5} \Delta T
    \end{align}
\end{subequations}
which reduce to equilibrium noise for $\Delta T \to 0$. For strong bias $\Delta T \gg \bar{T}$, we find 
\begin{subequations}
    \label{eq:lambda_us_sb}
    \begin{align}
        & \Lambda^{\rm (sb)}_{{\rm phPf-}3}(x) = \frac{12  \zeta(3)}{\pi^2} k_{\rm B}\Delta T,\label{eq:Lphpf3sbus}\\
        & \Lambda^{\rm (sb)}_{{\rm aPf-}2}(x) = \left( \log(4) + \frac{3 \zeta(3)}{\pi^2} \right) k_{\rm B}\Delta T, \\
        & \Lambda^{\rm (sb)}_{{\rm phPf-}2}(x) = \frac{\pi^4}{60 \zeta(3)} k_{\rm B}\Delta T.
    \end{align}
\end{subequations}

\subsection{\texorpdfstring{Downstream-upstream delta-$T$ noise symmetry of phPf-$3$ at poor thermal equilibration}{}}
\label{sec:fulTdepLambda}
As depicted in Fig.~\ref{fig:Lambda_dT}, the most striking feature is that the downstream and upstream noise kernels are uniquely equal for the phPf-$3$ interface (blue curves in Fig.~\ref{fig:Lambda_dT}). This equality is clearly manifest also in the asymptotic expressions: compare~\eqref{eq:Lphpf3wbds} and \eqref{eq:Lphpf3wbus}, respectively \eqref{eq:Lphpf3sbds} and \eqref{eq:Lphpf3sbus}.

Mathematically, this ``downstream-upstream symmetry'' follows from the fact that the exponent of the downstream and upstream sectors in the product of edge mode Green's functions [see Eq.~\eqref{eq:correlator}] are equal. More specifically, we have for the phPf-$3$ interface that
\begin{align}
\label{eq:phpf3Relation}
    & G_{\phi_1}(\tau,\bar{T}+\Delta T)G_{\phi_2}(\tau,\bar{T})G_\psi(\tau,\bar{T}+\Delta T)\notag \\
    & \propto G_{\phi_1}(\tau,\bar{T})G_{\phi_2}(\tau,\bar{T}+\Delta T)G_\psi(\tau,\bar{T}).
\end{align}
Here, the first and second line correspond to downstream and upstream bias conditions, respectively. Moreover, the proportionality factor in Eq.~\eqref{eq:phpf3Relation}, which includes powers of mode velocities and temperatures, crucially drops out when dividing the noise with the conductance to obtain $\Lambda$, see Eq.~\eqref{eq:Lambda}.
Relations similar to~\eqref{eq:phpf3Relation} do not hold for any other edge structure in Fig.~\ref{fig:interfaced_edge_structures}. Our proposal to test this symmetry is presented in the end of Sec.~\ref{sec:Device_proposal}. 

\subsection{Some integrals and their computation}
\label{sec:integrals}
In computing the noise kernel for a voltage biased phPf-$3$ interface, we face integrals
\begin{align}
    & S^{\rm loc} \propto \int_{-\infty}^\infty \frac{ \text{sech}^2(z)}{\left( \pi + 2 i z \right)^2} \,dz, \label{eq:Sloc1} \\
    & g^{\rm loc} \propto \int_{-\infty}^\infty \frac{ \text{sech}^2(z)}{\pi + 2 i z} \,dz \label{eq:gloc1} .
\end{align}
Integrals of this kind can be solved by using Mittag-Lefflers' theorem~\cite{TURNER201336}, which amounts to expanding the hyperbolic functions as
\begin{align}
    & \text{sech}^2(z) = - \sum_{k=0}^\infty \left[ \frac{1}{(z-A)^2} + \frac{1}{(z+A)^2} \right], 
\end{align}
where $A = i \frac{\pi}{2}(2 k +1)$. Similar expansions can be found for other hyperbolic functions, e.g., $\text{sech}(z)$. Inserting the series expansion back into Eqs.~\eqref{eq:Sloc1}-\eqref{eq:gloc1}, exchanging the order of integration and summation, we arrive at
\begin{align}
    S^{\rm loc} \propto \frac{\zeta(3)}{\pi^2} \quad {\rm and} \quad g^{\rm loc} \propto \frac{\pi}{6}.
\end{align}
For other integrals, complex contour integration and the residue theorem are more useful. For example, in the case of delta-$T$ noise at phPf-$2$ we encounter an integral
\begin{align}
    & S^{\rm loc} \propto \int_{-\infty}^\infty \frac{1}{\cosh(z)^3 (\pi + 2 i z)^2} \,dz .
\end{align}
Substituting $z \to 2 \pi t$ and manipulating the resulting expression leads to
\begin{align}
    & S^{\rm loc} \propto \frac{1}{8 \pi} \int_{-\infty}^\infty \frac{1}{\cosh(2 \pi t)^3 (\frac{1}{4} - i t)^2} \,dt .     
\end{align}
The right-hand-side can now be written as a derivative
\begin{align}
\label{eq:ex_int}
    S^{\rm loc} &\propto -\frac{1}{8 \pi} \int_{-\infty}^\infty \frac{\partial^2}{\partial a^2} \left( \frac{\ln(a - i t)}{\cosh(2 \pi t)^3} \right)\Bigg|_{a=1/4} \,dt \notag \\
    &\equiv-\frac{1}{8 \pi} \frac{\partial^2}{\partial a^2} J(a)\Big|_{a=1/4}.
\end{align}
We next write the function $J(a)$ as an integral along the closed rectangular contour $\gamma(z)$, defined by
\begin{align}
    -R \to R \to R + i \to -R + i \to -R, \ R \in \mathbb{R}^+,
\end{align}
in the complex $z$-plane as
\begin{align}
    \label{eq:svyatoslav}
    J(a) &= - \oint_{\gamma(z)} \frac{\ln\left(\Gamma(a - i z)\right)}{\cosh(2 \pi z)^3} dz\notag \\
    &= - 2 \pi i \sum_i \text{Res} \left( \frac{\ln\left(\Gamma(a - i z)\right)}{\cosh(2 \pi z)^3}, z_i\right),
\end{align}
where we used the residue theorem in the final step with the two third order poles $z_1 = i/4$ and $z_2= 3i/4$ enclosed by $\gamma(z)$. After some additional algebraic manipulations, we arrive at
\begin{align}
\label{eq:Ja_func}
    J(a) &= \frac{1}{8 \pi^2} \Big[ \psi^{(1)} \big(a + 3/4\big) - \psi^{(1)} \big(a + 1/4\big) \Big] \notag \\
    &+ \frac{1}{2} \Big[ \ln\big( \Gamma \left(a + 3/4\right) \big) - \ln\big( \Gamma \left(a + 1/4\right) \big) \Big],
\end{align}
where $\psi^{(1)}(z)=\partial^2_z \ln\Gamma[z]$ is the trigamma function.
Combining Eqs.~\eqref{eq:ex_int} and~\eqref{eq:Ja_func} gives
\begin{align}
    \frac{1}{8 \pi} \int_{-\infty}^\infty \frac{1}{\cosh(2 \pi t)^3 (\frac{1}{4} - i t)^2} \,dt  = \frac{17 \pi}{480}
\end{align}
The same method be can be used to produce the following identity
\begin{align}
    \int_{-\infty}^\infty \frac{1}{\cosh(2 \pi t)^3 (a - i t)^n} \,dt = \frac{(-1)^{n-1}}{(n-1)!} \frac{\,d^n J(a)}{\,d a^n}, \ n \in \mathbb{Z}^{+} .
\end{align}

\section{\texorpdfstring{Proof of the universality of $G^Q_{\rm H}$ for equal thermal equilibration on top and bottom edges}{}}
\label{sec:thermal_Hall_appendix}
Here, we prove Eq.~\eqref{eq:GQH}, namely that the thermal Hall conductance~\eqref{eq:GQ_H}  is universal as long as the two edges have the same degree of equilibration (even if it is poor).

We consider a generic edge structure with $k$ downstream (ds) and $N-k$ upstream (us) modes, and write the heat transport equation~\eqref{eq:energy_transport_equation} (with $\delta \vec{V}=0$, since we assume no voltage bias) as 
\begin{align}
    \label{eq:heat_transport_theta}
    &\partial_x \Vec{\theta}(x) = \mathcal{M}_T \Vec{\theta}(x).
\end{align}
Here, we defined $\Vec{\theta}(x)= \Vec{T}^2(x)$, and the matrix $\mathcal{M}_T$~\eqref{eq:MT} satisfies the heat current conservation law
\begin{align}
    \label{eq:heat_conservation}
    \sum_{j}(\mathcal{M}_T)_{ij} = 0, \quad \forall i.
\end{align}
The general solution of Eq.~\eqref{eq:heat_transport_theta} for an edge segment with length $L>0$ can be written as
\begin{align}
    \vec{\theta}(L) = e^{L \mathcal{M}_T}\vec{\theta}(0),
\end{align}
which we express in ds and us blockform as
\begin{align}
\label{eq:heat_blockform}
    & \begin{pmatrix} \vec{\theta}_{\rm ds}(L) \\ \vec{\theta}_{\rm us}(L) \end{pmatrix} = \left( \begin{array}{c|c} \mathcal{A} & \mathcal{B} \\[3pt] \hline \\[-1.5\medskipamount] \mathcal{C} & \mathcal{D} \end{array} \right) \begin{pmatrix} \vec{\theta}_{\rm ds}(0) \\ \vec{\theta}_{\rm us}(0) \end{pmatrix}.
    \end{align}
 Here, the block matrices $\mathcal{A} \in \mathbb{R}^{k \times k}$, $\mathcal{B} \in \mathbb{R}^{k \times (N-k)}$, $\mathcal{C} \in \mathbb{R}^{(N-k) \times k}$, and $\mathcal{D} \in \mathbb{R}^{(N-k) \times (N-k)}$. 
 In terms of these matrices, the heat conservation law~\eqref{eq:heat_conservation} translates to
 \begin{align}
     &\sum_{j}\mathcal{A}_{ij} + \sum_{j}\mathcal{B}_{ij} = 1, \quad \forall i,\label{eq:hc1}\\
     &\sum_{j}\mathcal{C}_{ij} + \sum_{j}\mathcal{D}_{ij} = 1, \quad \forall i \label{eq:hc2}.
 \end{align} 
 Next, we rearrange the terms in Eq.~\eqref{eq:heat_blockform} as
 \begin{align}
 \label{eq:heat_blockform_2}
    & \begin{pmatrix} \vec{\theta}_{\rm ds}(L) \\ \vec{\theta}_{\rm us}(0) \end{pmatrix} = \left( \begin{array}{c|c} \mathcal{A} - \mathcal{B} \mathcal{D}^{-1} \mathcal{C} & \mathcal{B} \mathcal{D}^{-1} \\[3pt] \hline \\[-1.5\medskipamount] -\mathcal{D}^{-1} \mathcal{C} & \mathcal{D}^{-1} \end{array} \right) \begin{pmatrix} \vec{\theta}_{\rm ds}(0) \\ \vec{\theta}_{\rm us}(L) \end{pmatrix} .
\end{align}  
Let us now consider the top edge, for which the boundary conditions read $\vec{\theta}_{\rm ds}(0)=(\bar{T}+\Delta T)^2\times (1,\hdots,1)^T_k$ and $\vec{\theta}_{\rm us}(L)=\bar{T}^2\times (1,\hdots,1)^T_{N-k}$ (see Fig.~\ref{fig:conductance_setup}). Plugging these quantitites into Eq.~\eqref{eq:heat_blockform_2}, we write the top edge heat current~\eqref{eq:JQ_top} as
\begin{align}
     J_Q^{\rm top} &= \frac{\kappa_0}{2}\left(\sum_{i:\chi_i=+1}n_i \theta^{i}_{\rm ds}(L)- \bar{T}^2 \sum_{i:\chi_i=-1}n_i\right)\notag\\
     & = \frac{\kappa_0}{2}\Big((\bar{T}+\Delta T)^2\sum_{i,j:\chi_i=+1}n_i (\mathcal{A} - \mathcal{B} \mathcal{D}^{-1} \mathcal{C})_{ij} \notag \\
     &+ \bar{T}^2 \sum_{i,j:\chi_i=+1}n_i(\mathcal{B} \mathcal{D}^{-1})_{ij} -  \bar{T}^2 \sum_{i:\chi_i=-1}n_i\Big),
\end{align}
where in the second line we used Eq.~\eqref{eq:heat_blockform_2}.
For the bottom edge, we have reversed boundary conditions $\vec{\theta}_{\rm ds}(0)=\bar{T}^2\times (1,\hdots,1)^T_k$ and $\vec{\theta}_{\rm us}(L)=(\bar{T}+\Delta T)^2\times (1,\hdots,1)^T_{N-k}$. The bottom edge heat current~\eqref{eq:JQ_bot} then reads
\begin{align}
     J_Q^{\rm bot} &= -\frac{\kappa_0}{2}\left(\sum_{i:\chi_i=+1}n_i \theta^{i}_{\rm ds}(L)- \bar{T}^2 \sum_{i:\chi_i=-1}n_i\right)\notag \\
     & = \frac{\kappa_0}{2}\Big(\bar{T}^2\sum_{i,j:\chi_i=+1}n_i (\mathcal{A} - \mathcal{B} \mathcal{D}^{-1} \mathcal{C})_{ij}\notag \\
     &+ (\bar{T}+\Delta T)^2 \sum_{i,j:\chi_i=+1}n_i(\mathcal{B} \mathcal{D}^{-1})_{ij} \notag \\
     &-(\bar{T}+\Delta T)^2 \sum_{i:\chi_i=-1}n_i\Big).
\end{align}
The crucial next step is the assumption that \textit{the degrees of thermal equilibration on the top and bottom edges are identical}. This translates to identical block matrices $\mathcal{A}$, $\mathcal{B}$, $\mathcal{C}$, and $\mathcal{D}$ for the two edges. Then, and only then, can we combine the two currents as
\begin{align}
\label{eq:J_net}
   & J_Q^{\rm top} - J_Q^{\rm bot} = \frac{\kappa_0}{2}\left((\bar{T}+\Delta T)^2 + \bar{T}^2\right) \notag \\
   &  \times \Big(\sum_{i,j:\chi_i=+1} n_i (\mathcal{A} - \mathcal{B} \mathcal{D}^{-1} \mathcal{C}+ \mathcal{B}\mathcal{D}^{-1} )_{ij} - \sum_{i:\chi_i=-1} n_i \Big).
\end{align}
Now, by using Eqs.~\eqref{eq:hc1}-\eqref{eq:hc2}, and identifying 
\begin{align}
    &\sum_{i:\chi_i=+1} n_i = c, \\
    &\sum_{i:\chi_i=-1} n_i = \bar{c},
\end{align}
the dependence on $\mathcal{A},\mathcal{B},\mathcal{C},$ and $\mathcal{D}$ cancels out, and we find that Eq.~\eqref{eq:J_net} reduces to
\begin{align}
    J_Q^{\rm top} - J_Q^{\rm bot} = \frac{\kappa_0}{2}\left((\bar{T}+\Delta T)^2 + \bar{T}^2\right) (c-\bar{c}).
\end{align}
Finally, inserting this expression into the definition of $G^Q_H$, given in Eq.~\eqref{eq:GQ_H}, gives our desired result~\eqref{eq:GQH}. Our proof generalizes the theoretical analysis for $\nu=2/3$ in Ref.~\cite{Melcer2023Jan} to any edge structure.

\section{\texorpdfstring{Computation of voltage and temperature profiles for the phPf-$3$ interface}{}}
\label{sec:Profile_Appendix}
We compute the noise and thermal conductance of all considered edge structure by using voltage and temperature profiles of the edge channels within the incoherent tunneling model introduced in Sec.~\ref{sec:Model_Setup}. Below, we present in detail their derivation for the phPf-$3$ interface using boundary conditions for charge current as well as down- and upstream delta-$T$ noise. Other edge structures are treated in a perfectly analogous manner. 

We begin by labelling the edge channels as $1,2,3$ from top to bottom according to Fig.~\ref{fig:interfaced_edge_structures}. Transport along the edge is characterized by the channel specific filling factors $\nu_i$, heat conductances $n_i$ and chiralities $\chi_i$, given by
\begin{enumerate}
    \item[1:] $\nu_1=1\phantom{/2}$, $n_1=1\phantom{/2}$, $\chi_1=+$,
    \item[2:] $\nu_2=1/2$, $n_2=1\phantom{/2}$, $\chi_2=-$,
    \item[3:] $\nu_3=0\phantom{/2}$, $n_3=1/2$, $\chi_3=+$.
\end{enumerate}
These values further specify the transport matrices $\mathcal{M}_V$ and $\mathcal{M}_T$ in Eqs.~\eqref{eq:MV} and~\eqref{eq:MT} as
\begin{align}
    \mathcal{M}_T &= \alpha \begin{pmatrix} -\chi_1 n_2 & \phantom{-}\chi_1 n_2 & 0 \\ \phantom{-}\chi_2 n_1 & -\chi_2 n_1 & 0 \\ 0 & 0 & 0 \end{pmatrix} \notag\\
    &+ \beta \begin{pmatrix} 0 & 0 & 0 \\ 0 & -\chi_2 n_3 & \phantom{-}\chi_2 n_3 \\ 0 & \phantom{-}\chi_3 n_2 & -\chi_3 n_2 \end{pmatrix} \label{eq:MTphpfm3}
    \intertext{and}
    \mathcal{M}_V &= \frac{1}{\ell^C_{\rm eq}} \begin{pmatrix} - \frac{\chi_1}{\nu_1} &  \phantom{-}\frac{\chi_1}{\nu_1} \\ \phantom{-}\frac{\chi_2}{\nu_2} & -\frac{\chi_2}{\nu_2}  \end{pmatrix}.
\end{align}

\subsection{\texorpdfstring{Charge current noise, $\Delta V \neq 0$, $\Delta T =0$}{}}
\begin{figure}[hb]
    \centering
    \includegraphics{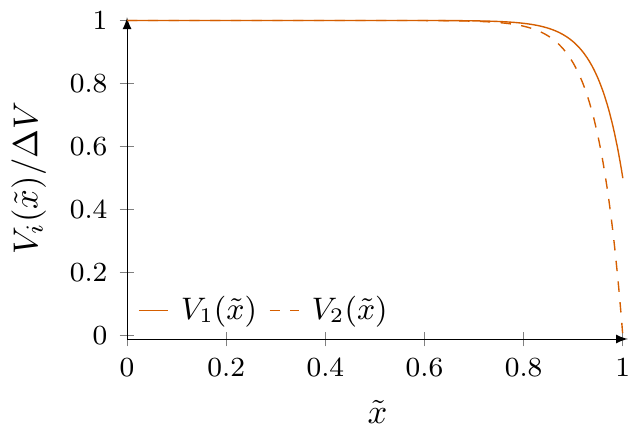}
    \caption{Channel resolved voltage profiles $V_1(x)$ and $V_2(x)$ of the charge carrying channels at phPf-$3$ using $\delta = 20$.}
    \label{fig:V-profile}
\end{figure}
\begin{figure*}[ht!]
    \centering
    \begin{tabular}{ccc}
        \subfloat[]{\includegraphics[scale=1]{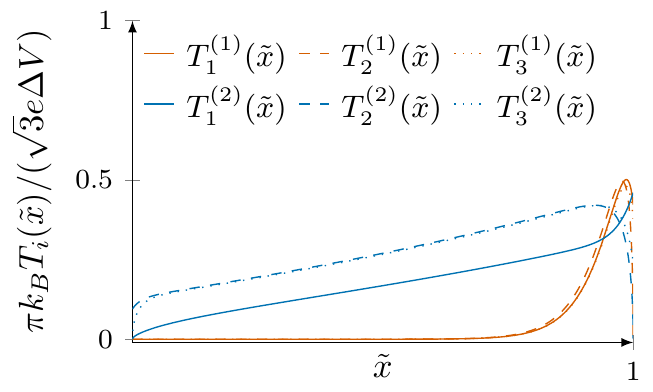}} & \hspace{1cm} & \subfloat[]{\includegraphics{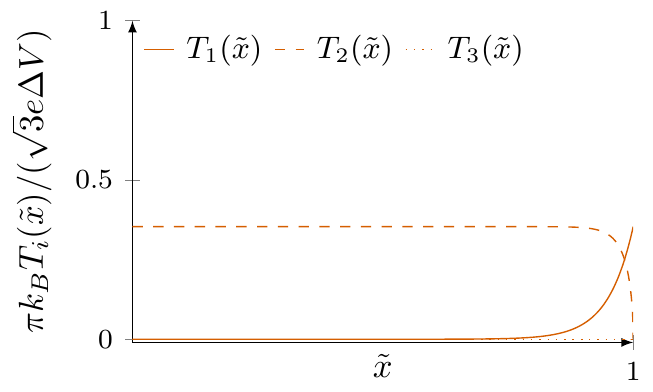}} \\
        \subfloat[]{\includegraphics[scale=1]{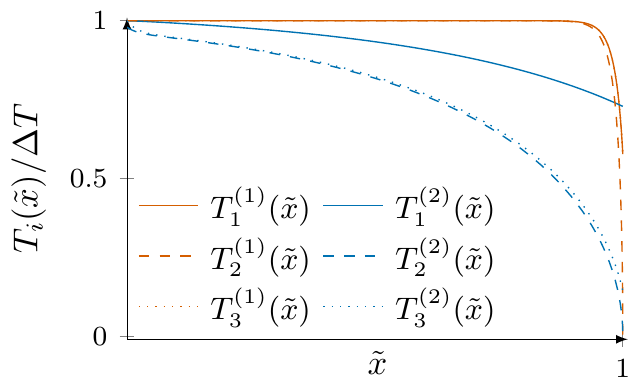}} & \hspace{1cm} & \subfloat[]{\includegraphics[scale=1]{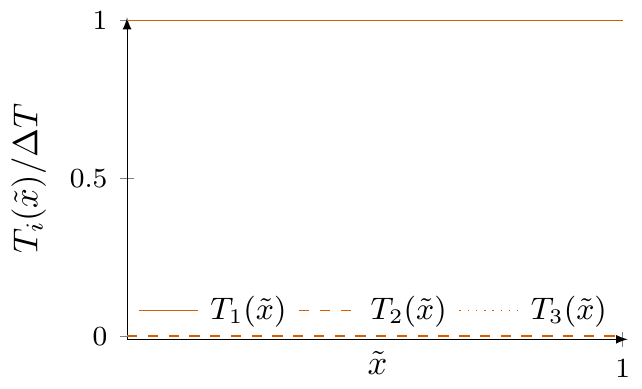}}
        \\
        \subfloat[]{\includegraphics[scale=1]{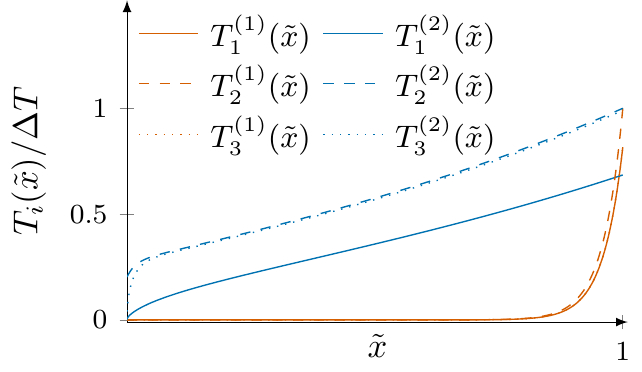}} & \hspace{1cm} & \subfloat[]{\includegraphics[scale=1]{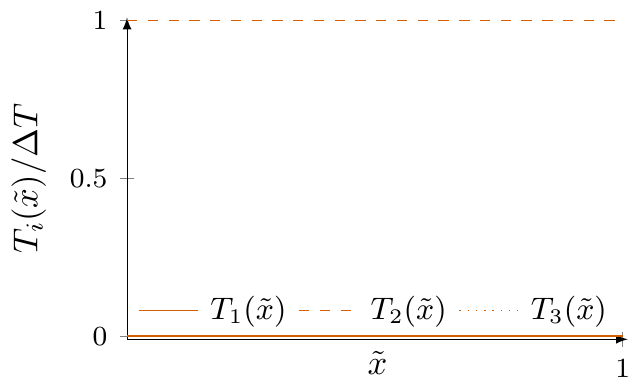}}
    \end{tabular}
    \caption{Channel resolved temperature profiles for channels $1$, $2$ and $3$ of the phPf-$3$ interface. The distance along the edge segment is parametrized $\Tilde{x}=x/L$. (a) Temperature profiles for voltage biased charge current noise for two sets of parameters: Superscript $^{(1)}$ labels profiles for the choice $(\alpha,\beta,\delta)=(100,100,20)$ (red lines ) and $^{(2)}$ are profiles for $ (\alpha,\beta,\delta)=(2,100,20)$ (blue lines). Both parameters sets correspond to finite thermal equilibration (b) Same as (a) but for poor thermal equilibration. (c)-(d) Same as (a)-(b) but for downstream delta-$T$ noise for strong temperature bias $\Delta T\gg \bar{T}$. (e)-(f) Same as (a)-(b) but for upstream delta-$T$ noise for strong temperature bias $\Delta T\gg \bar{T}$.}
    \label{fig:T-profiles}
\end{figure*}

For the charge current noise, the appropriate boundary conditions are
\begin{subequations}
    \begin{align}
        &T_i(0)&&=0, \ &&\text{for} \quad i \in \{1,2,3\} \\
        &V_i(0)&&=\Delta V, \ &&\text{for} \quad \chi_i = +1\\
        &V_i(1)&&=0, \ &&\text{for} \quad \chi_i = -1 .
    \end{align}
\end{subequations}
Here, we assumed that $e\Delta V \gg k_{\rm B}\bar{T}$ so that we may set the base temperature $\bar{T}\rightarrow 0$. The voltage profiles $\vec{V}(\Tilde{x})$ for the two charge carrying channels along the edge segment are obtained by solving Eq.~\eqref{eq:charge_transport_equation}. They are visualized in Fig.~\ref{fig:V-profile}. Clearly the voltage drop occurs only on the right-hand-side of the edge segment. This is the hot spot.  Note also that the voltage profiles are independent of the degree of thermal equilibration, since the processes leading to charge and thermal equilibration are considered to take place at different length scales. Knowledge of $\vec{V}(\Tilde{x})$ allows us to further compute the Joule-heating contribution in Eq.~\eqref{eq:Vdiff2}. For efficient thermal equilibration, we next solve Eq.~\eqref{eq:energy_transport_equation} and obtain temperature profiles which depend on the pairwise degrees of thermal equilibration: $\alpha = L/\ell^Q_{\rm eq,12}, \beta= L/\ell^Q_{\rm eq,23}$ [see Eq.~\eqref{eq:alpha_beta_2}], as well as the charge equilibration parameter $\delta=L/\ell^C_{\rm eq}$. Here, $L$ denotes the edge segment length and we define here $\Tilde{x} = x/L$ as a rescaled coordinate along the segment. The channel resolved temperature profiles are visualized in Fig.~\ref{fig:T-profiles}\textcolor{blue}{(a)} for two different sets of thermal equilibration lengths. We see that large $\alpha, \beta$ (red curves) are needed to have all channels at similar temperatures at the noise spot (i.e., for $\Tilde{x} \ll 1$) and thus to describe efficient thermal equilibration. This is contrasted by the blue curves for which $\alpha$ is not large and the channels are at different temperatures at the noise spot.
Absent thermal equilibration on the other hand corresponds to $\ell^Q_{\rm eq,ij} \to \infty$ and Eq.~\eqref{eq:energy_transport_equation} simplifies to
\begin{align}
    \partial_{\Tilde{x}} \vec{T^2}(\Tilde{x}) = \delta\vec{V}(\Tilde{x}).
\end{align}
The channel resolved temperature profiles for this solution are given in Fig.~\ref{fig:T-profiles}\textcolor{blue}{(b)}.

\subsection{\texorpdfstring{Downstream delta-$T$ noise, $\Delta V = 0$, $\Delta T \neq 0$}{}}
This type of noise corresponds to the boundary conditions
\begin{subequations}
    \begin{align}
        &V_i(0)&&=0=V_i(1), \ &&\text{for} \quad i \in \{1,2,3\}\\
        &T_i(0)&&=\bar{T}+\Delta T, \ &&\text{for} \quad \chi_i = + \\
        &T_i(1)&&=\bar{T}, \ &&\text{for} \quad \chi_i = - 
    \end{align}
\end{subequations}
which leads to trivial voltage profiles $V_i(\Tilde{x}) \equiv 0, \forall i$. As a consequence, the energy transport equation Eq.~\eqref{eq:energy_transport_equation} simplifies to
\begin{align}
    \label{eq:FdsdTeff}
    \partial_{\Tilde{x}}\vec{T^2}(\Tilde{x}) = \mathcal{M}_T \vec{T^2}(\Tilde{x}).
\end{align}
For efficient thermal equilibration, this equation is solved to obtain temperature profiles in terms of $(\alpha, \beta)$. Here, we consider the strong temperature bias limit, $\Delta T \gg \bar{T}$, and  consider the temperature profiles to leading order in $\Delta T/\bar{T}$. The resulting temperature profiles are shown in Fig.~\ref{fig:T-profiles}\textcolor{blue}{(c)}. In case of absent thermal equilibration, the energy transport equation reads
\begin{align}
    \label{eq:FdsdTabs} 
    \partial_{\Tilde{x}}\vec{T^2}(\Tilde{x}) = 0
\end{align}
Hence, the temperature profiles of the channels will be constant and follow immediately from the boundary conditions, see Fig.~\ref{fig:T-profiles}\textcolor{blue}{(d)}. 

\subsection{\texorpdfstring{Upstream delta-$T$ noise, $\Delta V = 0$, $\Delta T \neq 0$}{}}
For the upstream delta-$T$ noise, we use the boundary conditions
\begin{subequations}
    \begin{align}
        &V_i(0)&&=0=V_i(1), \ &&\text{for} \quad i \in \{1,2,3\}\\
        &T_i(0)&&=\bar{T}, \ &&\text{for} \quad \chi_i = + \\
        &T_i(1)&&=\bar{T}+\Delta T, \ &&\text{for} \quad \chi_i = - .
    \end{align}
\end{subequations}
Hence, the voltage profiles are also trivial $V_i(\Tilde{x}) \equiv 0, \forall i$ and we obtain the temperature profiles using Eq.~\eqref{eq:FdsdTeff} for efficient and Eq.~\eqref{eq:FdsdTabs} for absent thermal equilibration. The resulting temperature profiles are visualized in Fig.~\ref{fig:T-profiles}\textcolor{blue}{(e)} and Fig.~\ref{fig:T-profiles}\textcolor{blue}{(f)}, respectively.

\subsection{\texorpdfstring{Connection of temperature profiles to noise}{}}
The phPf-$3$ interface has $\nu_Q > 0$ so that for full thermal equilbiration, the heat transport is dominantly in the downstream direction [see Fig.~\ref{fig:interfaced_edge_structures}]. Any upstream heat transport is exponentially suppressed in $L$~\cite{Spanslatt2019}. For the situation of voltage biased charge current noise, in the regime of efficient thermal equilibration, we find non-vanishing temperatures only in the region $\Tilde{x} \approx 1$. This is seen from the red curves in Fig.~\ref{fig:T-profiles}\textcolor{blue}{(a)}. This characteristic region, called the \textit{hot spot}, is a result of Joule-heating close to the right (most downstream) contact and the dominantly downstream thermal transport. We see that the effective temperature at the \textit{noise spot} is very small: $T_{\rm ns} \approx 0$, and we expect small noise as well $S \simeq 0$. Indeed, by inserting the temperature profiles in the noise kernel $\Lambda(x) = k_{\rm B} T(\Tilde{x})$ in Eq.~\eqref{eq:noise_generation} and integrating, we obtain exponentially suppressed noise $S \sim \exp(-\delta)$, plotted in Fig.~\ref{fig:noisecomp}\textcolor{blue}{(a)}. 

For thermally equilibrated down- and upstream delta-$T$ noise, the boundary conditions for an edge with $\nu_Q > 0$ imply equal and constant temperatures of all channels at the noise spot [see Fig.~\ref{fig:T-profiles}\textcolor{blue}{(c)} and \textcolor{blue}{(e)}]. This leads to a hot spot  temperature $T_{\rm ns}=\bar{T} + \Delta T$ and $T_{\rm ns}=\bar{T}$ for down- and upstream delta-$T$ noise respectively. The corresponding noise profiles are shown in Fig.~\ref{fig:T-profiles}\textcolor{blue}{(c)}. Absent thermal equilibration on the other hand leads to constant but different temperatures of the channels at the noise spot for the three cases. We explore the out-of-equilibrium situation in the noise spot by using the Green's function method outlined in Sec.~\ref{sec:ChargeNoiseAbsEqdV} to compute noise kernels $\Lambda(\Tilde{x})$ with the full temperature profiles in case of charge current noise. This leads to the noise plots in Fig.~\ref{fig:noisecomp}\textcolor{blue}{(b)}. For thermally non-equilibrated down- and upstream delta-$T$ noise there is no additional source of heating along the edge. Following the same approach as for charge current noise in Sec.~\ref{sec:ChargeNoiseAbsEqdV}, we find a modified NJ noise for down- and upstream delta-$T$ noise, described by Eq.~\eqref{eq:S_ds_abs} and Eq.~\eqref{eq:S_us_abs} respectively. Adjusting $\mathcal{M}_V$, $\mathcal{M}_T$ as well as the boundary conditions according to the transport properties of the involved channels, the temperature and noise profiles of the other interfaces in Fig.~\ref{fig:noisecomp} are obtained following the same steps.

\section{Unit conversion without tears: from theorist to experimentalist units}
\label{sec:Conversion}
The voltage bias noise excess noise $S^{\rm exc}$ computed using our approach is expressed in units of $e^3 \Delta V/h$. Let us denote the noise in such units as $S^{\rm exc}_{\Delta V}$. To connect these units to more experimentally relevant units, we start by using the current-voltage relation for chiral edge transport 
\begin{align}
    & I_0 = \nu\frac{e^2}{h} \Delta V.
\end{align}
This relation holds for efficient charge equilibration, and is manifest experimentally by robust charge conductance quantization. By using this relation, we re-write the noise in terms of the source current $I_0$ with the notation
\begin{align}
    & S^{\rm exc}_{I_0} = \frac{S^{\rm exc}_{\Delta V}}{\Delta V} \frac{I_0}{\nu} \frac{h}{e^2}.
\end{align}
For a voltage bias $\Delta V$ given in $\mu V$, the excess noise can then be written as
\begin{align}
    & S^{\rm exc} \approx 6.20492 \left(\frac{S^{\rm exc}_{\Delta V}}{\Delta V e^3/h} \right) \cdot 10^{-30} \frac{A^2}{\mu V \ \rm Hz}.
\end{align}
Alternatively, for a bias current $I_0$ given in $nA$, we have
\begin{align}
    S^{\rm exc} \approx 1.602 \left(\frac{S^{\rm exc}_{\Delta V}}{\Delta V} \cdot \frac{I_0}{\nu}\right) \cdot 10^{-28} \frac{A^2}{nA \ \rm Hz} .
\end{align}


\begin{thebibliography}{101}%
\makeatletter
\providecommand \@ifxundefined [1]{%
 \@ifx{#1\undefined}
}%
\providecommand \@ifnum [1]{%
 \ifnum #1\expandafter \@firstoftwo
 \else \expandafter \@secondoftwo
 \fi
}%
\providecommand \@ifx [1]{%
 \ifx #1\expandafter \@firstoftwo
 \else \expandafter \@secondoftwo
 \fi
}%
\providecommand \natexlab [1]{#1}%
\providecommand \enquote  [1]{``#1''}%
\providecommand \bibnamefont  [1]{#1}%
\providecommand \bibfnamefont [1]{#1}%
\providecommand \citenamefont [1]{#1}%
\providecommand \href@noop [0]{\@secondoftwo}%
\providecommand \href [0]{\begingroup \@sanitize@url \@href}%
\providecommand \@href[1]{\@@startlink{#1}\@@href}%
\providecommand \@@href[1]{\endgroup#1\@@endlink}%
\providecommand \@sanitize@url [0]{\catcode `\\12\catcode `\$12\catcode
  `\&12\catcode `\#12\catcode `\^12\catcode `\_12\catcode `\%12\relax}%
\providecommand \@@startlink[1]{}%
\providecommand \@@endlink[0]{}%
\providecommand \url  [0]{\begingroup\@sanitize@url \@url }%
\providecommand \@url [1]{\endgroup\@href {#1}{\urlprefix }}%
\providecommand \urlprefix  [0]{URL }%
\providecommand \Eprint [0]{\href }%
\providecommand \doibase [0]{http://doi.org/}%
\providecommand \selectlanguage [0]{\@gobble}%
\providecommand \bibinfo  [0]{\@secondoftwo}%
\providecommand \bibfield  [0]{\@secondoftwo}%
\providecommand \translation [1]{[#1]}%
\providecommand \BibitemOpen [0]{}%
\providecommand \bibitemStop [0]{}%
\providecommand \bibitemNoStop [0]{.\EOS\space}%
\providecommand \EOS [0]{\spacefactor3000\relax}%
\providecommand \BibitemShut  [1]{\csname bibitem#1\endcsname}%
\let\auto@bib@innerbib\@empty
\bibitem [{\citenamefont {Stern}(2010)}]{Stern2010Mar}%
  \BibitemOpen
  \bibfield  {author} {\bibinfo {author} {\bibfnamefont {A.}~\bibnamefont
  {Stern}},\ }\bibfield  {title} {\emph {\bibinfo {title} {{Non-{A}belian
  states of matter}}},\ }\href {\doibase 10.1038/nature08915} {\bibfield
  {journal} {\bibinfo  {journal} {Nature}\ }\textbf {\bibinfo {volume} {464}},\
  \bibinfo {pages} {187} (\bibinfo {year} {2010})}\BibitemShut {NoStop}%
\bibitem [{\citenamefont {Kitaev}(2003)}]{Kitaev2003Jan}%
  \BibitemOpen
  \bibfield  {author} {\bibinfo {author} {\bibfnamefont {A.~{\relax Yu}.}\
  \bibnamefont {Kitaev}},\ }\bibfield  {title} {\emph {\bibinfo {title}
  {{Fault-tolerant quantum computation by anyons}}},\ }\href {\doibase
  10.1016/S0003-4916(02)00018-0} {\bibfield  {journal} {\bibinfo  {journal}
  {Ann. Phys.}\ }\textbf {\bibinfo {volume} {303}},\ \bibinfo {pages} {2}
  (\bibinfo {year} {2003})}\BibitemShut {NoStop}%
\bibitem [{\citenamefont {Nayak}\ \emph {et~al.}(2008)\citenamefont {Nayak},
  \citenamefont {Simon}, \citenamefont {Stern}, \citenamefont {Freedman},\ and\
  \citenamefont {Das~Sarma}}]{Nayak2008}%
  \BibitemOpen
  \bibfield  {author} {\bibinfo {author} {\bibfnamefont {C.}~\bibnamefont
  {Nayak}}, \bibinfo {author} {\bibfnamefont {S.~H.}\ \bibnamefont {Simon}},
  \bibinfo {author} {\bibfnamefont {A.}~\bibnamefont {Stern}}, \bibinfo
  {author} {\bibfnamefont {M.}~\bibnamefont {Freedman}}, \ and\ \bibinfo
  {author} {\bibfnamefont {S.}~\bibnamefont {Das~Sarma}},\ }\bibfield  {title}
  {\emph {\bibinfo {title} {Non-{A}belian anyons and topological quantum
  computation}},\ }\href {\doibase 10.1103/RevModPhys.80.1083} {\bibfield
  {journal} {\bibinfo  {journal} {Rev. Mod. Phys.}\ }\textbf {\bibinfo {volume}
  {80}},\ \bibinfo {pages} {1083} (\bibinfo {year} {2008})}\BibitemShut
  {NoStop}%
\bibitem [{\citenamefont {Tsui}\ \emph {et~al.}(1982)\citenamefont {Tsui},
  \citenamefont {Stormer},\ and\ \citenamefont {Gossard}}]{Stormer1982}%
  \BibitemOpen
  \bibfield  {author} {\bibinfo {author} {\bibfnamefont {D.~C.}\ \bibnamefont
  {Tsui}}, \bibinfo {author} {\bibfnamefont {H.~L.}\ \bibnamefont {Stormer}}, \
  and\ \bibinfo {author} {\bibfnamefont {A.~C.}\ \bibnamefont {Gossard}},\
  }\bibfield  {title} {\emph {\bibinfo {title} {Two-dimensional
  magnetotransport in the extreme quantum limit}},\ }\href {\doibase
  10.1103/PhysRevLett.48.1559} {\bibfield  {journal} {\bibinfo  {journal}
  {Phys. Rev. Lett.}\ }\textbf {\bibinfo {volume} {48}},\ \bibinfo {pages}
  {1559} (\bibinfo {year} {1982})}\BibitemShut {NoStop}%
\bibitem [{\citenamefont {Laughlin}(1983)}]{Laughlin1983}%
  \BibitemOpen
  \bibfield  {author} {\bibinfo {author} {\bibfnamefont {R.~B.}\ \bibnamefont
  {Laughlin}},\ }\bibfield  {title} {\emph {\bibinfo {title} {Anomalous quantum
  {H}all effect: An incompressible quantum fluid with fractionally charged
  excitations}},\ }\href {\doibase 10.1103/PhysRevLett.50.1395} {\bibfield
  {journal} {\bibinfo  {journal} {Phys. Rev. Lett.}\ }\textbf {\bibinfo
  {volume} {50}},\ \bibinfo {pages} {1395} (\bibinfo {year}
  {1983})}\BibitemShut {NoStop}%
\bibitem [{\citenamefont {Willett}\ \emph {et~al.}(1987)\citenamefont
  {Willett}, \citenamefont {Eisenstein}, \citenamefont {St\"ormer},
  \citenamefont {Tsui}, \citenamefont {Gossard},\ and\ \citenamefont
  {English}}]{Willet1987}%
  \BibitemOpen
  \bibfield  {author} {\bibinfo {author} {\bibfnamefont {R.}~\bibnamefont
  {Willett}}, \bibinfo {author} {\bibfnamefont {J.~P.}\ \bibnamefont
  {Eisenstein}}, \bibinfo {author} {\bibfnamefont {H.~L.}\ \bibnamefont
  {St\"ormer}}, \bibinfo {author} {\bibfnamefont {D.~C.}\ \bibnamefont {Tsui}},
  \bibinfo {author} {\bibfnamefont {A.~C.}\ \bibnamefont {Gossard}}, \ and\
  \bibinfo {author} {\bibfnamefont {J.~H.}\ \bibnamefont {English}},\
  }\bibfield  {title} {\emph {\bibinfo {title} {Observation of an
  even-denominator quantum number in the fractional quantum {H}all effect}},\
  }\href {\doibase 10.1103/PhysRevLett.59.1776} {\bibfield  {journal} {\bibinfo
   {journal} {Phys. Rev. Lett.}\ }\textbf {\bibinfo {volume} {59}},\ \bibinfo
  {pages} {1776} (\bibinfo {year} {1987})}\BibitemShut {NoStop}%
\bibitem [{\citenamefont {Moore}\ and\ \citenamefont {Read}(1991)}]{Moore1991}%
  \BibitemOpen
  \bibfield  {author} {\bibinfo {author} {\bibfnamefont {G.}~\bibnamefont
  {Moore}}\ and\ \bibinfo {author} {\bibfnamefont {N.}~\bibnamefont {Read}},\
  }\bibfield  {title} {\emph {\bibinfo {title} {Nonabelions in the fractional
  quantum {H}all effect}},\ }\href {\doibase
  https://doi.org/10.1016/0550-3213(91)90407-O} {\bibfield  {journal} {\bibinfo
   {journal} {Nuclear Physics B}\ }\textbf {\bibinfo {volume} {360}},\ \bibinfo
  {pages} {362 } (\bibinfo {year} {1991})}\BibitemShut {NoStop}%
\bibitem [{\citenamefont {Wen}(1991)}]{Wen1991Feb}%
  \BibitemOpen
  \bibfield  {author} {\bibinfo {author} {\bibfnamefont {X.~G.}\ \bibnamefont
  {Wen}},\ }\bibfield  {title} {\emph {\bibinfo {title} {{Non-{A}belian
  statistics in the fractional quantum {H}all states}}},\ }\href {\doibase
  10.1103/PhysRevLett.66.802} {\bibfield  {journal} {\bibinfo  {journal} {Phys.
  Rev. Lett.}\ }\textbf {\bibinfo {volume} {66}},\ \bibinfo {pages} {802}
  (\bibinfo {year} {1991})}\BibitemShut {NoStop}%
\bibitem [{\citenamefont {Read}\ and\ \citenamefont
  {Green}(2000)}]{ReadGreen2000}%
  \BibitemOpen
  \bibfield  {author} {\bibinfo {author} {\bibfnamefont {N.}~\bibnamefont
  {Read}}\ and\ \bibinfo {author} {\bibfnamefont {D.}~\bibnamefont {Green}},\
  }\bibfield  {title} {\emph {\bibinfo {title} {Paired states of fermions in
  two dimensions with breaking of parity and time-reversal symmetries and the
  fractional quantum {H}all effect}},\ }\href {\doibase
  10.1103/PhysRevB.61.10267} {\bibfield  {journal} {\bibinfo  {journal} {Phys.
  Rev. B}\ }\textbf {\bibinfo {volume} {61}},\ \bibinfo {pages} {10267}
  (\bibinfo {year} {2000})}\BibitemShut {NoStop}%
\bibitem [{\citenamefont {Ma}\ \emph {et~al.}(2022)\citenamefont {Ma},
  \citenamefont {Peterson}, \citenamefont {Scarola},\ and\ \citenamefont
  {Yang}}]{Ma2022Aug}%
  \BibitemOpen
  \bibfield  {author} {\bibinfo {author} {\bibfnamefont {K.~K.~W.}\
  \bibnamefont {Ma}}, \bibinfo {author} {\bibfnamefont {M.~R.}\ \bibnamefont
  {Peterson}}, \bibinfo {author} {\bibfnamefont {V.~W.}\ \bibnamefont
  {Scarola}}, \ and\ \bibinfo {author} {\bibfnamefont {K.}~\bibnamefont
  {Yang}},\ }\bibfield  {title} {\emph {\bibinfo {title} {{Fractional quantum
  {H}all effect at the filling factor $\nu=5/2$}}},\ }\href {\doibase
  10.48550/arXiv.2208.07908} {\bibfield  {journal} {\bibinfo  {journal}
  {arXiv}\ } (\bibinfo {year} {2022}),\ 10.48550/arXiv.2208.07908},\ \Eprint
  {http://arxiv.org/abs/2208.07908} {2208.07908} \BibitemShut {NoStop}%
\bibitem [{\citenamefont {Dutta}\ \emph
  {et~al.}(2022{\natexlab{a}})\citenamefont {Dutta}, \citenamefont {Umansky},
  \citenamefont {Banerjee},\ and\ \citenamefont {Heiblum}}]{dutta2022sep}%
  \BibitemOpen
  \bibfield  {author} {\bibinfo {author} {\bibfnamefont {B.}~\bibnamefont
  {Dutta}}, \bibinfo {author} {\bibfnamefont {V.}~\bibnamefont {Umansky}},
  \bibinfo {author} {\bibfnamefont {M.}~\bibnamefont {Banerjee}}, \ and\
  \bibinfo {author} {\bibfnamefont {M.}~\bibnamefont {Heiblum}},\ }\bibfield
  {title} {\emph {\bibinfo {title} {{Isolated ballistic non-{A}belian interface
  channel}}},\ }\href {\doibase 10.1126/science.abm6571} {\bibfield  {journal}
  {\bibinfo  {journal} {Science}\ }\textbf {\bibinfo {volume} {377}},\ \bibinfo
  {pages} {1198} (\bibinfo {year} {2022}{\natexlab{a}})}\BibitemShut {NoStop}%
\bibitem [{\citenamefont {Dutta}\ \emph
  {et~al.}(2022{\natexlab{b}})\citenamefont {Dutta}, \citenamefont {Yang},
  \citenamefont {Melcer}, \citenamefont {Kundu}, \citenamefont {Heiblum},
  \citenamefont {Umansky}, \citenamefont {Oreg}, \citenamefont {Stern},\ and\
  \citenamefont {Mross}}]{Dutta2022}%
  \BibitemOpen
  \bibfield  {author} {\bibinfo {author} {\bibfnamefont {B.}~\bibnamefont
  {Dutta}}, \bibinfo {author} {\bibfnamefont {W.}~\bibnamefont {Yang}},
  \bibinfo {author} {\bibfnamefont {R.}~\bibnamefont {Melcer}}, \bibinfo
  {author} {\bibfnamefont {H.~K.}\ \bibnamefont {Kundu}}, \bibinfo {author}
  {\bibfnamefont {M.}~\bibnamefont {Heiblum}}, \bibinfo {author} {\bibfnamefont
  {V.}~\bibnamefont {Umansky}}, \bibinfo {author} {\bibfnamefont
  {Y.}~\bibnamefont {Oreg}}, \bibinfo {author} {\bibfnamefont {A.}~\bibnamefont
  {Stern}}, \ and\ \bibinfo {author} {\bibfnamefont {D.}~\bibnamefont
  {Mross}},\ }\bibfield  {title} {\emph {\bibinfo {title} {Distinguishing
  between non-{A}belian topological orders in a quantum {H}all system}},\
  }\href {\doibase 10.1126/science.abg6116} {\bibfield  {journal} {\bibinfo
  {journal} {Science}\ }\textbf {\bibinfo {volume} {375}},\ \bibinfo {pages}
  {193} (\bibinfo {year} {2022}{\natexlab{b}})}\BibitemShut {NoStop}%
\bibitem [{\citenamefont {Levin}\ \emph {et~al.}(2007)\citenamefont {Levin},
  \citenamefont {Halperin},\ and\ \citenamefont {Rosenow}}]{Levin2007}%
  \BibitemOpen
  \bibfield  {author} {\bibinfo {author} {\bibfnamefont {M.}~\bibnamefont
  {Levin}}, \bibinfo {author} {\bibfnamefont {B.~I.}\ \bibnamefont {Halperin}},
  \ and\ \bibinfo {author} {\bibfnamefont {B.}~\bibnamefont {Rosenow}},\
  }\bibfield  {title} {\emph {\bibinfo {title} {Particle-hole symmetry and the
  {P}faffian state}},\ }\href {\doibase 10.1103/PhysRevLett.99.236806}
  {\bibfield  {journal} {\bibinfo  {journal} {Phys. Rev. Lett.}\ }\textbf
  {\bibinfo {volume} {99}},\ \bibinfo {pages} {236806} (\bibinfo {year}
  {2007})}\BibitemShut {NoStop}%
\bibitem [{\citenamefont {Lee}\ \emph {et~al.}(2007)\citenamefont {Lee},
  \citenamefont {Ryu}, \citenamefont {Nayak},\ and\ \citenamefont
  {Fisher}}]{Lee2007}%
  \BibitemOpen
  \bibfield  {author} {\bibinfo {author} {\bibfnamefont {S.-S.}\ \bibnamefont
  {Lee}}, \bibinfo {author} {\bibfnamefont {S.}~\bibnamefont {Ryu}}, \bibinfo
  {author} {\bibfnamefont {C.}~\bibnamefont {Nayak}}, \ and\ \bibinfo {author}
  {\bibfnamefont {M.~P.~A.}\ \bibnamefont {Fisher}},\ }\bibfield  {title}
  {\emph {\bibinfo {title} {Particle-hole symmetry and the
  $\ensuremath{\nu}=\frac{5}{2}$ quantum {H}all state}},\ }\href {\doibase
  10.1103/PhysRevLett.99.236807} {\bibfield  {journal} {\bibinfo  {journal}
  {Phys. Rev. Lett.}\ }\textbf {\bibinfo {volume} {99}},\ \bibinfo {pages}
  {236807} (\bibinfo {year} {2007})}\BibitemShut {NoStop}%
\bibitem [{\citenamefont {Fidkowski}\ \emph {et~al.}(2013)\citenamefont
  {Fidkowski}, \citenamefont {Chen},\ and\ \citenamefont
  {Vishwanath}}]{Fidkowski2013}%
  \BibitemOpen
  \bibfield  {author} {\bibinfo {author} {\bibfnamefont {L.}~\bibnamefont
  {Fidkowski}}, \bibinfo {author} {\bibfnamefont {X.}~\bibnamefont {Chen}}, \
  and\ \bibinfo {author} {\bibfnamefont {A.}~\bibnamefont {Vishwanath}},\
  }\bibfield  {title} {\emph {\bibinfo {title} {Non-{A}belian topological order
  on the surface of a 3d topological superconductor from an exactly solved
  model}},\ }\href {\doibase 10.1103/PhysRevX.3.041016} {\bibfield  {journal}
  {\bibinfo  {journal} {Phys. Rev. X}\ }\textbf {\bibinfo {volume} {3}},\
  \bibinfo {pages} {041016} (\bibinfo {year} {2013})}\BibitemShut {NoStop}%
\bibitem [{\citenamefont {Son}(2015)}]{Son2015}%
  \BibitemOpen
  \bibfield  {author} {\bibinfo {author} {\bibfnamefont {D.~T.}\ \bibnamefont
  {Son}},\ }\bibfield  {title} {\emph {\bibinfo {title} {Is the composite
  fermion a dirac particle?}},\ }\href {\doibase 10.1103/PhysRevX.5.031027}
  {\bibfield  {journal} {\bibinfo  {journal} {Phys. Rev. X}\ }\textbf {\bibinfo
  {volume} {5}},\ \bibinfo {pages} {031027} (\bibinfo {year}
  {2015})}\BibitemShut {NoStop}%
\bibitem [{\citenamefont {Zucker}\ and\ \citenamefont
  {Feldman}(2016)}]{Zucker2016}%
  \BibitemOpen
  \bibfield  {author} {\bibinfo {author} {\bibfnamefont {P.~T.}\ \bibnamefont
  {Zucker}}\ and\ \bibinfo {author} {\bibfnamefont {D.~E.}\ \bibnamefont
  {Feldman}},\ }\bibfield  {title} {\emph {\bibinfo {title} {Stabilization of
  the particle-hole {P}faffian order by {L}andau-level mixing and impurities
  that break particle-hole symmetry}},\ }\href {\doibase
  10.1103/PhysRevLett.117.096802} {\bibfield  {journal} {\bibinfo  {journal}
  {Phys. Rev. Lett.}\ }\textbf {\bibinfo {volume} {117}},\ \bibinfo {pages}
  {096802} (\bibinfo {year} {2016})}\BibitemShut {NoStop}%
\bibitem [{\citenamefont {Antoni\ifmmode~\acute{c}\else \'{c}\fi{}}\ \emph
  {et~al.}(2018)\citenamefont {Antoni\ifmmode~\acute{c}\else \'{c}\fi{}},
  \citenamefont {Vu\ifmmode \check{c}\else \v{c}\fi{}i\ifmmode \check{c}\else
  \v{c}\fi{}evi\ifmmode~\acute{c}\else \'{c}\fi{}},\ and\ \citenamefont
  {Milovanovi\ifmmode~\acute{c}\else \'{c}\fi{}}}]{Antonic2018}%
  \BibitemOpen
  \bibfield  {author} {\bibinfo {author} {\bibfnamefont {L.}~\bibnamefont
  {Antoni\ifmmode~\acute{c}\else \'{c}\fi{}}}, \bibinfo {author} {\bibfnamefont
  {J.}~\bibnamefont {Vu\ifmmode \check{c}\else \v{c}\fi{}i\ifmmode
  \check{c}\else \v{c}\fi{}evi\ifmmode~\acute{c}\else \'{c}\fi{}}}, \ and\
  \bibinfo {author} {\bibfnamefont {M.~V.}\ \bibnamefont
  {Milovanovi\ifmmode~\acute{c}\else \'{c}\fi{}}},\ }\bibfield  {title} {\emph
  {\bibinfo {title} {Paired states at 5/2: Particle-hole {P}faffian and
  particle-hole symmetry breaking}},\ }\href {\doibase
  10.1103/PhysRevB.98.115107} {\bibfield  {journal} {\bibinfo  {journal} {Phys.
  Rev. B}\ }\textbf {\bibinfo {volume} {98}},\ \bibinfo {pages} {115107}
  (\bibinfo {year} {2018})}\BibitemShut {NoStop}%
\bibitem [{\citenamefont {Morf}(1998)}]{Morf1998}%
  \BibitemOpen
  \bibfield  {author} {\bibinfo {author} {\bibfnamefont {R.~H.}\ \bibnamefont
  {Morf}},\ }\bibfield  {title} {\emph {\bibinfo {title} {{Transition from
  Quantum {H}all to Compressible States in the Second {L}andau Level: New Light
  on the $\nu=5/2$ Enigma}}},\ }\href {\doibase 10.1103/PhysRevLett.80.1505}
  {\bibfield  {journal} {\bibinfo  {journal} {Phys. Rev. Lett.}\ }\textbf
  {\bibinfo {volume} {80}},\ \bibinfo {pages} {1505} (\bibinfo {year}
  {1998})}\BibitemShut {NoStop}%
\bibitem [{\citenamefont {Storni}\ \emph {et~al.}(2010)\citenamefont {Storni},
  \citenamefont {Morf},\ and\ \citenamefont {Das~Sarma}}]{Storni2010}%
  \BibitemOpen
  \bibfield  {author} {\bibinfo {author} {\bibfnamefont {M.}~\bibnamefont
  {Storni}}, \bibinfo {author} {\bibfnamefont {R.~H.}\ \bibnamefont {Morf}}, \
  and\ \bibinfo {author} {\bibfnamefont {S.}~\bibnamefont {Das~Sarma}},\
  }\bibfield  {title} {\emph {\bibinfo {title} {{Fractional Quantum {H}all
  State at $\nu=5/2$ and the Moore-Read {P}faffian}}},\ }\href {\doibase
  10.1103/PhysRevLett.104.076803} {\bibfield  {journal} {\bibinfo  {journal}
  {Phys. Rev. Lett.}\ }\textbf {\bibinfo {volume} {104}},\ \bibinfo {pages}
  {076803} (\bibinfo {year} {2010})}\BibitemShut {NoStop}%
\bibitem [{\citenamefont {W{\ifmmode\acute{o}\else\'{o}\fi}js}\ \emph
  {et~al.}(2010)\citenamefont {W{\ifmmode\acute{o}\else\'{o}\fi}js},
  \citenamefont {T{\ifmmode\mbox{\H{o}}\else\H{o}\fi}ke},\ and\ \citenamefont
  {Jain}}]{Wojs2010Aug}%
  \BibitemOpen
  \bibfield  {author} {\bibinfo {author} {\bibfnamefont {A.}~\bibnamefont
  {W{\ifmmode\acute{o}\else\'{o}\fi}js}}, \bibinfo {author} {\bibfnamefont
  {C.}~\bibnamefont {T{\ifmmode\mbox{\H{o}}\else\H{o}\fi}ke}}, \ and\ \bibinfo
  {author} {\bibfnamefont {J.~K.}\ \bibnamefont {Jain}},\ }\bibfield  {title}
  {\emph {\bibinfo {title} {{{L}andau-Level Mixing and the Emergence of
  {P}faffian Excitations for the $5/2$ Fractional Quantum {H}all Effect}}},\
  }\href {\doibase 10.1103/PhysRevLett.105.096802} {\bibfield  {journal}
  {\bibinfo  {journal} {Phys. Rev. Lett.}\ }\textbf {\bibinfo {volume} {105}},\
  \bibinfo {pages} {096802} (\bibinfo {year} {2010})}\BibitemShut {NoStop}%
\bibitem [{\citenamefont {Rezayi}\ and\ \citenamefont
  {Simon}(2011)}]{Rezayi2011Mar}%
  \BibitemOpen
  \bibfield  {author} {\bibinfo {author} {\bibfnamefont {E.~H.}\ \bibnamefont
  {Rezayi}}\ and\ \bibinfo {author} {\bibfnamefont {S.~H.}\ \bibnamefont
  {Simon}},\ }\bibfield  {title} {\emph {\bibinfo {title} {{Breaking of
  Particle-Hole Symmetry by {L}andau Level Mixing in the $\ensuremath{\nu}=5/2$
  Quantized {H}all State}}},\ }\href {\doibase 10.1103/PhysRevLett.106.116801}
  {\bibfield  {journal} {\bibinfo  {journal} {Phys. Rev. Lett.}\ }\textbf
  {\bibinfo {volume} {106}},\ \bibinfo {pages} {116801} (\bibinfo {year}
  {2011})}\BibitemShut {NoStop}%
\bibitem [{\citenamefont {Zaletel}\ \emph {et~al.}(2015)\citenamefont
  {Zaletel}, \citenamefont {Mong}, \citenamefont {Pollmann},\ and\
  \citenamefont {Rezayi}}]{Zaletel2015Jan}%
  \BibitemOpen
  \bibfield  {author} {\bibinfo {author} {\bibfnamefont {M.~P.}\ \bibnamefont
  {Zaletel}}, \bibinfo {author} {\bibfnamefont {R.~S.~K.}\ \bibnamefont
  {Mong}}, \bibinfo {author} {\bibfnamefont {F.}~\bibnamefont {Pollmann}}, \
  and\ \bibinfo {author} {\bibfnamefont {E.~H.}\ \bibnamefont {Rezayi}},\
  }\bibfield  {title} {\emph {\bibinfo {title} {{Infinite density matrix
  renormalization group for multicomponent quantum {H}all systems}}},\ }\href
  {\doibase 10.1103/PhysRevB.91.045115} {\bibfield  {journal} {\bibinfo
  {journal} {Phys. Rev. B}\ }\textbf {\bibinfo {volume} {91}},\ \bibinfo
  {pages} {045115} (\bibinfo {year} {2015})}\BibitemShut {NoStop}%
\bibitem [{\citenamefont {Rezayi}(2017)}]{Rezayi2017}%
  \BibitemOpen
  \bibfield  {author} {\bibinfo {author} {\bibfnamefont {E.~H.}\ \bibnamefont
  {Rezayi}},\ }\bibfield  {title} {\emph {\bibinfo {title} {{L}andau level
  mixing and the ground state of the $\ensuremath{\nu}=5/2$ quantum {H}all
  effect}},\ }\href {\doibase 10.1103/PhysRevLett.119.026801} {\bibfield
  {journal} {\bibinfo  {journal} {Phys. Rev. Lett.}\ }\textbf {\bibinfo
  {volume} {119}},\ \bibinfo {pages} {026801} (\bibinfo {year}
  {2017})}\BibitemShut {NoStop}%
\bibitem [{\citenamefont {Simon}\ \emph {et~al.}(2020)\citenamefont {Simon},
  \citenamefont {Ippoliti}, \citenamefont {Zaletel},\ and\ \citenamefont
  {Rezayi}}]{Simon2020Jan}%
  \BibitemOpen
  \bibfield  {author} {\bibinfo {author} {\bibfnamefont {S.~H.}\ \bibnamefont
  {Simon}}, \bibinfo {author} {\bibfnamefont {M.}~\bibnamefont {Ippoliti}},
  \bibinfo {author} {\bibfnamefont {M.~P.}\ \bibnamefont {Zaletel}}, \ and\
  \bibinfo {author} {\bibfnamefont {E.~H.}\ \bibnamefont {Rezayi}},\ }\bibfield
   {title} {\emph {\bibinfo {title} {{Energetics of {P}faffian--anti-{P}faffian
  domains}}},\ }\href {\doibase 10.1103/PhysRevB.101.041302} {\bibfield
  {journal} {\bibinfo  {journal} {Phys. Rev. B}\ }\textbf {\bibinfo {volume}
  {101}},\ \bibinfo {pages} {041302} (\bibinfo {year} {2020})}\BibitemShut
  {NoStop}%
\bibitem [{\citenamefont {Radu}\ \emph {et~al.}(2008)\citenamefont {Radu},
  \citenamefont {Miller}, \citenamefont {Marcus}, \citenamefont {Kastner},
  \citenamefont {Pfeiffer},\ and\ \citenamefont {West}}]{Radu2008}%
  \BibitemOpen
  \bibfield  {author} {\bibinfo {author} {\bibfnamefont {I.~P.}\ \bibnamefont
  {Radu}}, \bibinfo {author} {\bibfnamefont {J.~B.}\ \bibnamefont {Miller}},
  \bibinfo {author} {\bibfnamefont {C.~M.}\ \bibnamefont {Marcus}}, \bibinfo
  {author} {\bibfnamefont {M.~A.}\ \bibnamefont {Kastner}}, \bibinfo {author}
  {\bibfnamefont {L.~N.}\ \bibnamefont {Pfeiffer}}, \ and\ \bibinfo {author}
  {\bibfnamefont {K.~W.}\ \bibnamefont {West}},\ }\bibfield  {title} {\emph
  {\bibinfo {title} {Quasi-particle properties from tunneling in the v = 5/2
  fractional quantum {H}all state}},\ }\href {\doibase 10.1126/science.1157560}
  {\bibfield  {journal} {\bibinfo  {journal} {Science}\ }\textbf {\bibinfo
  {volume} {320}},\ \bibinfo {pages} {899} (\bibinfo {year}
  {2008})}\BibitemShut {NoStop}%
\bibitem [{\citenamefont {Lin}\ \emph {et~al.}(2012)\citenamefont {Lin},
  \citenamefont {Dillard}, \citenamefont {Kastner}, \citenamefont {Pfeiffer},\
  and\ \citenamefont {West}}]{Lin2012}%
  \BibitemOpen
  \bibfield  {author} {\bibinfo {author} {\bibfnamefont {X.}~\bibnamefont
  {Lin}}, \bibinfo {author} {\bibfnamefont {C.}~\bibnamefont {Dillard}},
  \bibinfo {author} {\bibfnamefont {M.~A.}\ \bibnamefont {Kastner}}, \bibinfo
  {author} {\bibfnamefont {L.~N.}\ \bibnamefont {Pfeiffer}}, \ and\ \bibinfo
  {author} {\bibfnamefont {K.~W.}\ \bibnamefont {West}},\ }\bibfield  {title}
  {\emph {\bibinfo {title} {Measurements of quasiparticle tunneling in the
  $\ensuremath{\nu}=\frac{5}{2}$ fractional quantum {H}all state}},\ }\href
  {\doibase 10.1103/PhysRevB.85.165321} {\bibfield  {journal} {\bibinfo
  {journal} {Phys. Rev. B}\ }\textbf {\bibinfo {volume} {85}},\ \bibinfo
  {pages} {165321} (\bibinfo {year} {2012})}\BibitemShut {NoStop}%
\bibitem [{\citenamefont {Lin}\ \emph {et~al.}(2014)\citenamefont {Lin},
  \citenamefont {Du},\ and\ \citenamefont {Xie}}]{Lin2014}%
  \BibitemOpen
  \bibfield  {author} {\bibinfo {author} {\bibfnamefont {X.}~\bibnamefont
  {Lin}}, \bibinfo {author} {\bibfnamefont {R.}~\bibnamefont {Du}}, \ and\
  \bibinfo {author} {\bibfnamefont {X.}~\bibnamefont {Xie}},\ }\bibfield
  {title} {\emph {\bibinfo {title} {{Recent experimental progress of fractional
  quantum {H}all effect: 5/2 filling state and graphene}}},\ }\href {\doibase
  10.1093/nsr/nwu071} {\bibfield  {journal} {\bibinfo  {journal} {National
  Science Review}\ }\textbf {\bibinfo {volume} {1}},\ \bibinfo {pages} {564}
  (\bibinfo {year} {2014})}\BibitemShut {NoStop}%
\bibitem [{\citenamefont {Wen}(1990)}]{Wen1990a}%
  \BibitemOpen
  \bibfield  {author} {\bibinfo {author} {\bibfnamefont {X.~G.}\ \bibnamefont
  {Wen}},\ }\bibfield  {title} {\emph {\bibinfo {title} {Topological orders in
  rigid states}},\ }\href {\doibase 10.1142/S0217979290000139} {\bibfield
  {journal} {\bibinfo  {journal} {Int. J. Mod. Phys. B}\ }\textbf {\bibinfo
  {volume} {04}},\ \bibinfo {pages} {239} (\bibinfo {year} {1990})}\BibitemShut
  {NoStop}%
\bibitem [{\citenamefont {Francesco}\ \emph {et~al.}(2012)\citenamefont
  {Francesco}, \citenamefont {Mathieu},\ and\ \citenamefont
  {S{\'e}n{\'e}chal}}]{francesco2012conformal}%
  \BibitemOpen
  \bibfield  {author} {\bibinfo {author} {\bibfnamefont {P.}~\bibnamefont
  {Francesco}}, \bibinfo {author} {\bibfnamefont {P.}~\bibnamefont {Mathieu}},
  \ and\ \bibinfo {author} {\bibfnamefont {D.}~\bibnamefont
  {S{\'e}n{\'e}chal}},\ }\href@noop {} {\emph {\bibinfo {title} {Conformal
  field theory}}}\ (\bibinfo  {publisher} {Springer Science \& Business
  Media},\ \bibinfo {year} {2012})\BibitemShut {NoStop}%
\bibitem [{\citenamefont {Kane}\ and\ \citenamefont {Fisher}(1997)}]{Kane1997}%
  \BibitemOpen
  \bibfield  {author} {\bibinfo {author} {\bibfnamefont {C.~L.}\ \bibnamefont
  {Kane}}\ and\ \bibinfo {author} {\bibfnamefont {M.~P.~A.}\ \bibnamefont
  {Fisher}},\ }\bibfield  {title} {\emph {\bibinfo {title} {Quantized thermal
  transport in the fractional quantum {H}all effect}},\ }\href {\doibase
  10.1103/PhysRevB.55.15832} {\bibfield  {journal} {\bibinfo  {journal} {Phys.
  Rev. B}\ }\textbf {\bibinfo {volume} {55}},\ \bibinfo {pages} {15832}
  (\bibinfo {year} {1997})}\BibitemShut {NoStop}%
\bibitem [{\citenamefont {Cappelli}\ \emph {et~al.}(2002)\citenamefont
  {Cappelli}, \citenamefont {Huerta},\ and\ \citenamefont
  {Zemba}}]{Capelli2002}%
  \BibitemOpen
  \bibfield  {author} {\bibinfo {author} {\bibfnamefont {A.}~\bibnamefont
  {Cappelli}}, \bibinfo {author} {\bibfnamefont {M.}~\bibnamefont {Huerta}}, \
  and\ \bibinfo {author} {\bibfnamefont {G.~R.}\ \bibnamefont {Zemba}},\
  }\bibfield  {title} {\emph {\bibinfo {title} {Thermal transport in chiral
  conformal theories and hierarchical quantum {H}all states}},\ }\href
  {\doibase https://doi.org/10.1016/S0550-3213(02)00340-1} {\bibfield
  {journal} {\bibinfo  {journal} {Nuclear Physics B}\ }\textbf {\bibinfo
  {volume} {636}},\ \bibinfo {pages} {568 } (\bibinfo {year}
  {2002})}\BibitemShut {NoStop}%
\bibitem [{\citenamefont {Jezouin}\ \emph {et~al.}(2013)\citenamefont
  {Jezouin}, \citenamefont {Parmentier}, \citenamefont {Anthore}, \citenamefont
  {Gennser}, \citenamefont {Cavanna}, \citenamefont {Jin},\ and\ \citenamefont
  {Pierre}}]{Jezouin2013}%
  \BibitemOpen
  \bibfield  {author} {\bibinfo {author} {\bibfnamefont {S.}~\bibnamefont
  {Jezouin}}, \bibinfo {author} {\bibfnamefont {F.~D.}\ \bibnamefont
  {Parmentier}}, \bibinfo {author} {\bibfnamefont {A.}~\bibnamefont {Anthore}},
  \bibinfo {author} {\bibfnamefont {U.}~\bibnamefont {Gennser}}, \bibinfo
  {author} {\bibfnamefont {A.}~\bibnamefont {Cavanna}}, \bibinfo {author}
  {\bibfnamefont {Y.}~\bibnamefont {Jin}}, \ and\ \bibinfo {author}
  {\bibfnamefont {F.}~\bibnamefont {Pierre}},\ }\bibfield  {title} {\emph
  {\bibinfo {title} {Quantum limit of heat flow across a single electronic
  channel}},\ }\href {\doibase 10.1126/science.1241912} {\bibfield  {journal}
  {\bibinfo  {journal} {Science}\ }\textbf {\bibinfo {volume} {342}},\ \bibinfo
  {pages} {601} (\bibinfo {year} {2013})}\BibitemShut {NoStop}%
\bibitem [{\citenamefont {Banerjee}\ \emph {et~al.}(2017)\citenamefont
  {Banerjee}, \citenamefont {Heiblum}, \citenamefont {Rosenblatt},
  \citenamefont {Oreg}, \citenamefont {Feldman}, \citenamefont {Stern},\ and\
  \citenamefont {Umansky}}]{Banerjee2017}%
  \BibitemOpen
  \bibfield  {author} {\bibinfo {author} {\bibfnamefont {M.}~\bibnamefont
  {Banerjee}}, \bibinfo {author} {\bibfnamefont {M.}~\bibnamefont {Heiblum}},
  \bibinfo {author} {\bibfnamefont {A.}~\bibnamefont {Rosenblatt}}, \bibinfo
  {author} {\bibfnamefont {Y.}~\bibnamefont {Oreg}}, \bibinfo {author}
  {\bibfnamefont {D.~E.}\ \bibnamefont {Feldman}}, \bibinfo {author}
  {\bibfnamefont {A.}~\bibnamefont {Stern}}, \ and\ \bibinfo {author}
  {\bibfnamefont {V.}~\bibnamefont {Umansky}},\ }\bibfield  {title} {\emph
  {\bibinfo {title} {Observed quantization of anyonic heat flow}},\ }\href
  {https://doi.org/10.1038/nature22052} {\bibfield  {journal} {\bibinfo
  {journal} {Nature}\ }\textbf {\bibinfo {volume} {545}},\ \bibinfo {pages} {75
  EP } (\bibinfo {year} {2017})}\BibitemShut {NoStop}%
\bibitem [{\citenamefont {Banerjee}\ \emph {et~al.}(2018)\citenamefont
  {Banerjee}, \citenamefont {Heiblum}, \citenamefont {Umansky}, \citenamefont
  {Feldman}, \citenamefont {Oreg},\ and\ \citenamefont {Stern}}]{Banerjee2018}%
  \BibitemOpen
  \bibfield  {author} {\bibinfo {author} {\bibfnamefont {M.}~\bibnamefont
  {Banerjee}}, \bibinfo {author} {\bibfnamefont {M.}~\bibnamefont {Heiblum}},
  \bibinfo {author} {\bibfnamefont {V.}~\bibnamefont {Umansky}}, \bibinfo
  {author} {\bibfnamefont {D.~E.}\ \bibnamefont {Feldman}}, \bibinfo {author}
  {\bibfnamefont {Y.}~\bibnamefont {Oreg}}, \ and\ \bibinfo {author}
  {\bibfnamefont {A.}~\bibnamefont {Stern}},\ }\bibfield  {title} {\emph
  {\bibinfo {title} {Observation of half-integer thermal {H}all conductance}},\
  }\href {\doibase 10.1038/s41586-018-0184-1} {\bibfield  {journal} {\bibinfo
  {journal} {Nature}\ }\textbf {\bibinfo {volume} {559}},\ \bibinfo {pages}
  {205} (\bibinfo {year} {2018})}\BibitemShut {NoStop}%
\bibitem [{\citenamefont {Melcer}\ \emph {et~al.}(2022)\citenamefont {Melcer},
  \citenamefont {Dutta}, \citenamefont
  {Sp{\aa}nsl{\ifmmode\ddot{a}\else\"{a}\fi}tt}, \citenamefont {Park},
  \citenamefont {Mirlin},\ and\ \citenamefont {Umansky}}]{Melcer2022Jan}%
  \BibitemOpen
  \bibfield  {author} {\bibinfo {author} {\bibfnamefont {R.~A.}\ \bibnamefont
  {Melcer}}, \bibinfo {author} {\bibfnamefont {B.}~\bibnamefont {Dutta}},
  \bibinfo {author} {\bibfnamefont {C.}~\bibnamefont
  {Sp{\aa}nsl{\ifmmode\ddot{a}\else\"{a}\fi}tt}}, \bibinfo {author}
  {\bibfnamefont {J.}~\bibnamefont {Park}}, \bibinfo {author} {\bibfnamefont
  {A.~D.}\ \bibnamefont {Mirlin}}, \ and\ \bibinfo {author} {\bibfnamefont
  {V.}~\bibnamefont {Umansky}},\ }\bibfield  {title} {\emph {\bibinfo {title}
  {{Absent thermal equilibration on fractional quantum {H}all edges over
  macroscopic scale}}},\ }\href {\doibase 10.1038/s41467-022-28009-0}
  {\bibfield  {journal} {\bibinfo  {journal} {Nat. Commun.}\ }\textbf {\bibinfo
  {volume} {13}},\ \bibinfo {pages} {1} (\bibinfo {year} {2022})}\BibitemShut
  {NoStop}%
\bibitem [{\citenamefont {Melcer}\ \emph {et~al.}(2023)\citenamefont {Melcer},
  \citenamefont {Konyzheva}, \citenamefont {Heiblum},\ and\ \citenamefont
  {Umansky}}]{Melcer2023Jan}%
  \BibitemOpen
  \bibfield  {author} {\bibinfo {author} {\bibfnamefont {R.~A.}\ \bibnamefont
  {Melcer}}, \bibinfo {author} {\bibfnamefont {S.}~\bibnamefont {Konyzheva}},
  \bibinfo {author} {\bibfnamefont {M.}~\bibnamefont {Heiblum}}, \ and\
  \bibinfo {author} {\bibfnamefont {V.}~\bibnamefont {Umansky}},\ }\bibfield
  {title} {\emph {\bibinfo {title} {{Direct determination of the topological
  thermal conductance via local power measurement}}},\ }\href {\doibase
  10.1038/s41567-022-01885-5} {\bibfield  {journal} {\bibinfo  {journal} {Nat.
  Phys.}\ ,\ \bibinfo {pages} {1}} (\bibinfo {year} {2023})},\ \Eprint
  {http://arxiv.org/abs/https://doi.org/10.1038/s41567-022-01885-5}
  {https://doi.org/10.1038/s41567-022-01885-5} \BibitemShut {NoStop}%
\bibitem [{\citenamefont {Srivastav}\ \emph {et~al.}(2019)\citenamefont
  {Srivastav}, \citenamefont {Sahu}, \citenamefont {Watanabe}, \citenamefont
  {Taniguchi}, \citenamefont {Banerjee},\ and\ \citenamefont
  {Das}}]{Srivastav2019Jul}%
  \BibitemOpen
  \bibfield  {author} {\bibinfo {author} {\bibfnamefont {S.~K.}\ \bibnamefont
  {Srivastav}}, \bibinfo {author} {\bibfnamefont {M.~R.}\ \bibnamefont {Sahu}},
  \bibinfo {author} {\bibfnamefont {K.}~\bibnamefont {Watanabe}}, \bibinfo
  {author} {\bibfnamefont {T.}~\bibnamefont {Taniguchi}}, \bibinfo {author}
  {\bibfnamefont {S.}~\bibnamefont {Banerjee}}, \ and\ \bibinfo {author}
  {\bibfnamefont {A.}~\bibnamefont {Das}},\ }\bibfield  {title} {\emph
  {\bibinfo {title} {{Universal quantized thermal conductance in graphene}}},\
  }\href {\doibase 10.1126/sciadv.aaw5798} {\bibfield  {journal} {\bibinfo
  {journal} {Sci. Adv.}\ }\textbf {\bibinfo {volume} {5}},\ \bibinfo {pages}
  {eaaw5798} (\bibinfo {year} {2019})}\BibitemShut {NoStop}%
\bibitem [{\citenamefont {Srivastav}\ \emph {et~al.}(2021)\citenamefont
  {Srivastav}, \citenamefont {Kumar}, \citenamefont
  {Sp{\aa}nsl{\ifmmode\ddot{a}\else\"{a}\fi}tt}, \citenamefont {Watanabe},
  \citenamefont {Taniguchi}, \citenamefont {Mirlin}, \citenamefont {Gefen},\
  and\ \citenamefont {Das}}]{Srivastav2021May}%
  \BibitemOpen
  \bibfield  {author} {\bibinfo {author} {\bibfnamefont {S.~K.}\ \bibnamefont
  {Srivastav}}, \bibinfo {author} {\bibfnamefont {R.}~\bibnamefont {Kumar}},
  \bibinfo {author} {\bibfnamefont {C.}~\bibnamefont
  {Sp{\aa}nsl{\ifmmode\ddot{a}\else\"{a}\fi}tt}}, \bibinfo {author}
  {\bibfnamefont {K.}~\bibnamefont {Watanabe}}, \bibinfo {author}
  {\bibfnamefont {T.}~\bibnamefont {Taniguchi}}, \bibinfo {author}
  {\bibfnamefont {A.~D.}\ \bibnamefont {Mirlin}}, \bibinfo {author}
  {\bibfnamefont {Y.}~\bibnamefont {Gefen}}, \ and\ \bibinfo {author}
  {\bibfnamefont {A.}~\bibnamefont {Das}},\ }\bibfield  {title} {\emph
  {\bibinfo {title} {{Vanishing Thermal Equilibration for Hole-Conjugate
  Fractional Quantum {H}all States in Graphene}}},\ }\href {\doibase
  10.1103/PhysRevLett.126.216803} {\bibfield  {journal} {\bibinfo  {journal}
  {Phys. Rev. Lett.}\ }\textbf {\bibinfo {volume} {126}},\ \bibinfo {pages}
  {216803} (\bibinfo {year} {2021})}\BibitemShut {NoStop}%
\bibitem [{\citenamefont {Le~Breton}\ \emph {et~al.}(2022)\citenamefont
  {Le~Breton}, \citenamefont {Delagrange}, \citenamefont {Hong}, \citenamefont
  {Garg}, \citenamefont {Watanabe}, \citenamefont {Taniguchi}, \citenamefont
  {Ribeiro-Palau}, \citenamefont {Roulleau}, \citenamefont {Roche},\ and\
  \citenamefont {Parmentier}}]{LeBreton2022Sep}%
  \BibitemOpen
  \bibfield  {author} {\bibinfo {author} {\bibfnamefont {G.}~\bibnamefont
  {Le~Breton}}, \bibinfo {author} {\bibfnamefont {R.}~\bibnamefont
  {Delagrange}}, \bibinfo {author} {\bibfnamefont {Y.}~\bibnamefont {Hong}},
  \bibinfo {author} {\bibfnamefont {M.}~\bibnamefont {Garg}}, \bibinfo {author}
  {\bibfnamefont {K.}~\bibnamefont {Watanabe}}, \bibinfo {author}
  {\bibfnamefont {T.}~\bibnamefont {Taniguchi}}, \bibinfo {author}
  {\bibfnamefont {R.}~\bibnamefont {Ribeiro-Palau}}, \bibinfo {author}
  {\bibfnamefont {P.}~\bibnamefont {Roulleau}}, \bibinfo {author}
  {\bibfnamefont {P.}~\bibnamefont {Roche}}, \ and\ \bibinfo {author}
  {\bibfnamefont {F.~D.}\ \bibnamefont {Parmentier}},\ }\bibfield  {title}
  {\emph {\bibinfo {title} {{Heat Equilibration of Integer and Fractional
  Quantum Hall Edge Modes in Graphene}}},\ }\href {\doibase
  10.1103/PhysRevLett.129.116803} {\bibfield  {journal} {\bibinfo  {journal}
  {Phys. Rev. Lett.}\ }\textbf {\bibinfo {volume} {129}},\ \bibinfo {pages}
  {116803} (\bibinfo {year} {2022})}\BibitemShut {NoStop}%
\bibitem [{\citenamefont {Srivastav}\ \emph {et~al.}(2022)\citenamefont
  {Srivastav}, \citenamefont {Kumar}, \citenamefont
  {Sp{\aa}nsl{\ifmmode\ddot{a}\else\"{a}\fi}tt}, \citenamefont {Watanabe},
  \citenamefont {Taniguchi}, \citenamefont {Mirlin}, \citenamefont {Gefen},\
  and\ \citenamefont {Das}}]{Srivastav2022Sep}%
  \BibitemOpen
  \bibfield  {author} {\bibinfo {author} {\bibfnamefont {S.~K.}\ \bibnamefont
  {Srivastav}}, \bibinfo {author} {\bibfnamefont {R.}~\bibnamefont {Kumar}},
  \bibinfo {author} {\bibfnamefont {C.}~\bibnamefont
  {Sp{\aa}nsl{\ifmmode\ddot{a}\else\"{a}\fi}tt}}, \bibinfo {author}
  {\bibfnamefont {K.}~\bibnamefont {Watanabe}}, \bibinfo {author}
  {\bibfnamefont {T.}~\bibnamefont {Taniguchi}}, \bibinfo {author}
  {\bibfnamefont {A.~D.}\ \bibnamefont {Mirlin}}, \bibinfo {author}
  {\bibfnamefont {Y.}~\bibnamefont {Gefen}}, \ and\ \bibinfo {author}
  {\bibfnamefont {A.}~\bibnamefont {Das}},\ }\bibfield  {title} {\emph
  {\bibinfo {title} {{Determination of topological edge quantum numbers of
  fractional quantum {H}all phases by thermal conductance measurements}}},\
  }\href {\doibase 10.1038/s41467-022-32956-z} {\bibfield  {journal} {\bibinfo
  {journal} {Nat. Commun.}\ }\textbf {\bibinfo {volume} {13}},\ \bibinfo
  {pages} {1} (\bibinfo {year} {2022})}\BibitemShut {NoStop}%
\bibitem [{\citenamefont {Kane}\ and\ \citenamefont
  {Fisher}(1995{\natexlab{a}})}]{Kane1995}%
  \BibitemOpen
  \bibfield  {author} {\bibinfo {author} {\bibfnamefont {C.~L.}\ \bibnamefont
  {Kane}}\ and\ \bibinfo {author} {\bibfnamefont {M.~P.~A.}\ \bibnamefont
  {Fisher}},\ }\bibfield  {title} {\emph {\bibinfo {title} {Contacts and
  edge-state equilibration in the fractional quantum {H}all effect}},\ }\href
  {\doibase 10.1103/PhysRevB.52.17393} {\bibfield  {journal} {\bibinfo
  {journal} {Phys. Rev. B}\ }\textbf {\bibinfo {volume} {52}},\ \bibinfo
  {pages} {17393} (\bibinfo {year} {1995}{\natexlab{a}})}\BibitemShut {NoStop}%
\bibitem [{\citenamefont {Protopopov}\ \emph {et~al.}(2017)\citenamefont
  {Protopopov}, \citenamefont {Gefen},\ and\ \citenamefont
  {Mirlin}}]{Protopopov2017}%
  \BibitemOpen
  \bibfield  {author} {\bibinfo {author} {\bibfnamefont {I.}~\bibnamefont
  {Protopopov}}, \bibinfo {author} {\bibfnamefont {Y.}~\bibnamefont {Gefen}}, \
  and\ \bibinfo {author} {\bibfnamefont {A.}~\bibnamefont {Mirlin}},\
  }\bibfield  {title} {\emph {\bibinfo {title} {Transport in a disordered
  $\ensuremath{\nu}=2/3$ fractional quantum {H}all junction}},\ }\href
  {\doibase https://doi.org/10.1016/j.aop.2017.07.015} {\bibfield  {journal}
  {\bibinfo  {journal} {Annals of Physics}\ }\textbf {\bibinfo {volume}
  {385}},\ \bibinfo {pages} {287 } (\bibinfo {year} {2017})}\BibitemShut
  {NoStop}%
\bibitem [{\citenamefont {Nosiglia}\ \emph {et~al.}(2018)\citenamefont
  {Nosiglia}, \citenamefont {Park}, \citenamefont {Rosenow},\ and\
  \citenamefont {Gefen}}]{Nosiglia2018}%
  \BibitemOpen
  \bibfield  {author} {\bibinfo {author} {\bibfnamefont {C.}~\bibnamefont
  {Nosiglia}}, \bibinfo {author} {\bibfnamefont {J.}~\bibnamefont {Park}},
  \bibinfo {author} {\bibfnamefont {B.}~\bibnamefont {Rosenow}}, \ and\
  \bibinfo {author} {\bibfnamefont {Y.}~\bibnamefont {Gefen}},\ }\bibfield
  {title} {\emph {\bibinfo {title} {Incoherent transport on the
  $\ensuremath{\nu}=2/3$ quantum {H}all edge}},\ }\href {\doibase
  10.1103/PhysRevB.98.115408} {\bibfield  {journal} {\bibinfo  {journal} {Phys.
  Rev. B}\ }\textbf {\bibinfo {volume} {98}},\ \bibinfo {pages} {115408}
  (\bibinfo {year} {2018})}\BibitemShut {NoStop}%
\bibitem [{\citenamefont {Aharon-Steinberg}\ \emph {et~al.}(2019)\citenamefont
  {Aharon-Steinberg}, \citenamefont {Oreg},\ and\ \citenamefont
  {Stern}}]{Aharon2019}%
  \BibitemOpen
  \bibfield  {author} {\bibinfo {author} {\bibfnamefont {A.}~\bibnamefont
  {Aharon-Steinberg}}, \bibinfo {author} {\bibfnamefont {Y.}~\bibnamefont
  {Oreg}}, \ and\ \bibinfo {author} {\bibfnamefont {A.}~\bibnamefont {Stern}},\
  }\bibfield  {title} {\emph {\bibinfo {title} {Phenomenological theory of heat
  transport in the fractional quantum {H}all effect}},\ }\href {\doibase
  10.1103/PhysRevB.99.041302} {\bibfield  {journal} {\bibinfo  {journal} {Phys.
  Rev. B}\ }\textbf {\bibinfo {volume} {99}},\ \bibinfo {pages} {041302}
  (\bibinfo {year} {2019})}\BibitemShut {NoStop}%
\bibitem [{\citenamefont {Ma}\ and\ \citenamefont {Feldman}(2020)}]{Ma2020Jul}%
  \BibitemOpen
  \bibfield  {author} {\bibinfo {author} {\bibfnamefont {K.~K.~W.}\
  \bibnamefont {Ma}}\ and\ \bibinfo {author} {\bibfnamefont {D.~E.}\
  \bibnamefont {Feldman}},\ }\bibfield  {title} {\emph {\bibinfo {title}
  {{Thermal Equilibration on the Edges of Topological Liquids}}},\ }\href
  {\doibase 10.1103/PhysRevLett.125.016801} {\bibfield  {journal} {\bibinfo
  {journal} {Phys. Rev. Lett.}\ }\textbf {\bibinfo {volume} {125}},\ \bibinfo
  {pages} {016801} (\bibinfo {year} {2020})}\BibitemShut {NoStop}%
\bibitem [{\citenamefont {Lafont}\ \emph {et~al.}(2019)\citenamefont {Lafont},
  \citenamefont {Rosenblatt}, \citenamefont {Heiblum},\ and\ \citenamefont
  {Umansky}}]{Lafont2019}%
  \BibitemOpen
  \bibfield  {author} {\bibinfo {author} {\bibfnamefont {F.}~\bibnamefont
  {Lafont}}, \bibinfo {author} {\bibfnamefont {A.}~\bibnamefont {Rosenblatt}},
  \bibinfo {author} {\bibfnamefont {M.}~\bibnamefont {Heiblum}}, \ and\
  \bibinfo {author} {\bibfnamefont {V.}~\bibnamefont {Umansky}},\ }\bibfield
  {title} {\emph {\bibinfo {title} {Counter-propagating charge transport in the
  quantum {H}all effect regime}},\ }\href {\doibase 10.1126/science.aar3766}
  {\bibfield  {journal} {\bibinfo  {journal} {Science}\ }\textbf {\bibinfo
  {volume} {363}},\ \bibinfo {pages} {54} (\bibinfo {year} {2019})}\BibitemShut
  {NoStop}%
\bibitem [{\citenamefont {Cohen}\ \emph {et~al.}(2019)\citenamefont {Cohen},
  \citenamefont {Ronen}, \citenamefont {Yang}, \citenamefont {Banitt},
  \citenamefont {Park}, \citenamefont {Heiblum}, \citenamefont {Mirlin},
  \citenamefont {Gefen},\ and\ \citenamefont {Umansky}}]{Cohen2019}%
  \BibitemOpen
  \bibfield  {author} {\bibinfo {author} {\bibfnamefont {Y.}~\bibnamefont
  {Cohen}}, \bibinfo {author} {\bibfnamefont {Y.}~\bibnamefont {Ronen}},
  \bibinfo {author} {\bibfnamefont {W.}~\bibnamefont {Yang}}, \bibinfo {author}
  {\bibfnamefont {D.}~\bibnamefont {Banitt}}, \bibinfo {author} {\bibfnamefont
  {J.}~\bibnamefont {Park}}, \bibinfo {author} {\bibfnamefont {M.}~\bibnamefont
  {Heiblum}}, \bibinfo {author} {\bibfnamefont {A.~D.}\ \bibnamefont {Mirlin}},
  \bibinfo {author} {\bibfnamefont {Y.}~\bibnamefont {Gefen}}, \ and\ \bibinfo
  {author} {\bibfnamefont {V.}~\bibnamefont {Umansky}},\ }\bibfield  {title}
  {\emph {\bibinfo {title} {Synthesizing a \ensuremath{\nu}=2/3 fractional
  quantum {H}all effect edge state from counter-propagating \ensuremath{\nu}=1
  and \ensuremath{\nu}=1/3 states}},\ }\href {\doibase
  10.1038/s41467-019-09920-5} {\bibfield  {journal} {\bibinfo  {journal}
  {Nature Communications}\ }\textbf {\bibinfo {volume} {10}},\ \bibinfo {pages}
  {1920} (\bibinfo {year} {2019})}\BibitemShut {NoStop}%
\bibitem [{\citenamefont {Simon}(2018)}]{Simon2018}%
  \BibitemOpen
  \bibfield  {author} {\bibinfo {author} {\bibfnamefont {S.~H.}\ \bibnamefont
  {Simon}},\ }\bibfield  {title} {\emph {\bibinfo {title} {Interpretation of
  thermal conductance of the $\ensuremath{\nu}=5/2$ edge}},\ }\href {\doibase
  10.1103/PhysRevB.97.121406} {\bibfield  {journal} {\bibinfo  {journal} {Phys.
  Rev. B}\ }\textbf {\bibinfo {volume} {97}},\ \bibinfo {pages} {121406}
  (\bibinfo {year} {2018})}\BibitemShut {NoStop}%
\bibitem [{\citenamefont {Feldman}(2018)}]{Feldman2018}%
  \BibitemOpen
  \bibfield  {author} {\bibinfo {author} {\bibfnamefont {D.~E.}\ \bibnamefont
  {Feldman}},\ }\bibfield  {title} {\emph {\bibinfo {title} {Comment on
  ``interpretation of thermal conductance of the $\ensuremath{\nu}=5/2$
  edge''}},\ }\href {\doibase 10.1103/PhysRevB.98.167401} {\bibfield  {journal}
  {\bibinfo  {journal} {Phys. Rev. B}\ }\textbf {\bibinfo {volume} {98}},\
  \bibinfo {pages} {167401} (\bibinfo {year} {2018})}\BibitemShut {NoStop}%
\bibitem [{\citenamefont {Asasi}\ and\ \citenamefont
  {Mulligan}(2020)}]{Asasi2020}%
  \BibitemOpen
  \bibfield  {author} {\bibinfo {author} {\bibfnamefont {H.}~\bibnamefont
  {Asasi}}\ and\ \bibinfo {author} {\bibfnamefont {M.}~\bibnamefont
  {Mulligan}},\ }\bibfield  {title} {\emph {\bibinfo {title} {{Partial
  equilibration of anti-{P}faffian edge modes at $\ensuremath{\nu}=5/2$}}},\
  }\href {\doibase 10.1103/PhysRevB.102.205104} {\bibfield  {journal} {\bibinfo
   {journal} {Phys. Rev. B}\ }\textbf {\bibinfo {volume} {102}},\ \bibinfo
  {pages} {205104} (\bibinfo {year} {2020})}\BibitemShut {NoStop}%
\bibitem [{\citenamefont {Simon}\ and\ \citenamefont
  {Rosenow}(2020)}]{Simon2020}%
  \BibitemOpen
  \bibfield  {author} {\bibinfo {author} {\bibfnamefont {S.~H.}\ \bibnamefont
  {Simon}}\ and\ \bibinfo {author} {\bibfnamefont {B.}~\bibnamefont
  {Rosenow}},\ }\bibfield  {title} {\emph {\bibinfo {title} {Partial
  equilibration of the anti-{P}faffian edge due to majorana disorder}},\ }\href
  {\doibase 10.1103/PhysRevLett.124.126801} {\bibfield  {journal} {\bibinfo
  {journal} {Phys. Rev. Lett.}\ }\textbf {\bibinfo {volume} {124}},\ \bibinfo
  {pages} {126801} (\bibinfo {year} {2020})}\BibitemShut {NoStop}%
\bibitem [{\citenamefont {Park}\ \emph {et~al.}(2020)\citenamefont {Park},
  \citenamefont {Sp{\aa}nsl{\ifmmode\ddot{a}\else\"{a}\fi}tt}, \citenamefont
  {Gefen},\ and\ \citenamefont {Mirlin}}]{Park2020Oct}%
  \BibitemOpen
  \bibfield  {author} {\bibinfo {author} {\bibfnamefont {J.}~\bibnamefont
  {Park}}, \bibinfo {author} {\bibfnamefont {C.}~\bibnamefont
  {Sp{\aa}nsl{\ifmmode\ddot{a}\else\"{a}\fi}tt}}, \bibinfo {author}
  {\bibfnamefont {Y.}~\bibnamefont {Gefen}}, \ and\ \bibinfo {author}
  {\bibfnamefont {A.~D.}\ \bibnamefont {Mirlin}},\ }\bibfield  {title} {\emph
  {\bibinfo {title} {{Noise on the non-{A}belian $\ensuremath{\nu}=5/2$
  Fractional Quantum {H}all Edge}}},\ }\href {\doibase
  10.1103/PhysRevLett.125.157702} {\bibfield  {journal} {\bibinfo  {journal}
  {Phys. Rev. Lett.}\ }\textbf {\bibinfo {volume} {125}},\ \bibinfo {pages}
  {157702} (\bibinfo {year} {2020})}\BibitemShut {NoStop}%
\bibitem [{\citenamefont {Mross}\ \emph {et~al.}(2018)\citenamefont {Mross},
  \citenamefont {Oreg}, \citenamefont {Stern}, \citenamefont {Margalit},\ and\
  \citenamefont {Heiblum}}]{Mross2018}%
  \BibitemOpen
  \bibfield  {author} {\bibinfo {author} {\bibfnamefont {D.~F.}\ \bibnamefont
  {Mross}}, \bibinfo {author} {\bibfnamefont {Y.}~\bibnamefont {Oreg}},
  \bibinfo {author} {\bibfnamefont {A.}~\bibnamefont {Stern}}, \bibinfo
  {author} {\bibfnamefont {G.}~\bibnamefont {Margalit}}, \ and\ \bibinfo
  {author} {\bibfnamefont {M.}~\bibnamefont {Heiblum}},\ }\bibfield  {title}
  {\emph {\bibinfo {title} {Theory of disorder-induced half-integer thermal
  {H}all conductance}},\ }\href {\doibase 10.1103/PhysRevLett.121.026801}
  {\bibfield  {journal} {\bibinfo  {journal} {Phys. Rev. Lett.}\ }\textbf
  {\bibinfo {volume} {121}},\ \bibinfo {pages} {026801} (\bibinfo {year}
  {2018})}\BibitemShut {NoStop}%
\bibitem [{\citenamefont {Wang}\ \emph {et~al.}(2018)\citenamefont {Wang},
  \citenamefont {Vishwanath},\ and\ \citenamefont {Halperin}}]{Wang2018}%
  \BibitemOpen
  \bibfield  {author} {\bibinfo {author} {\bibfnamefont {C.}~\bibnamefont
  {Wang}}, \bibinfo {author} {\bibfnamefont {A.}~\bibnamefont {Vishwanath}}, \
  and\ \bibinfo {author} {\bibfnamefont {B.~I.}\ \bibnamefont {Halperin}},\
  }\bibfield  {title} {\emph {\bibinfo {title} {Topological order from disorder
  and the quantized {H}all thermal metal: Possible applications to the
  $\ensuremath{\nu}=5/2$ state}},\ }\href {\doibase 10.1103/PhysRevB.98.045112}
  {\bibfield  {journal} {\bibinfo  {journal} {Phys. Rev. B}\ }\textbf {\bibinfo
  {volume} {98}},\ \bibinfo {pages} {045112} (\bibinfo {year}
  {2018})}\BibitemShut {NoStop}%
\bibitem [{\citenamefont {Lian}\ and\ \citenamefont {Wang}(2018)}]{Biao2018}%
  \BibitemOpen
  \bibfield  {author} {\bibinfo {author} {\bibfnamefont {B.}~\bibnamefont
  {Lian}}\ and\ \bibinfo {author} {\bibfnamefont {J.}~\bibnamefont {Wang}},\
  }\bibfield  {title} {\emph {\bibinfo {title} {Theory of the disordered
  $\ensuremath{\nu}=\frac{5}{2}$ quantum thermal {H}all state: Emergent
  symmetry and phase diagram}},\ }\href {\doibase 10.1103/PhysRevB.97.165124}
  {\bibfield  {journal} {\bibinfo  {journal} {Phys. Rev. B}\ }\textbf {\bibinfo
  {volume} {97}},\ \bibinfo {pages} {165124} (\bibinfo {year}
  {2018})}\BibitemShut {NoStop}%
\bibitem [{\citenamefont {Zhu}\ \emph {et~al.}(2020)\citenamefont {Zhu},
  \citenamefont {Sheng},\ and\ \citenamefont {Yang}}]{Zhu2020}%
  \BibitemOpen
  \bibfield  {author} {\bibinfo {author} {\bibfnamefont {W.}~\bibnamefont
  {Zhu}}, \bibinfo {author} {\bibfnamefont {D.~N.}\ \bibnamefont {Sheng}}, \
  and\ \bibinfo {author} {\bibfnamefont {K.}~\bibnamefont {Yang}},\ }\bibfield
  {title} {\emph {\bibinfo {title} {{Topological Interface between {P}faffian
  and Anti-{P}faffian Order in $\ensuremath{\nu}=5/2$ Quantum {H}all
  Effect}}},\ }\href {\doibase 10.1103/PhysRevLett.125.146802} {\bibfield
  {journal} {\bibinfo  {journal} {Phys. Rev. Lett.}\ }\textbf {\bibinfo
  {volume} {125}},\ \bibinfo {pages} {146802} (\bibinfo {year}
  {2020})}\BibitemShut {NoStop}%
\bibitem [{\citenamefont {Hsin}\ \emph {et~al.}(2020)\citenamefont {Hsin},
  \citenamefont {Lin}, \citenamefont {Paquette},\ and\ \citenamefont
  {Wang}}]{Hsin2020}%
  \BibitemOpen
  \bibfield  {author} {\bibinfo {author} {\bibfnamefont {P.-S.}\ \bibnamefont
  {Hsin}}, \bibinfo {author} {\bibfnamefont {Y.-H.}\ \bibnamefont {Lin}},
  \bibinfo {author} {\bibfnamefont {N.~M.}\ \bibnamefont {Paquette}}, \ and\
  \bibinfo {author} {\bibfnamefont {J.}~\bibnamefont {Wang}},\ }\bibfield
  {title} {\emph {\bibinfo {title} {{Effective field theory for fractional
  quantum {H}all systems near $\ensuremath{\nu}=5/2$}}},\ }\href {\doibase
  10.1103/PhysRevResearch.2.043242} {\bibfield  {journal} {\bibinfo  {journal}
  {Phys. Rev. Res.}\ }\textbf {\bibinfo {volume} {2}},\ \bibinfo {pages}
  {043242} (\bibinfo {year} {2020})}\BibitemShut {NoStop}%
\bibitem [{\citenamefont {Fulga}\ \emph {et~al.}(2020)\citenamefont {Fulga},
  \citenamefont {Oreg}, \citenamefont {Mirlin}, \citenamefont {Stern},\ and\
  \citenamefont {Mross}}]{Fulga2020}%
  \BibitemOpen
  \bibfield  {author} {\bibinfo {author} {\bibfnamefont {I.~C.}\ \bibnamefont
  {Fulga}}, \bibinfo {author} {\bibfnamefont {Y.}~\bibnamefont {Oreg}},
  \bibinfo {author} {\bibfnamefont {A.~D.}\ \bibnamefont {Mirlin}}, \bibinfo
  {author} {\bibfnamefont {A.}~\bibnamefont {Stern}}, \ and\ \bibinfo {author}
  {\bibfnamefont {D.~F.}\ \bibnamefont {Mross}},\ }\bibfield  {title} {\emph
  {\bibinfo {title} {{Temperature Enhancement of Thermal {H}all Conductance
  Quantization}}},\ }\href {\doibase 10.1103/PhysRevLett.125.236802} {\bibfield
   {journal} {\bibinfo  {journal} {Phys. Rev. Lett.}\ }\textbf {\bibinfo
  {volume} {125}},\ \bibinfo {pages} {236802} (\bibinfo {year}
  {2020})}\BibitemShut {NoStop}%
\bibitem [{\citenamefont {Das}\ \emph {et~al.}(2022)\citenamefont {Das},
  \citenamefont {Das},\ and\ \citenamefont {Mandal}}]{Das2022Jun}%
  \BibitemOpen
  \bibfield  {author} {\bibinfo {author} {\bibfnamefont {S.}~\bibnamefont
  {Das}}, \bibinfo {author} {\bibfnamefont {S.}~\bibnamefont {Das}}, \ and\
  \bibinfo {author} {\bibfnamefont {S.~S.}\ \bibnamefont {Mandal}},\ }\bibfield
   {title} {\emph {\bibinfo {title} {{An Anomalous Reentrant 5/2 Quantum Hall
  Phase at Moderate Landau-Level-Mixing Strength}}},\ }\href {\doibase
  10.48550/arXiv.2206.04419} {\bibfield  {journal} {\bibinfo  {journal}
  {arXiv}\ } (\bibinfo {year} {2022}),\ 10.48550/arXiv.2206.04419},\ \Eprint
  {http://arxiv.org/abs/2206.04419} {2206.04419} \BibitemShut {NoStop}%
\bibitem [{\citenamefont {Giamarchi}\ and\ \citenamefont
  {Schulz}(1988)}]{Giamarchi1988Jan}%
  \BibitemOpen
  \bibfield  {author} {\bibinfo {author} {\bibfnamefont {T.}~\bibnamefont
  {Giamarchi}}\ and\ \bibinfo {author} {\bibfnamefont {H.~J.}\ \bibnamefont
  {Schulz}},\ }\bibfield  {title} {\emph {\bibinfo {title} {{Anderson
  localization and interactions in one-dimensional metals}}},\ }\href {\doibase
  10.1103/PhysRevB.37.325} {\bibfield  {journal} {\bibinfo  {journal} {Phys.
  Rev. B}\ }\textbf {\bibinfo {volume} {37}},\ \bibinfo {pages} {325} (\bibinfo
  {year} {1988})}\BibitemShut {NoStop}%
\bibitem [{\citenamefont {Gornyi}\ \emph {et~al.}(2007)\citenamefont {Gornyi},
  \citenamefont {Mirlin},\ and\ \citenamefont {Polyakov}}]{Gornyi2007Feb}%
  \BibitemOpen
  \bibfield  {author} {\bibinfo {author} {\bibfnamefont {I.~V.}\ \bibnamefont
  {Gornyi}}, \bibinfo {author} {\bibfnamefont {A.~D.}\ \bibnamefont {Mirlin}},
  \ and\ \bibinfo {author} {\bibfnamefont {D.~G.}\ \bibnamefont {Polyakov}},\
  }\bibfield  {title} {\emph {\bibinfo {title} {Electron transport in a
  disordered luttinger liquid}},\ }\href {\doibase 10.1103/PhysRevB.75.085421}
  {\bibfield  {journal} {\bibinfo  {journal} {Phys. Rev. B}\ }\textbf {\bibinfo
  {volume} {75}},\ \bibinfo {pages} {085421} (\bibinfo {year}
  {2007})}\BibitemShut {NoStop}%
\bibitem [{\citenamefont {Sp{\aa}nsl{\ifmmode\ddot{a}\else\"{a}\fi}tt}\ \emph
  {et~al.}(2023)\citenamefont {Sp{\aa}nsl{\ifmmode\ddot{a}\else\"{a}\fi}tt},
  \citenamefont {Stern},\ and\ \citenamefont {Mirlin}}]{Spanslatt2023Feb}%
  \BibitemOpen
  \bibfield  {author} {\bibinfo {author} {\bibfnamefont {C.}~\bibnamefont
  {Sp{\aa}nsl{\ifmmode\ddot{a}\else\"{a}\fi}tt}}, \bibinfo {author}
  {\bibfnamefont {A.}~\bibnamefont {Stern}}, \ and\ \bibinfo {author}
  {\bibfnamefont {A.~D.}\ \bibnamefont {Mirlin}},\ }\bibfield  {title} {\emph
  {\bibinfo {title} {{Transport Signatures of Fractional Quantum Hall Binding
  Transitions}}},\ }\href {\doibase 10.48550/arXiv.2302.05781} {\bibfield
  {journal} {\bibinfo  {journal} {arXiv}\ } (\bibinfo {year} {2023}),\
  10.48550/arXiv.2302.05781},\ \Eprint {http://arxiv.org/abs/2302.05781}
  {2302.05781} \BibitemShut {NoStop}%
\bibitem [{\citenamefont {Yutushui}\ \emph {et~al.}(2022)\citenamefont
  {Yutushui}, \citenamefont {Stern},\ and\ \citenamefont
  {Mross}}]{Yutushui2022Jan}%
  \BibitemOpen
  \bibfield  {author} {\bibinfo {author} {\bibfnamefont {M.}~\bibnamefont
  {Yutushui}}, \bibinfo {author} {\bibfnamefont {A.}~\bibnamefont {Stern}}, \
  and\ \bibinfo {author} {\bibfnamefont {D.~F.}\ \bibnamefont {Mross}},\
  }\bibfield  {title} {\emph {\bibinfo {title} {{Identifying the
  $\ensuremath{\nu}=\frac{5}{2}$ Topological Order through Charge Transport
  Measurements}}},\ }\href {\doibase 10.1103/PhysRevLett.128.016401} {\bibfield
   {journal} {\bibinfo  {journal} {Phys. Rev. Lett.}\ }\textbf {\bibinfo
  {volume} {128}},\ \bibinfo {pages} {016401} (\bibinfo {year}
  {2022})}\BibitemShut {NoStop}%
\bibitem [{\citenamefont {Park}\ \emph {et~al.}(2019)\citenamefont {Park},
  \citenamefont {Mirlin}, \citenamefont {Rosenow},\ and\ \citenamefont
  {Gefen}}]{Park2019}%
  \BibitemOpen
  \bibfield  {author} {\bibinfo {author} {\bibfnamefont {J.}~\bibnamefont
  {Park}}, \bibinfo {author} {\bibfnamefont {A.~D.}\ \bibnamefont {Mirlin}},
  \bibinfo {author} {\bibfnamefont {B.}~\bibnamefont {Rosenow}}, \ and\
  \bibinfo {author} {\bibfnamefont {Y.}~\bibnamefont {Gefen}},\ }\bibfield
  {title} {\emph {\bibinfo {title} {Noise on complex quantum {H}all edges:
  Chiral anomaly and heat diffusion}},\ }\href {\doibase
  10.1103/PhysRevB.99.161302} {\bibfield  {journal} {\bibinfo  {journal} {Phys.
  Rev. B}\ }\textbf {\bibinfo {volume} {99}},\ \bibinfo {pages} {161302}
  (\bibinfo {year} {2019})}\BibitemShut {NoStop}%
\bibitem [{\citenamefont {Sp\aa{}nsl\"att}\ \emph {et~al.}(2019)\citenamefont
  {Sp\aa{}nsl\"att}, \citenamefont {Park}, \citenamefont {Gefen},\ and\
  \citenamefont {Mirlin}}]{Spanslatt2019}%
  \BibitemOpen
  \bibfield  {author} {\bibinfo {author} {\bibfnamefont {C.}~\bibnamefont
  {Sp\aa{}nsl\"att}}, \bibinfo {author} {\bibfnamefont {J.}~\bibnamefont
  {Park}}, \bibinfo {author} {\bibfnamefont {Y.}~\bibnamefont {Gefen}}, \ and\
  \bibinfo {author} {\bibfnamefont {A.~D.}\ \bibnamefont {Mirlin}},\ }\bibfield
   {title} {\emph {\bibinfo {title} {Topological classification of shot noise
  on fractional quantum hall edges}},\ }\href {\doibase
  10.1103/PhysRevLett.123.137701} {\bibfield  {journal} {\bibinfo  {journal}
  {Phys. Rev. Lett.}\ }\textbf {\bibinfo {volume} {123}},\ \bibinfo {pages}
  {137701} (\bibinfo {year} {2019})}\BibitemShut {NoStop}%
\bibitem [{\citenamefont {Sp\aa{}nsl\"att}\ \emph {et~al.}(2020)\citenamefont
  {Sp\aa{}nsl\"att}, \citenamefont {Park}, \citenamefont {Gefen},\ and\
  \citenamefont {Mirlin}}]{Spanslatt2020}%
  \BibitemOpen
  \bibfield  {author} {\bibinfo {author} {\bibfnamefont {C.}~\bibnamefont
  {Sp\aa{}nsl\"att}}, \bibinfo {author} {\bibfnamefont {J.}~\bibnamefont
  {Park}}, \bibinfo {author} {\bibfnamefont {Y.}~\bibnamefont {Gefen}}, \ and\
  \bibinfo {author} {\bibfnamefont {A.~D.}\ \bibnamefont {Mirlin}},\ }\bibfield
   {title} {\emph {\bibinfo {title} {Conductance plateaus and shot noise in
  fractional quantum {H}all point contacts}},\ }\href {\doibase
  10.1103/PhysRevB.101.075308} {\bibfield  {journal} {\bibinfo  {journal}
  {Phys. Rev. B}\ }\textbf {\bibinfo {volume} {101}},\ \bibinfo {pages}
  {075308} (\bibinfo {year} {2020})}\BibitemShut {NoStop}%
\bibitem [{Com()}]{Comment}%
  \BibitemOpen
  \href@noop {} {}\bibinfo {note} {Note that when interfacing the $\nu=5/2$
  edge with $n=3$, the downstream and upstream directions are swapped. The sign
  of $\nu_Q$ is then adjusted so that $\nu_Q>0$ corresponds to the direction of
  the equilibrated charge flow.}\BibitemShut {Stop}%
\bibitem [{\citenamefont {Wan}\ \emph {et~al.}(2006)\citenamefont {Wan},
  \citenamefont {Yang},\ and\ \citenamefont {Rezayi}}]{Wan2006}%
  \BibitemOpen
  \bibfield  {author} {\bibinfo {author} {\bibfnamefont {X.}~\bibnamefont
  {Wan}}, \bibinfo {author} {\bibfnamefont {K.}~\bibnamefont {Yang}}, \ and\
  \bibinfo {author} {\bibfnamefont {E.~H.}\ \bibnamefont {Rezayi}},\ }\bibfield
   {title} {\emph {\bibinfo {title} {Edge excitations and non-{A}belian
  statistics in the moore-read state: A numerical study in the presence of
  coulomb interaction and edge confinement}},\ }\href {\doibase
  10.1103/PhysRevLett.97.256804} {\bibfield  {journal} {\bibinfo  {journal}
  {Phys. Rev. Lett.}\ }\textbf {\bibinfo {volume} {97}},\ \bibinfo {pages}
  {256804} (\bibinfo {year} {2006})}\BibitemShut {NoStop}%
\bibitem [{\citenamefont {Wan}\ \emph {et~al.}(2008)\citenamefont {Wan},
  \citenamefont {Hu}, \citenamefont {Rezayi},\ and\ \citenamefont
  {Yang}}]{Wan2008}%
  \BibitemOpen
  \bibfield  {author} {\bibinfo {author} {\bibfnamefont {X.}~\bibnamefont
  {Wan}}, \bibinfo {author} {\bibfnamefont {Z.-X.}\ \bibnamefont {Hu}},
  \bibinfo {author} {\bibfnamefont {E.~H.}\ \bibnamefont {Rezayi}}, \ and\
  \bibinfo {author} {\bibfnamefont {K.}~\bibnamefont {Yang}},\ }\bibfield
  {title} {\emph {\bibinfo {title} {Fractional quantum {H}all effect at
  $\nu=5/2$: Ground states, non-{A}belian quasiholes, and edge modes in a
  microscopic model}},\ }\href {\doibase 10.1103/PhysRevB.77.165316} {\bibfield
   {journal} {\bibinfo  {journal} {Phys. Rev. B}\ }\textbf {\bibinfo {volume}
  {77}},\ \bibinfo {pages} {165316} (\bibinfo {year} {2008})}\BibitemShut
  {NoStop}%
\bibitem [{\citenamefont {Overbosch}\ and\ \citenamefont
  {Wen}(2008)}]{Overbosch2008}%
  \BibitemOpen
  \bibfield  {author} {\bibinfo {author} {\bibfnamefont {B.}~\bibnamefont
  {Overbosch}}\ and\ \bibinfo {author} {\bibfnamefont {X.-G.}\ \bibnamefont
  {Wen}},\ }\bibfield  {title} {\emph {\bibinfo {title} {Phase transitions on
  the edge of the $\nu= 5/2$ {P}faffian and anti-{P}faffian quantum {H}all
  state}},\ }\href {https://arxiv.org/abs/0804.2087} {\bibfield  {journal}
  {\bibinfo  {journal} {arXiv preprint arXiv:0804.2087}\ } (\bibinfo {year}
  {2008})}\BibitemShut {NoStop}%
\bibitem [{\citenamefont {Zhang}\ \emph {et~al.}(2014)\citenamefont {Zhang},
  \citenamefont {Wu}, \citenamefont {Hutasoit},\ and\ \citenamefont
  {Jain}}]{Zhang2014}%
  \BibitemOpen
  \bibfield  {author} {\bibinfo {author} {\bibfnamefont {Y.}~\bibnamefont
  {Zhang}}, \bibinfo {author} {\bibfnamefont {Y.-H.}\ \bibnamefont {Wu}},
  \bibinfo {author} {\bibfnamefont {J.~A.}\ \bibnamefont {Hutasoit}}, \ and\
  \bibinfo {author} {\bibfnamefont {J.~K.}\ \bibnamefont {Jain}},\ }\bibfield
  {title} {\emph {\bibinfo {title} {Theoretical investigation of edge
  reconstruction in the $\nu=5/2$ and $7/3$ fractional quantum {H}all
  states}},\ }\href {\doibase 10.1103/PhysRevB.90.165104} {\bibfield  {journal}
  {\bibinfo  {journal} {Phys. Rev. B}\ }\textbf {\bibinfo {volume} {90}},\
  \bibinfo {pages} {165104} (\bibinfo {year} {2014})}\BibitemShut {NoStop}%
\bibitem [{\citenamefont {Manna}\ \emph {et~al.}(2022)\citenamefont {Manna},
  \citenamefont {Das}, \citenamefont {Goldstein},\ and\ \citenamefont
  {Gefen}}]{Manna2022Dec}%
  \BibitemOpen
  \bibfield  {author} {\bibinfo {author} {\bibfnamefont {S.}~\bibnamefont
  {Manna}}, \bibinfo {author} {\bibfnamefont {A.}~\bibnamefont {Das}}, \bibinfo
  {author} {\bibfnamefont {M.}~\bibnamefont {Goldstein}}, \ and\ \bibinfo
  {author} {\bibfnamefont {Y.}~\bibnamefont {Gefen}},\ }\bibfield  {title}
  {\emph {\bibinfo {title} {{Full Classification of Transport on an
  Equilibrated $5/2$ Edge}}},\ }\href {\doibase 10.48550/arXiv.2212.05732}
  {\bibfield  {journal} {\bibinfo  {journal} {arXiv}\ } (\bibinfo {year}
  {2022}),\ 10.48550/arXiv.2212.05732},\ \Eprint
  {http://arxiv.org/abs/2212.05732} {2212.05732} \BibitemShut {NoStop}%
\bibitem [{\citenamefont {Fujisawa}(2022)}]{Fujisawa2022Apr}%
  \BibitemOpen
  \bibfield  {author} {\bibinfo {author} {\bibfnamefont {T.}~\bibnamefont
  {Fujisawa}},\ }\bibfield  {title} {\emph {\bibinfo {title} {{Nonequilibrium
  Charge Dynamics of Tomonaga{\textendash}Luttinger Liquids in Quantum {H}all
  Edge Channels}}},\ }\href {\doibase 10.1002/andp.202100354} {\bibfield
  {journal} {\bibinfo  {journal} {Ann. Phys.}\ }\textbf {\bibinfo {volume}
  {534}},\ \bibinfo {pages} {2100354} (\bibinfo {year} {2022})}\BibitemShut
  {NoStop}%
\bibitem [{\citenamefont {Kumar}\ \emph {et~al.}(2022)\citenamefont {Kumar},
  \citenamefont {Srivastav}, \citenamefont
  {Sp{\aa}nsl{\ifmmode\ddot{a}\else\"{a}\fi}tt}, \citenamefont {Watanabe},
  \citenamefont {Taniguchi}, \citenamefont {Gefen}, \citenamefont {Mirlin},\
  and\ \citenamefont {Das}}]{Kumar2022Jan}%
  \BibitemOpen
  \bibfield  {author} {\bibinfo {author} {\bibfnamefont {R.}~\bibnamefont
  {Kumar}}, \bibinfo {author} {\bibfnamefont {S.~K.}\ \bibnamefont
  {Srivastav}}, \bibinfo {author} {\bibfnamefont {C.}~\bibnamefont
  {Sp{\aa}nsl{\ifmmode\ddot{a}\else\"{a}\fi}tt}}, \bibinfo {author}
  {\bibfnamefont {K.}~\bibnamefont {Watanabe}}, \bibinfo {author}
  {\bibfnamefont {T.}~\bibnamefont {Taniguchi}}, \bibinfo {author}
  {\bibfnamefont {Y.}~\bibnamefont {Gefen}}, \bibinfo {author} {\bibfnamefont
  {A.~D.}\ \bibnamefont {Mirlin}}, \ and\ \bibinfo {author} {\bibfnamefont
  {A.}~\bibnamefont {Das}},\ }\bibfield  {title} {\emph {\bibinfo {title}
  {{Observation of ballistic upstream modes at fractional quantum {H}all edges
  of graphene}}},\ }\href {\doibase 10.1038/s41467-021-27805-4} {\bibfield
  {journal} {\bibinfo  {journal} {Nat. Commun.}\ }\textbf {\bibinfo {volume}
  {13}},\ \bibinfo {pages} {1} (\bibinfo {year} {2022})}\BibitemShut {NoStop}%
\bibitem [{\citenamefont {Hein}(2022)}]{Hein2022}%
  \BibitemOpen
  \bibfield  {author} {\bibinfo {author} {\bibfnamefont {M.}~\bibnamefont
  {Hein}},\ }\emph {\bibinfo {title} {Impact of Equilibration on the Heat
  Conductance and Noise of non-{A}belian fractional Quantum {H}all Edges}},\
  \href {https://odr.chalmers.se/handle/20.500.12380/305490} {Master's
  thesis},\ \bibinfo  {school} {Chalmers University of Technology} (\bibinfo
  {year} {2022}),\ \bibinfo {note} {online; Accessed \today}\BibitemShut
  {NoStop}%
\bibitem [{\citenamefont {Sp{\aa}nsl{\ifmmode\ddot{a}\else\"{a}\fi}tt}\ \emph
  {et~al.}(2021)\citenamefont {Sp{\aa}nsl{\ifmmode\ddot{a}\else\"{a}\fi}tt},
  \citenamefont {Gefen}, \citenamefont {Gornyi},\ and\ \citenamefont
  {Polyakov}}]{Spanslatt2021Sep}%
  \BibitemOpen
  \bibfield  {author} {\bibinfo {author} {\bibfnamefont {C.}~\bibnamefont
  {Sp{\aa}nsl{\ifmmode\ddot{a}\else\"{a}\fi}tt}}, \bibinfo {author}
  {\bibfnamefont {Y.}~\bibnamefont {Gefen}}, \bibinfo {author} {\bibfnamefont
  {I.~V.}\ \bibnamefont {Gornyi}}, \ and\ \bibinfo {author} {\bibfnamefont
  {D.~G.}\ \bibnamefont {Polyakov}},\ }\bibfield  {title} {\emph {\bibinfo
  {title} {{Contacts, equilibration, and interactions in fractional quantum
  {H}all edge transport}}},\ }\href {\doibase 10.1103/PhysRevB.104.115416}
  {\bibfield  {journal} {\bibinfo  {journal} {Phys. Rev. B}\ }\textbf {\bibinfo
  {volume} {104}},\ \bibinfo {pages} {115416} (\bibinfo {year}
  {2021})}\BibitemShut {NoStop}%
\bibitem [{\citenamefont {Nyquist}(1928)}]{Nyquist1928Jul}%
  \BibitemOpen
  \bibfield  {author} {\bibinfo {author} {\bibfnamefont {H.}~\bibnamefont
  {Nyquist}},\ }\bibfield  {title} {\emph {\bibinfo {title} {{Thermal Agitation
  of Electric Charge in Conductors}}},\ }\href {\doibase
  10.1103/PhysRev.32.110} {\bibfield  {journal} {\bibinfo  {journal} {Phys.
  Rev.}\ }\textbf {\bibinfo {volume} {32}},\ \bibinfo {pages} {110} (\bibinfo
  {year} {1928})}\BibitemShut {NoStop}%
\bibitem [{\citenamefont {Johnson}(1928)}]{Johnson1928Jul}%
  \BibitemOpen
  \bibfield  {author} {\bibinfo {author} {\bibfnamefont {J.~B.}\ \bibnamefont
  {Johnson}},\ }\bibfield  {title} {\emph {\bibinfo {title} {{Thermal Agitation
  of Electricity in Conductors}}},\ }\href {\doibase 10.1103/PhysRev.32.97}
  {\bibfield  {journal} {\bibinfo  {journal} {Phys. Rev.}\ }\textbf {\bibinfo
  {volume} {32}},\ \bibinfo {pages} {97} (\bibinfo {year} {1928})}\BibitemShut
  {NoStop}%
\bibitem [{\citenamefont {Lumbroso}\ \emph {et~al.}(2018)\citenamefont
  {Lumbroso}, \citenamefont {Simine}, \citenamefont {Nitzan}, \citenamefont
  {Segal},\ and\ \citenamefont {Tal}}]{Lumbroso2018}%
  \BibitemOpen
  \bibfield  {author} {\bibinfo {author} {\bibfnamefont {O.~S.}\ \bibnamefont
  {Lumbroso}}, \bibinfo {author} {\bibfnamefont {L.}~\bibnamefont {Simine}},
  \bibinfo {author} {\bibfnamefont {A.}~\bibnamefont {Nitzan}}, \bibinfo
  {author} {\bibfnamefont {D.}~\bibnamefont {Segal}}, \ and\ \bibinfo {author}
  {\bibfnamefont {O.}~\bibnamefont {Tal}},\ }\bibfield  {title} {\emph
  {\bibinfo {title} {Electronic noise due to temperature differences in
  atomic-scale junctions}},\ }\href {\doibase 10.1038/s41586-018-0592-2}
  {\bibfield  {journal} {\bibinfo  {journal} {Nature}\ }\textbf {\bibinfo
  {volume} {562}},\ \bibinfo {pages} {240} (\bibinfo {year}
  {2018})}\BibitemShut {NoStop}%
\bibitem [{\citenamefont {Rech}\ \emph {et~al.}(2020)\citenamefont {Rech},
  \citenamefont {Jonckheere}, \citenamefont {Gr\'emaud},\ and\ \citenamefont
  {Martin}}]{RechMartinPRL20}%
  \BibitemOpen
  \bibfield  {author} {\bibinfo {author} {\bibfnamefont {J.}~\bibnamefont
  {Rech}}, \bibinfo {author} {\bibfnamefont {T.}~\bibnamefont {Jonckheere}},
  \bibinfo {author} {\bibfnamefont {B.}~\bibnamefont {Gr\'emaud}}, \ and\
  \bibinfo {author} {\bibfnamefont {T.}~\bibnamefont {Martin}},\ }\bibfield
  {title} {\emph {\bibinfo {title} {Negative delta-{$T$} noise in the
  fractional quantum {H}all effect}},\ }\href {\doibase
  10.1103/PhysRevLett.125.086801} {\bibfield  {journal} {\bibinfo  {journal}
  {Phys. Rev. Lett.}\ }\textbf {\bibinfo {volume} {125}},\ \bibinfo {pages}
  {086801} (\bibinfo {year} {2020})}\BibitemShut {NoStop}%
\bibitem [{\citenamefont {Schiller}\ \emph {et~al.}(2022)\citenamefont
  {Schiller}, \citenamefont {Oreg},\ and\ \citenamefont
  {Snizhko}}]{Schiller2022}%
  \BibitemOpen
  \bibfield  {author} {\bibinfo {author} {\bibfnamefont {N.}~\bibnamefont
  {Schiller}}, \bibinfo {author} {\bibfnamefont {Y.}~\bibnamefont {Oreg}}, \
  and\ \bibinfo {author} {\bibfnamefont {K.}~\bibnamefont {Snizhko}},\
  }\bibfield  {title} {\emph {\bibinfo {title} {Extracting the scaling
  dimension of quantum {H}all quasiparticles from current correlations}},\
  }\href {\doibase 10.1103/PhysRevB.105.165150} {\bibfield  {journal} {\bibinfo
   {journal} {Phys. Rev. B}\ }\textbf {\bibinfo {volume} {105}},\ \bibinfo
  {pages} {165150} (\bibinfo {year} {2022})}\BibitemShut {NoStop}%
\bibitem [{\citenamefont {Zhang}\ \emph {et~al.}(2022)\citenamefont {Zhang},
  \citenamefont {Gornyi},\ and\ \citenamefont {Sp\aa{}nsl\"att}}]{Zhang2022}%
  \BibitemOpen
  \bibfield  {author} {\bibinfo {author} {\bibfnamefont {G.}~\bibnamefont
  {Zhang}}, \bibinfo {author} {\bibfnamefont {I.~V.}\ \bibnamefont {Gornyi}}, \
  and\ \bibinfo {author} {\bibfnamefont {C.}~\bibnamefont {Sp\aa{}nsl\"att}},\
  }\bibfield  {title} {\emph {\bibinfo {title} {Delta-{$T$} noise for weak
  tunneling in one-dimensional systems: Interactions versus quantum
  statistics}},\ }\href {\doibase 10.1103/PhysRevB.105.195423} {\bibfield
  {journal} {\bibinfo  {journal} {Phys. Rev. B}\ }\textbf {\bibinfo {volume}
  {105}},\ \bibinfo {pages} {195423} (\bibinfo {year} {2022})}\BibitemShut
  {NoStop}%
\bibitem [{\citenamefont {Rebora}\ \emph {et~al.}(2022)\citenamefont {Rebora},
  \citenamefont {Rech}, \citenamefont {Ferraro}, \citenamefont {Jonckheere},
  \citenamefont {Martin},\ and\ \citenamefont {Sassetti}}]{Rebora2022Dec}%
  \BibitemOpen
  \bibfield  {author} {\bibinfo {author} {\bibfnamefont {G.}~\bibnamefont
  {Rebora}}, \bibinfo {author} {\bibfnamefont {J.}~\bibnamefont {Rech}},
  \bibinfo {author} {\bibfnamefont {D.}~\bibnamefont {Ferraro}}, \bibinfo
  {author} {\bibfnamefont {T.}~\bibnamefont {Jonckheere}}, \bibinfo {author}
  {\bibfnamefont {T.}~\bibnamefont {Martin}}, \ and\ \bibinfo {author}
  {\bibfnamefont {M.}~\bibnamefont {Sassetti}},\ }\bibfield  {title} {\emph
  {\bibinfo {title} {{Delta-$T$ noise for fractional quantum Hall states at
  different filling factor}}},\ }\href {\doibase
  10.1103/PhysRevResearch.4.043191} {\bibfield  {journal} {\bibinfo  {journal}
  {Phys. Rev. Res.}\ }\textbf {\bibinfo {volume} {4}},\ \bibinfo {pages}
  {043191} (\bibinfo {year} {2022})}\BibitemShut {NoStop}%
\bibitem [{\citenamefont {Kane}\ \emph {et~al.}(1994)\citenamefont {Kane},
  \citenamefont {Fisher},\ and\ \citenamefont {Polchinski}}]{Kane1994}%
  \BibitemOpen
  \bibfield  {author} {\bibinfo {author} {\bibfnamefont {C.~L.}\ \bibnamefont
  {Kane}}, \bibinfo {author} {\bibfnamefont {M.~P.~A.}\ \bibnamefont {Fisher}},
  \ and\ \bibinfo {author} {\bibfnamefont {J.}~\bibnamefont {Polchinski}},\
  }\bibfield  {title} {\emph {\bibinfo {title} {Randomness at the edge: Theory
  of quantum {H}all transport at filling \ensuremath{\nu}=2/3}},\ }\href
  {\doibase 10.1103/PhysRevLett.72.4129} {\bibfield  {journal} {\bibinfo
  {journal} {Phys. Rev. Lett.}\ }\textbf {\bibinfo {volume} {72}},\ \bibinfo
  {pages} {4129} (\bibinfo {year} {1994})}\BibitemShut {NoStop}%
\bibitem [{\citenamefont {Kane}\ and\ \citenamefont
  {Fisher}(1995{\natexlab{b}})}]{Kane1995b}%
  \BibitemOpen
  \bibfield  {author} {\bibinfo {author} {\bibfnamefont {C.~L.}\ \bibnamefont
  {Kane}}\ and\ \bibinfo {author} {\bibfnamefont {M.~P.~A.}\ \bibnamefont
  {Fisher}},\ }\bibfield  {title} {\emph {\bibinfo {title} {Impurity scattering
  and transport of fractional quantum {H}all edge states}},\ }\href {\doibase
  10.1103/PhysRevB.51.13449} {\bibfield  {journal} {\bibinfo  {journal} {Phys.
  Rev. B}\ }\textbf {\bibinfo {volume} {51}},\ \bibinfo {pages} {13449}
  (\bibinfo {year} {1995}{\natexlab{b}})}\BibitemShut {NoStop}%
\bibitem [{\citenamefont {Moore}\ and\ \citenamefont {Wen}(1998)}]{Moore1998}%
  \BibitemOpen
  \bibfield  {author} {\bibinfo {author} {\bibfnamefont {J.~E.}\ \bibnamefont
  {Moore}}\ and\ \bibinfo {author} {\bibfnamefont {X.-G.}\ \bibnamefont
  {Wen}},\ }\bibfield  {title} {\emph {\bibinfo {title} {Classification of
  disordered phases of quantum {H}all edge states}},\ }\href {\doibase
  10.1103/PhysRevB.57.10138} {\bibfield  {journal} {\bibinfo  {journal} {Phys.
  Rev. B}\ }\textbf {\bibinfo {volume} {57}},\ \bibinfo {pages} {10138}
  (\bibinfo {year} {1998})}\BibitemShut {NoStop}%
\bibitem [{\citenamefont {Safi}\ and\ \citenamefont {Schulz}(1995)}]{Safi1995}%
  \BibitemOpen
  \bibfield  {author} {\bibinfo {author} {\bibfnamefont {I.}~\bibnamefont
  {Safi}}\ and\ \bibinfo {author} {\bibfnamefont {H.~J.}\ \bibnamefont
  {Schulz}},\ }\bibfield  {title} {\emph {\bibinfo {title} {Transport in an
  inhomogeneous interacting one-dimensional system}},\ }\href {\doibase
  10.1103/PhysRevB.52.R17040} {\bibfield  {journal} {\bibinfo  {journal} {Phys.
  Rev. B}\ }\textbf {\bibinfo {volume} {52}},\ \bibinfo {pages} {R17040}
  (\bibinfo {year} {1995})}\BibitemShut {NoStop}%
\bibitem [{\citenamefont {Krive}(1998)}]{Krive1998}%
  \BibitemOpen
  \bibfield  {author} {\bibinfo {author} {\bibfnamefont {I.~V.}\ \bibnamefont
  {Krive}},\ }\bibfield  {title} {\emph {\bibinfo {title} {Thermal transport
  through {L}uttinger liquid constriction}},\ }\href {\doibase
  10.1063/1.593605} {\bibfield  {journal} {\bibinfo  {journal} {Low Temperature
  Physics}\ }\textbf {\bibinfo {volume} {24}},\ \bibinfo {pages} {377}
  (\bibinfo {year} {1998})}\BibitemShut {NoStop}%
\bibitem [{\citenamefont {Ki}\ \emph {et~al.}(2014)\citenamefont {Ki},
  \citenamefont {Fal{'}ko}, \citenamefont {Abanin},\ and\ \citenamefont
  {Morpurgo}}]{Ki2014Apr}%
  \BibitemOpen
  \bibfield  {author} {\bibinfo {author} {\bibfnamefont {D.-K.}\ \bibnamefont
  {Ki}}, \bibinfo {author} {\bibfnamefont {V.~I.}\ \bibnamefont {Fal{'}ko}},
  \bibinfo {author} {\bibfnamefont {D.~A.}\ \bibnamefont {Abanin}}, \ and\
  \bibinfo {author} {\bibfnamefont {A.~F.}\ \bibnamefont {Morpurgo}},\
  }\bibfield  {title} {\emph {\bibinfo {title} {Observation of even denominator
  fractional quantum {H}all effect in suspended bilayer graphene}},\ }\href
  {\doibase 10.1021/nl5003922} {\bibfield  {journal} {\bibinfo  {journal} {Nano
  Lett.}\ }\textbf {\bibinfo {volume} {14}},\ \bibinfo {pages} {2135} (\bibinfo
  {year} {2014})}\BibitemShut {NoStop}%
\bibitem [{\citenamefont {Kim}\ \emph {et~al.}(2015)\citenamefont {Kim},
  \citenamefont {Lee}, \citenamefont {Jung}, \citenamefont
  {Sk{\ifmmode\acute{a}\else\'{a}\fi}kalov{\ifmmode\acute{a}\else\'{a}\fi}},
  \citenamefont {Taniguchi}, \citenamefont {Watanabe}, \citenamefont {Kim},\
  and\ \citenamefont {Smet}}]{Kim2015Nov}%
  \BibitemOpen
  \bibfield  {author} {\bibinfo {author} {\bibfnamefont {Y.}~\bibnamefont
  {Kim}}, \bibinfo {author} {\bibfnamefont {D.~S.}\ \bibnamefont {Lee}},
  \bibinfo {author} {\bibfnamefont {S.}~\bibnamefont {Jung}}, \bibinfo {author}
  {\bibfnamefont {V.}~\bibnamefont
  {Sk{\ifmmode\acute{a}\else\'{a}\fi}kalov{\ifmmode\acute{a}\else\'{a}\fi}}},
  \bibinfo {author} {\bibfnamefont {T.}~\bibnamefont {Taniguchi}}, \bibinfo
  {author} {\bibfnamefont {K.}~\bibnamefont {Watanabe}}, \bibinfo {author}
  {\bibfnamefont {J.~S.}\ \bibnamefont {Kim}}, \ and\ \bibinfo {author}
  {\bibfnamefont {J.~H.}\ \bibnamefont {Smet}},\ }\bibfield  {title} {\emph
  {\bibinfo {title} {{Fractional Quantum {H}all States in Bilayer Graphene
  Probed by Transconductance Fluctuations}}},\ }\href {\doibase
  10.1021/acs.nanolett.5b02876} {\bibfield  {journal} {\bibinfo  {journal}
  {Nano Lett.}\ }\textbf {\bibinfo {volume} {15}},\ \bibinfo {pages} {7445}
  (\bibinfo {year} {2015})}\BibitemShut {NoStop}%
\bibitem [{\citenamefont {Zibrov}\ \emph {et~al.}(2017)\citenamefont {Zibrov},
  \citenamefont {Kometter}, \citenamefont {Zhou}, \citenamefont {Spanton},
  \citenamefont {Taniguchi}, \citenamefont {Watanabe}, \citenamefont
  {Zaletel},\ and\ \citenamefont {Young}}]{Zibrov2017Sep}%
  \BibitemOpen
  \bibfield  {author} {\bibinfo {author} {\bibfnamefont {A.~A.}\ \bibnamefont
  {Zibrov}}, \bibinfo {author} {\bibfnamefont {C.}~\bibnamefont {Kometter}},
  \bibinfo {author} {\bibfnamefont {H.}~\bibnamefont {Zhou}}, \bibinfo {author}
  {\bibfnamefont {E.~M.}\ \bibnamefont {Spanton}}, \bibinfo {author}
  {\bibfnamefont {T.}~\bibnamefont {Taniguchi}}, \bibinfo {author}
  {\bibfnamefont {K.}~\bibnamefont {Watanabe}}, \bibinfo {author}
  {\bibfnamefont {M.~P.}\ \bibnamefont {Zaletel}}, \ and\ \bibinfo {author}
  {\bibfnamefont {A.~F.}\ \bibnamefont {Young}},\ }\bibfield  {title} {\emph
  {\bibinfo {title} {{Tunable interacting composite fermion phases in a
  half-filled bilayer-graphene {L}andau level}}},\ }\href {\doibase
  10.1038/nature23893} {\bibfield  {journal} {\bibinfo  {journal} {Nature}\
  }\textbf {\bibinfo {volume} {549}},\ \bibinfo {pages} {360} (\bibinfo {year}
  {2017})}\BibitemShut {NoStop}%
\bibitem [{\citenamefont {Zibrov}\ \emph {et~al.}(2018)\citenamefont {Zibrov},
  \citenamefont {Spanton}, \citenamefont {Zhou}, \citenamefont {Kometter},
  \citenamefont {Taniguchi}, \citenamefont {Watanabe},\ and\ \citenamefont
  {Young}}]{Zibrov2018Sep}%
  \BibitemOpen
  \bibfield  {author} {\bibinfo {author} {\bibfnamefont {A.~A.}\ \bibnamefont
  {Zibrov}}, \bibinfo {author} {\bibfnamefont {E.~M.}\ \bibnamefont {Spanton}},
  \bibinfo {author} {\bibfnamefont {H.}~\bibnamefont {Zhou}}, \bibinfo {author}
  {\bibfnamefont {C.}~\bibnamefont {Kometter}}, \bibinfo {author}
  {\bibfnamefont {T.}~\bibnamefont {Taniguchi}}, \bibinfo {author}
  {\bibfnamefont {K.}~\bibnamefont {Watanabe}}, \ and\ \bibinfo {author}
  {\bibfnamefont {A.~F.}\ \bibnamefont {Young}},\ }\bibfield  {title} {\emph
  {\bibinfo {title} {{Even-denominator fractional quantum {H}all states at an
  isospin transition in monolayer graphene}}},\ }\href {\doibase
  10.1038/s41567-018-0190-0} {\bibfield  {journal} {\bibinfo  {journal} {Nat.
  Phys.}\ }\textbf {\bibinfo {volume} {14}},\ \bibinfo {pages} {930} (\bibinfo
  {year} {2018})}\BibitemShut {NoStop}%
\bibitem [{\citenamefont {Li}\ \emph {et~al.}(2017)\citenamefont {Li},
  \citenamefont {Tan}, \citenamefont {Chen}, \citenamefont {Zeng},
  \citenamefont {Taniguchi}, \citenamefont {Watanabe}, \citenamefont {Hone},\
  and\ \citenamefont {Dean}}]{Li2017}%
  \BibitemOpen
  \bibfield  {author} {\bibinfo {author} {\bibfnamefont {J.~I.~A.}\
  \bibnamefont {Li}}, \bibinfo {author} {\bibfnamefont {C.}~\bibnamefont
  {Tan}}, \bibinfo {author} {\bibfnamefont {S.}~\bibnamefont {Chen}}, \bibinfo
  {author} {\bibfnamefont {Y.}~\bibnamefont {Zeng}}, \bibinfo {author}
  {\bibfnamefont {T.}~\bibnamefont {Taniguchi}}, \bibinfo {author}
  {\bibfnamefont {K.}~\bibnamefont {Watanabe}}, \bibinfo {author}
  {\bibfnamefont {J.}~\bibnamefont {Hone}}, \ and\ \bibinfo {author}
  {\bibfnamefont {C.~R.}\ \bibnamefont {Dean}},\ }\bibfield  {title} {\emph
  {\bibinfo {title} {Even-denominator fractional quantum {H}all states in
  bilayer graphene}},\ }\href {\doibase 10.1126/science.aao2521} {\bibfield
  {journal} {\bibinfo  {journal} {Science}\ }\textbf {\bibinfo {volume}
  {358}},\ \bibinfo {pages} {648} (\bibinfo {year} {2017})}\BibitemShut
  {NoStop}%
\bibitem [{\citenamefont {Kim}\ \emph {et~al.}(2019)\citenamefont {Kim},
  \citenamefont {Balram}, \citenamefont {Taniguchi}, \citenamefont {Watanabe},
  \citenamefont {Jain},\ and\ \citenamefont {Smet}}]{Kim2019Feb}%
  \BibitemOpen
  \bibfield  {author} {\bibinfo {author} {\bibfnamefont {Y.}~\bibnamefont
  {Kim}}, \bibinfo {author} {\bibfnamefont {A.~C.}\ \bibnamefont {Balram}},
  \bibinfo {author} {\bibfnamefont {T.}~\bibnamefont {Taniguchi}}, \bibinfo
  {author} {\bibfnamefont {K.}~\bibnamefont {Watanabe}}, \bibinfo {author}
  {\bibfnamefont {J.~K.}\ \bibnamefont {Jain}}, \ and\ \bibinfo {author}
  {\bibfnamefont {J.~H.}\ \bibnamefont {Smet}},\ }\bibfield  {title} {\emph
  {\bibinfo {title} {Even denominator fractional quantum {H}all states in
  higher {L}andau levels of graphene}},\ }\href {\doibase
  10.1038/s41567-018-0355-x} {\bibfield  {journal} {\bibinfo  {journal} {Nature
  Physics}\ }\textbf {\bibinfo {volume} {15}},\ \bibinfo {pages} {154}
  (\bibinfo {year} {2019})}\BibitemShut {NoStop}%
\bibitem [{\citenamefont {Huang}\ \emph {et~al.}(2022)\citenamefont {Huang},
  \citenamefont {Fu}, \citenamefont {Hickey}, \citenamefont {Alem},
  \citenamefont {Lin}, \citenamefont {Watanabe}, \citenamefont {Taniguchi},\
  and\ \citenamefont {Zhu}}]{Huang2022Jul}%
  \BibitemOpen
  \bibfield  {author} {\bibinfo {author} {\bibfnamefont {K.}~\bibnamefont
  {Huang}}, \bibinfo {author} {\bibfnamefont {H.}~\bibnamefont {Fu}}, \bibinfo
  {author} {\bibfnamefont {D.~R.}\ \bibnamefont {Hickey}}, \bibinfo {author}
  {\bibfnamefont {N.}~\bibnamefont {Alem}}, \bibinfo {author} {\bibfnamefont
  {X.}~\bibnamefont {Lin}}, \bibinfo {author} {\bibfnamefont {K.}~\bibnamefont
  {Watanabe}}, \bibinfo {author} {\bibfnamefont {T.}~\bibnamefont {Taniguchi}},
  \ and\ \bibinfo {author} {\bibfnamefont {J.}~\bibnamefont {Zhu}},\ }\bibfield
   {title} {\emph {\bibinfo {title} {Valley isospin controlled fractional
  quantum {H}all states in bilayer graphene}},\ }\href {\doibase
  10.1103/PhysRevX.12.031019} {\bibfield  {journal} {\bibinfo  {journal} {Phys.
  Rev. X}\ }\textbf {\bibinfo {volume} {12}},\ \bibinfo {pages} {031019}
  (\bibinfo {year} {2022})}\BibitemShut {NoStop}%
\bibitem [{\citenamefont {Read}\ and\ \citenamefont {Rezayi}(1999)}]{Read1999}%
  \BibitemOpen
  \bibfield  {author} {\bibinfo {author} {\bibfnamefont {N.}~\bibnamefont
  {Read}}\ and\ \bibinfo {author} {\bibfnamefont {E.}~\bibnamefont {Rezayi}},\
  }\bibfield  {title} {\emph {\bibinfo {title} {Beyond paired quantum {H}all
  states: Parafermions and incompressible states in the first excited {L}andau
  level}},\ }\href {\doibase 10.1103/PhysRevB.59.8084} {\bibfield  {journal}
  {\bibinfo  {journal} {Phys. Rev. B}\ }\textbf {\bibinfo {volume} {59}},\
  \bibinfo {pages} {8084} (\bibinfo {year} {1999})}\BibitemShut {NoStop}%
\bibitem [{\citenamefont {Xia}\ \emph {et~al.}(2004)\citenamefont {Xia},
  \citenamefont {Pan}, \citenamefont {Vicente}, \citenamefont {Adams},
  \citenamefont {Sullivan}, \citenamefont {Stormer}, \citenamefont {Tsui},
  \citenamefont {Pfeiffer}, \citenamefont {Baldwin},\ and\ \citenamefont
  {West}}]{Xia2004}%
  \BibitemOpen
  \bibfield  {author} {\bibinfo {author} {\bibfnamefont {J.~S.}\ \bibnamefont
  {Xia}}, \bibinfo {author} {\bibfnamefont {W.}~\bibnamefont {Pan}}, \bibinfo
  {author} {\bibfnamefont {C.~L.}\ \bibnamefont {Vicente}}, \bibinfo {author}
  {\bibfnamefont {E.~D.}\ \bibnamefont {Adams}}, \bibinfo {author}
  {\bibfnamefont {N.~S.}\ \bibnamefont {Sullivan}}, \bibinfo {author}
  {\bibfnamefont {H.~L.}\ \bibnamefont {Stormer}}, \bibinfo {author}
  {\bibfnamefont {D.~C.}\ \bibnamefont {Tsui}}, \bibinfo {author}
  {\bibfnamefont {L.~N.}\ \bibnamefont {Pfeiffer}}, \bibinfo {author}
  {\bibfnamefont {K.~W.}\ \bibnamefont {Baldwin}}, \ and\ \bibinfo {author}
  {\bibfnamefont {K.~W.}\ \bibnamefont {West}},\ }\bibfield  {title} {\emph
  {\bibinfo {title} {Electron correlation in the second {L}andau level: A
  competition between many nearly degenerate quantum phases}},\ }\href
  {\doibase 10.1103/PhysRevLett.93.176809} {\bibfield  {journal} {\bibinfo
  {journal} {Phys. Rev. Lett.}\ }\textbf {\bibinfo {volume} {93}},\ \bibinfo
  {pages} {176809} (\bibinfo {year} {2004})}\BibitemShut {NoStop}%
\bibitem [{\citenamefont {Kumar}\ \emph {et~al.}(2010)\citenamefont {Kumar},
  \citenamefont {Cs\'athy}, \citenamefont {Manfra}, \citenamefont {Pfeiffer},\
  and\ \citenamefont {West}}]{Kumar2010}%
  \BibitemOpen
  \bibfield  {author} {\bibinfo {author} {\bibfnamefont {A.}~\bibnamefont
  {Kumar}}, \bibinfo {author} {\bibfnamefont {G.~A.}\ \bibnamefont {Cs\'athy}},
  \bibinfo {author} {\bibfnamefont {M.~J.}\ \bibnamefont {Manfra}}, \bibinfo
  {author} {\bibfnamefont {L.~N.}\ \bibnamefont {Pfeiffer}}, \ and\ \bibinfo
  {author} {\bibfnamefont {K.~W.}\ \bibnamefont {West}},\ }\bibfield  {title}
  {\emph {\bibinfo {title} {Nonconventional odd-denominator fractional quantum
  {H}all states in the second {L}andau level}},\ }\href {\doibase
  10.1103/PhysRevLett.105.246808} {\bibfield  {journal} {\bibinfo  {journal}
  {Phys. Rev. Lett.}\ }\textbf {\bibinfo {volume} {105}},\ \bibinfo {pages}
  {246808} (\bibinfo {year} {2010})}\BibitemShut {NoStop}%
\bibitem [{\citenamefont {Martin}(2005)}]{Martin2005}%
  \BibitemOpen
  \bibfield  {author} {\bibinfo {author} {\bibfnamefont {T.}~\bibnamefont
  {Martin}},\ }\bibfield  {title} {\emph {\bibinfo {title} {Noise in mesoscopic
  physics}},\ }in\ \href@noop {} {\emph {\bibinfo {booktitle} {Proceedings of
  the Les Houches Summer School, Session LXXXI}}},\ \bibinfo {editor} {edited
  by\ \bibinfo {editor} {\bibfnamefont {H.}~\bibnamefont {Bouchiat}}, \bibinfo
  {editor} {\bibfnamefont {Y.}~\bibnamefont {Gefen}}, \bibinfo {editor}
  {\bibfnamefont {S.}~\bibnamefont {Gu\'eron}}, \bibinfo {editor} {\bibfnamefont
  {G.}~\bibnamefont {Montambaux}}, \ and\ \bibinfo {editor} {\bibfnamefont
  {J.}~\bibnamefont {Dalibard}}}\ (\bibinfo  {publisher} {Elsevier},\ \bibinfo
  {address} {New York},\ \bibinfo {year} {2005})\BibitemShut {NoStop}%
\bibitem [{\citenamefont {Turner}(2013)}]{TURNER201336}%
  \BibitemOpen
  \bibfield  {author} {\bibinfo {author} {\bibfnamefont {L.~E.}\ \bibnamefont
  {Turner}},\ }\bibfield  {title} {\emph {\bibinfo {title} {The
  {M}ittag-{L}effler theorem: The origin, evolution, and reception of a
  mathematical result, 1876--1884}},\ }\href {\doibase
  https://doi.org/10.1016/j.hm.2012.10.002} {\bibfield  {journal} {\bibinfo
  {journal} {Historia Mathematica}\ }\textbf {\bibinfo {volume} {40}},\
  \bibinfo {pages} {36} (\bibinfo {year} {2013})}\BibitemShut {NoStop}%
\end{thebibliography}
%

\end{document}